%% file: data_paper_r3.tex
\RequirePackage{rotating}
\documentclass[useAMS,usenatbib,A4]{mn2e}
\usepackage{epsfig,aas_macros}
\usepackage{subfig}
\usepackage{amssymb}
\usepackage{latexsym}
\usepackage{amsmath}
\usepackage{placeins}
\usepackage{rotating}
\usepackage{enumerate}
\usepackage{flafter}
\title[Ionised Gas in Galaxy cluster Cores 1: The Sample]{Optical Emission Line Nebulae in Galaxy Cluster Cores 1: The Morphological, Kinematic and Spectral Properties of the Sample}

\author[Hamer et al.]
{\parbox[h]{\textwidth}
{S. L. Hamer,$^{1}$\thanks{E-mail: stephen.hamer.astro@gmail.com} A. C. Edge,$^{3}$ A. M. Swinbank,$^{3}$ R. J. Wilman,$^{4}$ F. Combes,$^{2}$ P. Salom\'e,$^{2}$ A. C. Fabian,$^{5}$ C. S. Crawford,$^{5}$ H. R. Russell,$^{5}$ J. Hlavacek-Larrondo,$^{6,7,8}$ B. McNamara,$^{9}$ M. N. Bremer,$^{10}$} \\
\vspace*{6pt}\\
$^{1}$CRAL, Observatoire de Lyon, CNRS, Universit\'e Lyon 1, 9 Avenue Charles Andr\'e, 69561 Saint Genis-Laval, France\\
$^{2}$LERMA Observatoire de Paris, CNRS, 61 rue de l'Observatoire, 75014 Paris, France\\
$^{3}$Institute for Computational Cosmology, Department of Physics, Durham University, South Road, Durham DH1 3LE, UK\\
$^{4}$Department of Physics, Durham University, South Road, Durham DH1 3LE, UK\\
$^{5}$Institute of Astronomy, University of Cambridge, Madingley Road, Cambridge CB1 0HA, UK \\
$^{6}$Kavli Institute for Particle Astrophysics and Cosmology, Stanford University, 382 Via Pueblo Mall, Stanford, CA 94305-4060, USA \\
$^{7}$Department of Physics, Stanford University, 452 Lomita Mall, Stanford, CA 94305-4085, USA \\
$^{8}$D\'epartement de Physique, Universit\'e de Montr\'eal, C.P. 6128, Succ. Centre-Ville, Montr\'eal, Qu\'ebec H3C 3J7, Canada \\
$^{9}$Department of Physics and Astronomy, University of Waterloo, Waterloo, ON N2L 3G1, Canada \\
$^{10}$H H Wills Physics Laboratory, Tyndall Avenue, Bristol BS8 1TL, UK \\
}

\begin{document}

\date{Accepted 8 March 2016. Received 26 Febuary 2016; in original form 29 August 2014}

\pagerange{\pageref{firstpage}--\pageref{lastpage}} \pubyear{2016}

\maketitle

\begin{abstract}

\noindent We present an Integral Field Unit survey of 73 galaxy clusters and groups with the 
{\em VIsible Multi Object Spectrograph} (VIMOS) on VLT.  We exploit the data to 
determine the H$\alpha$ gas dynamics on kpc-scales to study the feedback 
processes occurring within the dense cluster cores. We determine the kinematic state 
of the ionised gas and show that the majority of systems ($\sim$ 2/3) have relatively ordered 
velocity fields on kpc scales that are similar to the kinematics of rotating discs 
and are decoupled from the stellar kinematics 
of the Brightest Cluster Galaxy.  
The majority of the H$\alpha$ flux ($>$ 50\,\%) is typically associated with these 
ordered kinematics and most systems show relatively simple morphologies 
suggesting they have not been disturbed by a recent merger or interaction.  
Approximately 20\,\% of the sample (13/73) have disturbed morphologies which can typically 
be attributed to AGN activity disrupting the gas. Only one system shows 
any evidence of an interaction with another cluster member.  A spectral analysis of 
the gas suggests that the ionisation of the gas within cluster cores is dominated 
by non stellar processes, possibly originating from the intracluster medium itself.

\end{abstract}


\input{interoduction_r2}
\FloatBarrier
\input{analysis_r2}

\FloatBarrier
\input{observations_r2}
\FloatBarrier
\input{channels_r2}
\FloatBarrier
\input{results_r2}

\FloatBarrier
\input{spectra_r2}

\FloatBarrier
\input{conclusions_r2}

\FloatBarrier

\section*{Acknowledgements}

SLH acknowledges support from the European Research Council for Advanced Grant Program num 339659--MUSICOS. FC acknowledges the support from the European Research Council for Advanced Grant Program num 267399--Momentum. AMS acknowledges an STFC Advanced Fellowship. ACE acknowledges support from STFC grant ST/I001573/1.

Based on observations made with ESO Telescopes at the La Silla or Paranal Observatories under programme ID 080.A-0224 and 082.B-0671





\bibliographystyle{mn2e}
\bibliography{bib}




\label{lastpage}

\end{document}

%% file: interoduction_r2.tex
\section{Introduction}
\label{chap:intro}



One of the key issues for our understanding of galaxy formation and evolution is the 
mechanism through which feedback from a galaxy affects the cooling of gas surrounding it.
Simulations of galaxy formation which ignore non-gravitational heating 
\citep[e.g.][]{kat93} produce a galaxy population with an excess of massive galaxies 
when compared to the observed Universe.  This occurs as a result of gas cooling being 
constrained only by its density, which results in rapid cooling and an over production 
of stars in the dense clumps where galaxies form.  To address this problem the injection 
of energy into the gas from non-gravitational processes within the galaxies is invoked, 
a process dubbed `feedback'.  By including feedback in the simulations the over 
production of massive galaxies is stopped resulting in a galaxy population in closer 
agreement to that which is observed \citep{bow06,cro06}. 

While feedback is widely accepted as the means to slow the cooling of gas, the form that 
it takes is a highly debated topic. 
At face value X-ray observations of the central regions of massive galaxy clusters 
show intense X-ray emission suggesting that the Intra Cluster Medium (ICM) is undergoing 
significant radiative cooling \citep{fab81}.  This rapidly cooling gas should condense 
into cold gas clouds and/or form stars on short timescales relative to the age of the 
cluster \citep{fab94}.
However, the cold gas mass \citep{bd94,mj94,ode94,edg01,sc03} and star formation rates observed 
\citep{mo89} are too low to be consistent with the mass of gas that should be cooling 
from the ICM.  
Early observations with XMM-Newtons highly 
sensitive Reflection Grating Spectrometer (RGS) failed to detect the X-ray spectral 
features of gas cooling at $\sim$1 keV \citep{pet01,tam01,pet03}. 
The apparent lack of gas at 
X-ray temperatures of $<$~1 keV suggests that some process is truncating the cooling 
to lower temperatures and preventing much of the gas from cooling further  
\citep[see the review by][]{pf06}. 
However, 
as the ICM core is still radiating away most of 
its energy through X-ray emission at a high rate 
there must be some 
process acting in the cores of cooling flow clusters which is continually reheating the gas.
Thus ``feedback'' from non-gravitational processes needs to be invoked 
 in order to inject energy into the ICM and balance the effects of 
X-ray cooling. Many possible contributors to ICM heating have been suggested such as 
starbursts \citep{vei05}, shocks from mergers \citep{mv07}, sloshing of gas 
\citep{zuh10}, conduction from the surrounding ICM \citep{vf04} and mechanical or 
radiative feedback from an AGN \citep{mcn05,mcn07}.

For the most rapidly cooling cluster cores, the BCGs ubiquitously exhibit optical line 
emission \citep{hec89,cav08}. \citet{cra99} found significant line emission in 32~per cent 
of a sample of 201 BCGs selected from the Brightest Cluster Survey (BCS) X-ray selected 
sample \citep{ebe98}. This line emitting gas at 10$^4$ K traces filamentary structures
around the BCG \citep{hat05,hat06,mcd11} 
and direct comparison has shown qualitatively 
similar structures in the 10$^7$ K X-ray \citep{fab08} 
and 30 K molecular \citep{sal11} 
gas. This structural similarity suggests that the gas phases are linked with gas from 
the ICM cooling through the warm phase before quickly
condensing out into cold molecular gas clouds.  However, the masses of the gas at these 
temperatures are not consistent with that predicted from cooling \citep{jst87}. 
Once the gas is in molecular clouds, 
ionising radiation from within the cluster core can easily re-ionise and excite the 
surface of the gas clouds allowing the ionised gas to act as a proxy for the molecular gas.  

Early studies of the brightest handful of BCGs showed a variety of complex spatial 
and velocity structures present in the ionised gas \citep{cra02,hat06,wil06,edw09}.  
These early studies 
are highly suggestive, with some objects showing rotation, offset emission and most 
having largely uniform [NII]/H$\alpha$ ratios.  They also show that the most 
kinetically disturbed objects tend to be those which show 
evidence of a recent interaction.  
The presence of strong radio sources from the BCG at the centre of clusters 
\citep{bur94,hog15} further complicates the situation. In several clusters the  
intricate filamentary structures formed by the optical line emission appear to be 
spatially and dynamically linked to the expanding radio lobes \citep{con01,hat06}, 
the most clear example of this is NGC 1275 in the core of the Perseus cluster. 

In order to place constraints on the BCG population in general 
we have used the VIsible Multi Object Spectrograph 
(VIMOS) on the Very Large Telescope (VLT) to obtain 
spatially resolved spectroscopy around the redshifted H$\alpha$ emission of 73 BCGs, 
spanning a range in radio power,
selected to show evidence of extended line emission structures. 
By using this representative sample we will attempt to address three key 
questions pertinent to our understanding of cluster cores:
\begin{enumerate}[i)]
\item {\it What fraction of line emitting BGCs are highly disturbed?}  Previous IFU observations of just a handful of objects indicated the presence of gas at high velocity with structures and kinematics  suggesting possible interactions with  other cluster members \citep{wil06}.
\item {\it What role does the Brightest Cluster Galaxy play in the cooling of gas from the ICM?} Most clusters of galaxies have Brightest Cluster Galaxies whose positions are strongly correlated with the cluster core \citep{per98,san09,hud10}. 
In \citet{ham12} we identified 3 objects (2 from this sample) in which the presence of ionised and molecular gas at the location of the core suggested that cooling continued despite being offset from the BCG.  The systems studied in \citet{ham12} had very large physical offsets but are rare.  
\item {\it What role does the cold gas play in the feedback process and how is feedback fuelled?}  The degree of fine tuning required for feedback to be well regulated suggests a link between the non gravitational processes in the BCG and the gas cooling from the ICM.  In \citet{ham14} we showed that Hydra-A contains a cold gas disc which may be responsible for moving gas into the vicinity of the AGN where it can fuel the powerful jets which are injecting mechanical energy into the ICM.  If such kinematic structures are common then discs may be an important component in regulating the feedback process. 
\end{enumerate}


In Section \ref{chap:meth} we present details of the observations, data reduction and 
analysis techniques used throughout this study.  
We then
 present the observational parameters (Section \ref{sec:obs}) and study the maps from 
individual channels of the IFU in Section \ref{sec:chan}. 
Section \ref{chap:results} then presents the analysis of the morphology and kinematics
of the sample. 
In Section \ref{sec:spec} we study the physical condition of the gas in the
ionised nebula through its spectral properties. In Section \ref{chap:sum} we present a 
discussion of our findings with regard to the questions posed earlier in this 
Introduction. Finally we present a summary of our findings and the conclusions 
we draw from them in Section \ref{sec:con}

Throughout this study we assume a standard cosmology with $\Omega_{m}=$ 0.27, 
$\Lambda=$ 0.73 and $H_{o}=$ 71 km s$^{-1}$ Mpc$^{-1}$. 

%% file: analysis_r2.tex
\section{The Sample}
\label{chap:meth}



\subsection{Sample Selection and observation}

This sample was drawn from the ROSAT selected Brightest Cluster Survey 
\citep[BCS,][]{ebe98} and 
the 
ESO X-ray Cluster Elliptical Spectral Survey (EXCESS) 
which provided 
long slit spectra of 201 and 446 BCGs respectively. 
The EXCESS spectra were then used to identify clusters with optical 
line emission, the H$\alpha$ emission line was identified in 30\% of the sample which 
is consistent with the findings from the BCS \citep{cra99}.

Only objects with 
an integrated H$\alpha$ emission greater than 1\,$\times$\,10$^{-15}$ erg\,cm$^{-2}$\,s$^{-1}$ 
were selected.
The second selection 
criteria was that the H$\alpha$ emission showed an extent greater 
than two arcsec in the long slit observations 
to ensure only objects with line emission extended on scales greater than 
the expected seeing were observed.


\begin{figure*}
\psfig{figure=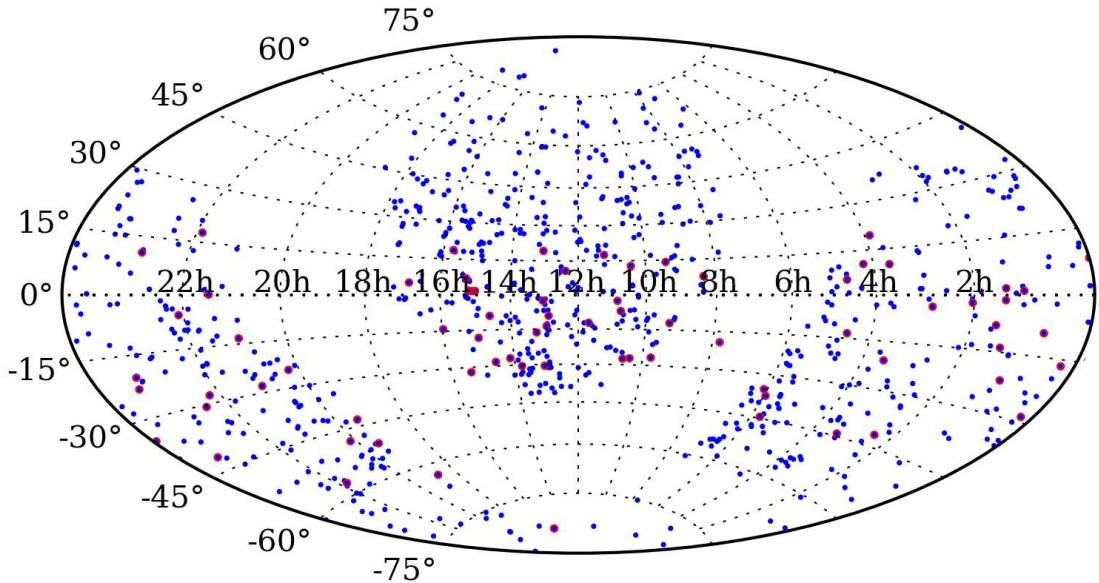,width=15cm}
\caption[Plot showing the positions of all objects in the parent sample and identifying those selected for observation with VIMOS.]{Here we show a plot of the sky identifying the locations of the sample.  The blue points show the objects in both the BCS and EXCESS samples, the objects selected and observed with VIMOS are circled in red.}
\label{fig:sky}
\end{figure*}



The VIMOS instrument on the 8.2--m 
VLT was the primary optical IFU used throughout this work.  The 
IFU is made up of 6400 fibres with a scale on the sky which can vary from 0.33'' to
0.67'' per fibre corresponding to a field of view ranging from 13''$\times$13'' 
to 54''$\times$54''.  It consists of four 
arms which split the field of view into four identically sized quadrants. Each  
quadrant feeds to a separate grism which disperses the incident light onto its own CCD.  
There are six grisms equipped on each arm which provide overlapping coverage from 360 to 
1000 nm with a spectral resolution ranging from R$\sim$200--2500.  
Due to being located 
in the southern hemisphere potential VIMOS targets are limited to equatorial southern 
objects 
(those with a declination of $<$+20).
        
The wavelength coverage allows easy access to the bright H$\alpha$ line 
in BCGs out to a redshift of $z$\,=\,0.25 while the field of view is more than sufficient to 
contain the emission in all but the most local clusters.
78 objects matched the selection criteria of which 
five already had optical IFU observations.  The final X-ray selected sample includes all
extended H$\alpha$ bright BCGs out 
to a redshift $z$\,=\,0.25 and represents an increase in the number of systems 
studied with IFUs of more than a factor of five.   
The presence of extended line emission suggests that all objects in the sample have 
low central entropy values \citep{cav08} and are thus consistent with rapidly cooling 
cores. 
Figure \ref{fig:sky} shows the 
locations of all objects from the parent samples and identifies those which were 
selected to be observed with VIMOS.

At high spectral resolution 
the kinematics of the gas can be mapped to a high degree of accuracy with each spectral 
resolution element (0.6\,\AA)  corresponding to $\sim$ 30 km s$^{-1}$ (at the wavelengths 
observed) with broad resolved lines allowing 
accurate centroiding to a fraction of this.  The spectral resolution also allows for the 
independent fitting of the H$\alpha$ and [NII] lines meaning a measure of the ionisation 
state of the gas can be obtained from the line ratios.  The wavelength coverage 
also allows for a detection of other important spectral features should they be present 
including the [SII] doublet (allowing for a measure of the electron density), the sodium 
D absorption feature (which can be used to study the stellar component of the BCG) and 
the H$\beta$ 
and [OIII] lines in some objects. 

VIMOS was used to obtain optical IFU spectroscopy of the 73 BCGs in 
the sample between October 2007 and September 2008.  For each object a set of three 
600s exposures were performed with a pointing dither of 1.5$''$ included between each to 
account for bad pixels. The {\em HR\_Orange Grism} and {\em GG435} filter (spectral 
resolution 
of $\sim$\,$\lambda/\Delta\lambda$\,$\sim$\,2650 over a wavelength range 5250--7400 \AA)
were used to observe H$\alpha$($\lambda_{rest}6562.8$\,\AA) in clusters with a redshift
below 0.13. For clusters with a redshift greater than this the {\em HR\_Red 
Grism} and {\em GG475} filter (spectral resolution of 
$\sim$\,$\lambda/\Delta\lambda$\,$\sim$\,3100 over a wavelength range 6450--8600 \AA) 
were used to sample the H$\alpha$.  The setup used was selected such that the 
spectra obtained would cover a spectral range which should include the principle lines 
of H$\alpha$, [OI], [NII] and [SII] assuming they are present in each objects spectrum.
As this project was proposed as a bad weather backup project,
the observations were taken in a range of conditions, with a median seeing 
of $\sim$\,1.5$''$ but with a range $\sim$\,0.5--2.6$''$.  
Both the {\em HR\_Orange} and the {\em HR\_Red} modes offer a 27$''$ $\times$ 27$''$ field of 
view which is covered by 1600 fibres.  Each fibre is coupled to a micro-lens to give 
near-continuous sky coverage with each fibre observing a region of 0.67$''$ in diameter. 
Details of the observations are summarised in Table \ref{tab:sam1}.

\subsection{Data Reduction}

The raw data were reduced using the VIMOS specific recipes in the ESO Recipe 
Execution Tool {\sc esorex}.  This package performed the basic data reduction on each 
data cube (a total of 12 for each object, one for each quadrant from the three pointings)
including bias subtraction, flat fielding, and the wavelength and flux calibration.
The wavelength calibration is achieved by comparing a separate exposure of an arc lamp
which produces emission lines at known wavelengths and comparing these observations 
to a catalogue of known line positions. At each spatial element the lines are identified 
and their position compared to that in the catalogue, differences between pixels are then 
corrected allowing the procedure to account for instrumental distortions.  The flux 
calibration is done using observations of a standard star (one for each quadrant of the 
IFU) which is compared to a standard spectrum of the star to determine the efficiency 
and response curves of the spectrograph. A standard extinction table is then used to 
calculate the flux losses due to the atmosphere at a given airmass at each wavelength, 
after which the scientific spectrum is multiplied by the response curve to produce the 
final flux calibrated spectrum.  Differences in the seeing conditions and airmass of the 
observations and standard star are the main contributing factor affecting the flux 
calibrations.  To minimise these effects standard stars were observed separately for 
each object during the observations to minimise these differences.  Excluding such 
systematic errors the method achieves an accuracy of better than 0.5\%.  Full details of 
the data reduction pipeline and processes can be found in the VIMOS Pipeline User Manual
(VLT-MAN-ESO-19500-3355). 
  
To subtract the sky point--like objects were masked to remove any stars within the 
field.  Such objects were defined as having a roughly 2D-Gaussian intensity profile 
with a FWHM on the order of the seeing and were removed out to the diameter of the 
seeing.  The BCG was then removed by masking all pixels within an isophote of half its 
peak intensity.  Using a lower 
threshold (40\%) had a negligible affect on the sky level (the medium over all 
sampled spatial elements remained the same to 6 decimal places) for all but the most 
local objects (z less than $\sim$0.01).  By contrast choosing a higher threshold 
(60\%) resulted in a median increase in the sky level across all objects of 3\% 
suggesting that sufficient light from the BCG is being sampled to affect the sky 
measurement.  Thus the 50\% level was chosen empirically as the best compromise to allow a 
good sampling of sky pixels while removing the majority of the BCGs light. 
The small field of view means that objects other than the BCG which are not point like 
were rare in our sample, however some observations did contain them.  These objects 
where removed in the same way as the BCG, using the same threshold and their peak 
intensity to produce the mask.
The sky level 
for each quadrant was calculated by taking the median value of the remaining pixels 
at each wavelength increment.  
The sky spectrum was subtracted from each pixel in 
the four quadrants before they were combined into a single data cube. 

For low 
redshift objects ($z$\,$\lesssim$\,0.02) the BCG easily fills a single quadrant so a higher flux 
threshold for removing the BCG must be used which can result in a substantial 
over subtraction of the sky. 
For such objects subtraction of a straight continuum baseline prior to the sky 
subtraction proved a more reliable means of isolating the emission lines.  The continuum 
baseline was calculated from the spectrum at each spatial resolution element by taking 
the median 
of the emission either side of the lines of interest. The spectral range used to 
calculate the median was 120 \AA\, in length and began 60 \AA\, from the emission line on 
either side.  The continuum baseline was then produced by interpolating over the 
wavelength range 
to be fitted between the medians calculated at either side of the emission lines. 
Once this baseline had been subtracted the sky was subtracted in the same 
manner as the higher redshift objects before the lines are fit.  The drawbacks of this 
method however are that it is only valid for a small spectral range so only one line 
complex can be studied in a given reduction and it also eliminates any information carried 
in the objects continuum emission.  

Finally the three 
exposures were median combined for each pointing in order to eliminate cosmic rays. The 
exception to this was RXCJ2014.8-2430 which only had two pointings so the mean of the 
two observations was used to produce the final cube.  The final mosaics provided cubes 
with a $\sim$\,30$''$\,$\times$\,30$''$ field of view once the dithering offsets are
taken into account.


\subsection{IFU Data Cube analysis}
\label{sec:fitting}
We took the IFU data cubes produced by the 
reduction and fit spectral models to the key 
lines at each resolution element. The primary diagnostic line structure 
was the H$\alpha$\,($\lambda_{rest}$ 6562.8\,\AA) 
and [NII]\,($\lambda_{rest}$ 6548.1/6583\,\AA) complex. 

\begin{figure*}
\psfig{figure=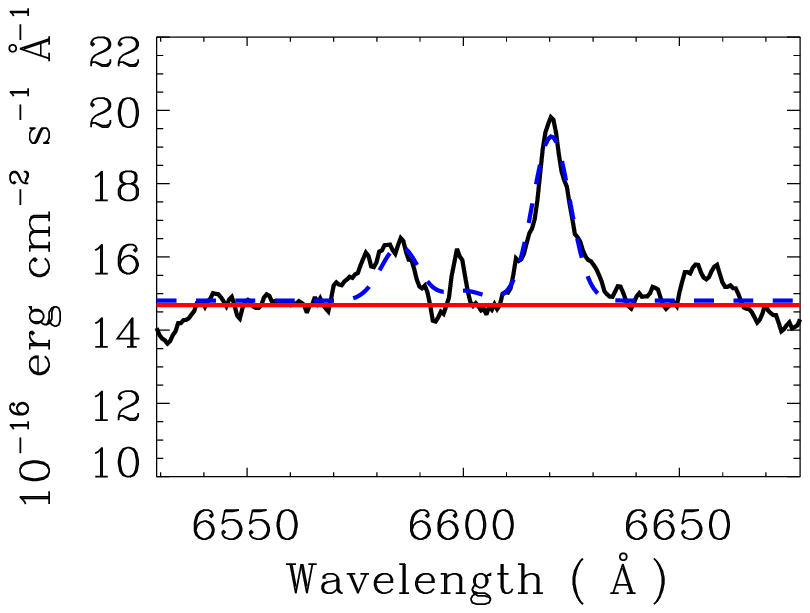,width=4.5cm}
\hspace{0.3cm}
\psfig{figure=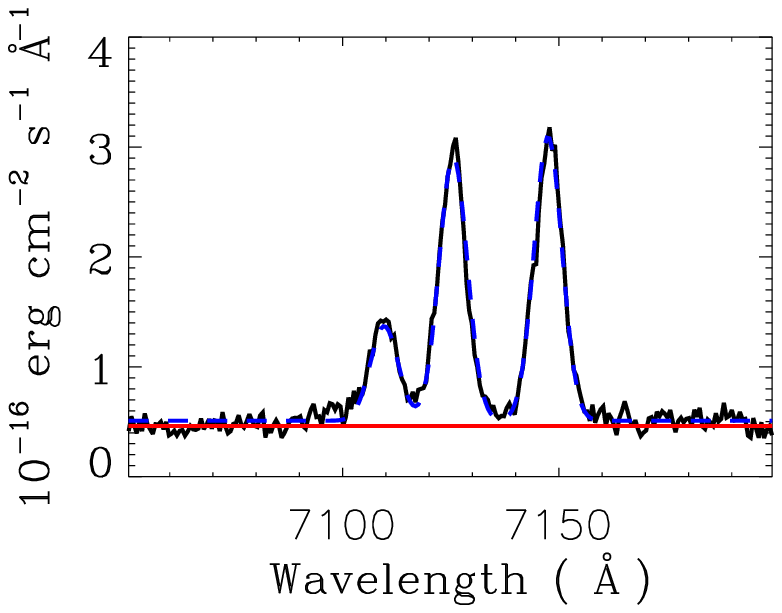,width=4cm}
\hspace{0.3cm}
\psfig{figure=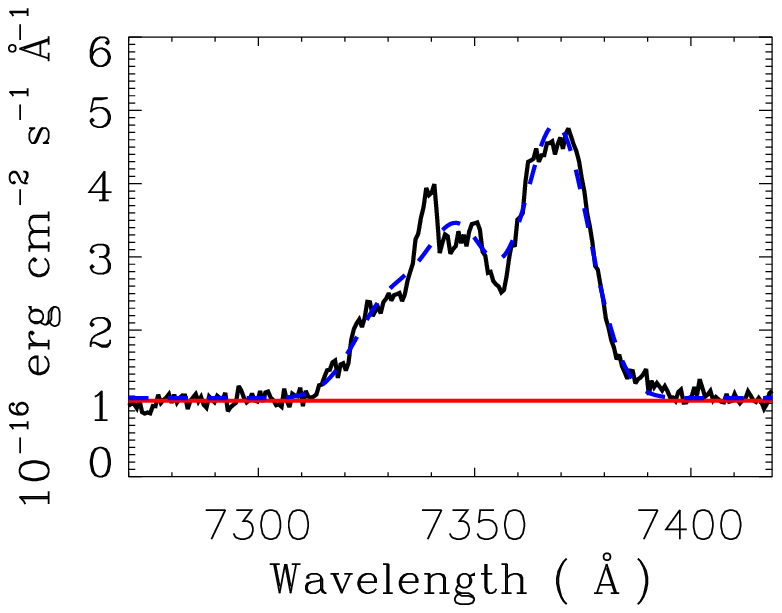,width=4cm}
\hspace{0.3cm}
\psfig{figure=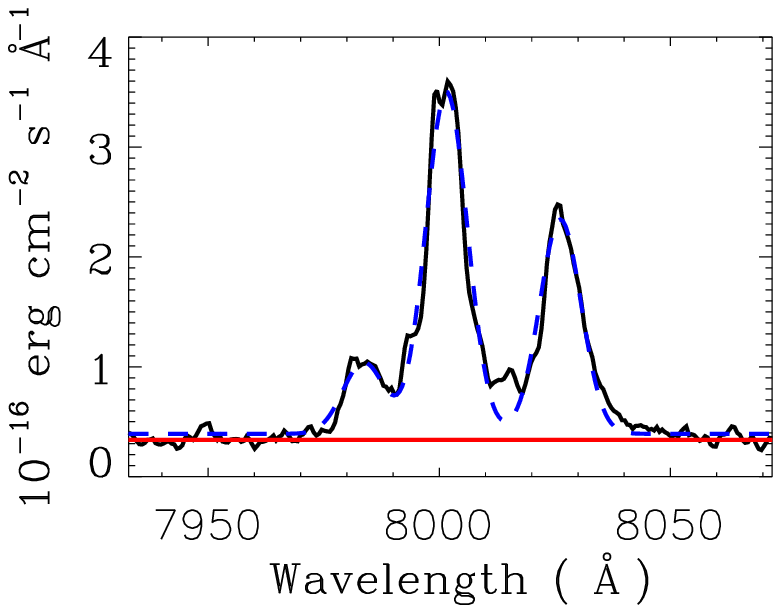,width=4cm}
\hspace{0.5cm}
\psfig{figure=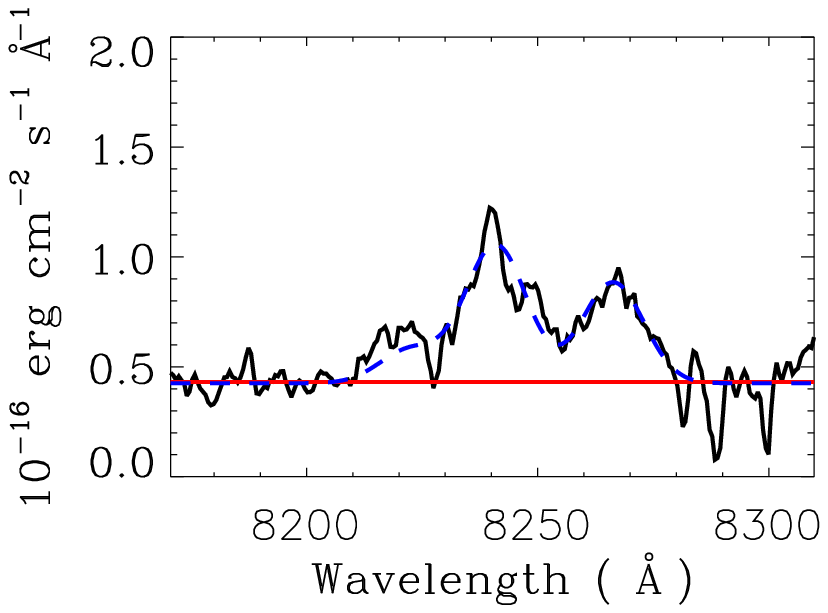,width=4cm}
\caption[Plot showing spectrum of the central region of a selection of objects limited to the wavelength range of the fits]{Here we show several spectra showing the H$\alpha$ and [NII] lines which represent the sample. The spectra are of the central region of five objects and the wavelength coverage is the same as the region used by the fitting routine.  From left to right, top to bottom the redshifts of the objects are $\approx$ 0 (NGC 5846),  $\approx$ 0.08 (Abell 478), $\approx$ 0.13 (Abell 1348), $\approx$ 0.21 (Abell 3017) and $\approx$ 0.26 (Abell 3444) so are representative of the whole sample.  The red solid line shows the flat continuum baseline fit and the blue dashed line shows the best fit spectral model.  The variation of flux, [NII]/H$\alpha$ ratio and line width are apparent from these plots.  
}
\label{fig:cont}
\end{figure*}

To fit these lines we extracted 
a region of the spectrum in each resolution element (lenslet) that was 240 spectral 
resolution elements in length ($\sim$ 144 \AA) centred at the wavelength of the
H$\alpha$ line at the cluster 
redshift.  We then fitted Gaussian emission-line profiles to these extracted regions 
using a $\chi^2$ minimisation procedure.  To the H$\alpha$--[NII] triplet of lines we
fit three separate Gaussian profiles on top of a flat baseline to account for variations 
in the average continuum flux and compare this to just a flat continuum baseline fit
(Figure \ref{fig:cont}). Each of the three emission line had the form 

\begin{equation}
\rm F(\lambda) = \frac{F_0}{2.\pi.\sigma} \times e^{\frac{-0.5.(\lambda-\lambda_z)^2}{\sigma^2}}
\label{eqn:lineprof}  
\end{equation}

with F$_0$ the total flux in the line, $\sigma$ the velocity dispersion of the line 
(FWHM/2(2ln2)$^{0.5}$), $\lambda$ the wavelength at a given position and $\lambda_z$ the 
wavelength of the line corrected for redshift.  The model spectrum contains 3 such lines 
with fixed (at redshift 0) separations, the H$\alpha$-[NII]$_{\lambda6583/\lambda6548.1}$ triplet.
Each line was produced independently and the resulting spectral models where added 
together on top of a flat baseline to produce the final spectral model.

The fitting routine was 
given an array of first guess variables which were then used to produce the fit.  For 
the H$\alpha$--[NII] triplet these variables were as follows:-

\begin{enumerate}

\item The redshift of the emission.  The redshift was used to determine the value of $\lambda_z$ in Equation \ref{eqn:lineprof} such that $\lambda_z$ = $\lambda_0 \times (1+z)$. Where $\lambda_0$ is the rest wavelength of the line and z is the redshift. The first guess for the redshift at each position was set to the systemic redshift of the BCG.

\item  The total flux in the H$\alpha$ line. This was used as F$_0$ for the H$\alpha$ line in Equation \ref{eqn:lineprof}. The first guess was set to the average flux at the wavelength of the line across the whole cube.

\item The total flux in the [NII]$_{\lambda6583}$ line. This was used as F$_0$ for the [NII]$_{\lambda6583}$ line in Equation \ref{eqn:lineprof}, F$_0$ for the [NII]$_{\lambda6548.1}$ line was fixed at one third of this value. The first guess was set to be equal to the H$\alpha$ flux.

\item The FWHM of the line in angstroms. This was used to calculate $\sigma$ in Equation \ref{eqn:lineprof} such that $\sigma$ = FWHM/2(2ln2)$^{0.5}$. The first guess set to the 
width of a skyline, $\sim$\,4.7\,\AA. 

\item The baseline flux level.  This was not used to produce any of the three line models but was added to the total model spectrum after they had been produced to account for variations in the median level of the continuum between lenslets.

\end{enumerate}

The redshift and width of the [NII] and H$\alpha$ lines were set to be the same to 
reduce the number of free parameters in the model.
 After producing the first guess the redshift, H$\alpha$ flux, [NII]$_{\lambda6583}$ flux, 
linewidth and continuum baseline level were allowed to vary using a 
least squared fitting routine \citep{mor93} until the model minimised to best fit the 
data. During the minimisation the parameters 
were constrained to ensure only the emission lines of interest were recovered and reduce 
the processing time required to achieve a fit.  We constrained the H$\alpha$ and 
[NII]$_{\lambda6583}$ flux to positive values (emission) and the sigma 
linewidth to $<$ 1000\,km\,s$^{-1}$ (FWHM$\sim$2355\,km\,s$^{-1}$). These profiles were 
initially fitted to 
the spectrum from each 0.67$''$  lenslet and adaptively binned to $\sim$\,2.0$''$ 
(3 $\times$ 3 lenslets) in regions 
with lower H$\alpha$ flux, as such the fits to the low surface brightness emission 
have a lower spatial resolution than the brightest regions. 

We accepted these fits as representing the presence of spectral lines in the data when 
they provided an improvement in signal to noise of 7 (7 times the standard deviation, 
or 7$\sigma$) when compared to 
a continuum baseline only fit.  When an acceptable fit was found the parameters of the best 
fitting model were stored and the 1$\sigma$ errors 
were calculated by varying each parameter slightly from the best fit and allowing 
all others to re--minimise.  For each pixel where an acceptable fit was found we 
attempted 
to fit a second model which included an additional line component with a redshift, 
intensity and width independent of the initial fit.  This extra component was allowed 
to become much broader than the initial fit (it was left unconstrained) and was included 
as a means to test the 
spectra for additional velocity components and broad line features.  This model was then 
accepted when it provided an improvement in signal to noise of 15 (15 times the standard 
deviation, or 15$\sigma$) when compared to a continuum baseline only fit.
For pixels where a second component was found to be significant a 
model including a second component for all lines was fitted to the data and the 
errors were recalculated to account for the new model.  This process detected only two 
objects with multiple velocity components and eight with broad components in their core 
regions.  
As multiple velocity components do not make up a significant fraction of the sample 
we choose not to address the nature and origin of the additional components in this paper.
The significance of these models was tested on a number of objects using an F--test 
of additional parameters. The basic H$\alpha$--[NII] line model when compared to a 
baseline only model was found to always 
have a significance level of $\alpha$\,$<$\,0.001 when required to fit a signal to noise 
of 7$\sigma$.  Likewise the threshold of a signal to noise of 15$\sigma$ was selected 
so as the additional line components had a significance level of $\alpha$\,$<$\,0.001 
when compared to the standard H$\alpha$--[NII] line model.

The parameters of the fits were then used to produce maps of the line flux, line-of-sight 
velocity and full width at half-maximum (FWHM) deconvolved for instrumental 
resolution. The velocity zero point was determined as the median of the measured 
velocities after removing the highest 2.5\% and lowest 2.5\% of the values.  A retrospective 
clipping method was used to remove pixels in which the fitting routine had returned 
parameters with non physical properties (velocities in excess of $\pm$\,1000\,km\,s$^{-1}$ 
relative to the mean which are unlikely to be associated with the BCG).  
Continuum images of the region covered by the 
observations were produced by taking the median from each lenslet over a region of the 
spectra (here after referred to as collapsing the cube) containing no emission lines or 
sky line residuals. These where produced too allow comparison between the position of 
continuum objects and the structure of the H$\alpha$ emission to identify possible 
offset peaks \citep{ham12} or interactions with other cluster members.

%% file: observations_r2.tex
\subsection{Observational Parameters}
\label{sec:obs}

Table~\ref{tab:sam1} outlines the key 
observational parameters of each object in the sample.  The redshift quoted in Table~\ref{tab:sam1} 
is the median redshift of the H$\alpha$ line in the object.  Most of the 
objects in the sample had all pointings performed on the same day, or at least in similar 
conditions. However, RXCJ2014.8-2430 had one of its pointings performed in very poor 
conditions. As such we have only used two of the pointings to make the final cube.  
RXCJ2101.8-2802 had pointings which were taken on different days, with 4 pointings 
in total, all of which were in consistent conditions and showed no other problems in the 
spectra so we choose to use all 4 pointings, giving this object a slightly longer 
integration time.  We calculated the mean seeing of all observations to be $\sim$ 1.5 
arcsec.  Before the full fitting routine was run on an object its spectrum was 
studied to identify the presence of additional emission lines (lines other than 
H$\alpha$ and [NII]),  these additional lines are listed in Table. \ref{tab:sam1}.  We 
base our classifications on the x-ray properties of each object (x-ray luminosity and 
where available x-ray temperature) using the classification of \citet{bah99} 
(Groups\,$<$\,L$_x$\,=\,10$^{43}$\,erg\,s$^{-1}$\,$<$\,Clusters and 
Groups\,$<$\,T$_x$\,=\,2\,keV\,$<$\,Clusters). By this definition our sample consists 
of 61 brightest cluster galaxies and 12 brightest group galaxies. 


\begin{table*}
\begin{center}
\scriptsize
\centerline{\sc Table \ref{tab:sam1}.}
\centerline{\sc The observational parameters listed for each object in the VIMOS sample}
\smallskip
\begin{tabular}{c c c c c c c c c c c}
\hline
\noalign{\smallskip} 
Cluster & Redshift & Grating & Exposure & Mean & Lines  & RA & Dec & L$_x$ & T$_x$ & Classification \\ 
 & & & (s) & Seeing('') & Detected & & & (10$^{43}$\,erg\,s$^{-1}$) & (keV) & \\[0.5ex]
\hline
Abell 1060 & 0.01263 & HR-orange & 3$\times$600 & 0.79 &  &  10:36:42.78 &  -27:31:39.66 & 3.34 & 3.2 & Cluster \\
\noalign{\smallskip} 
Abell 1084 & 0.13301 & HR-red    & 3$\times$600 & 0.68 & [OI] &  10:44:32.69 &  -07:04:05.21  & 48.99 & 7.3 & Cluster \\
\noalign{\smallskip} 
Abell 11   & 0.14910 & HR-red    & 3$\times$600 & 0.77 & [OI], [SII] & 00:12:33.58 &  -16:28:05.17  & 18.2 & --- & Cluster \\
\noalign{\smallskip} 
Abell 1111 & 0.16518 & HR-red    & 3$\times$600 & 0.85 & [OI], [SII] & 10:50:36.26 &  -02:36:13.37  & 27.30 & --- & Cluster  \\
\noalign{\smallskip} 
Abell 1204 & 0.17057 & HR-red    & 3$\times$600 & 0.81 & [OI] &  11:13:20.08 &  17:35:37.88  & 72.6 & 3.78 & Cluster \\
\noalign{\smallskip} 
Abell 133  & 0.05665 & HR-orange & 3$\times$600 & 0.93 & [OI], [SII] &  01:02:41.49 &  -21:52:48.62 & 16.17 & 3.71 & Cluster \\
\noalign{\smallskip} 
Abell 1348 & 0.11985 & HR-orange & 3$\times$600 & 0.87 & H$\beta$, [OIII], &  11:41:24.12 &  -12:16:35.90  & 21.09 & 5.5 & Cluster   \\
 & & & & & [NI], [OI] & & & & & \\
\noalign{\smallskip} 
.... & & & & \\
\hline
\end{tabular}  
\caption[The  observational parameters of the VIMOS sample]{The observational parameters listed for each object in the VIMOS sample.  The first column states the cluster designation as used throughout this paper.  The redshift is the median redshift acquired from the fits to the H$\alpha$-[NII] complex, the median is calculated from the fits after dropping the top and bottom 2.5\% of the redshifts. The grating is the VIMOS grating used for the observations and the exposure lists the total integration time on target.  The seeing is calculated as the mean of the DIMM seeing across all exposures for each object.  Lines detected lists all lines apart from the H$\alpha$ and [NII] lines that were detected by visual inspection, the H$\alpha$--abs line is included for systems which show some evidence of H$\alpha$ absorption. The right ascension and declination are measured at the centre of the field of view for each observation. X-ray luminosities were taken from the surveys of \citet{boh04,ebe98,ebe00} and temperatures are taken from \citet{ebe96,ebe98,ebe00,cav09}. The classifications are based on the x-ray properties following \citet{bah99}. The full table can be found in Appendix \ref{app:tab1}}
\label{tab:sam1}
\end{center}
\end{table*}

%% file: channels_r2.tex
\section{The Raw Data: Channel Maps}
\label{sec:chan}

We begin by presenting channel maps of several representative objects produced from 
the fully analysed data cubes which 
show the H$\alpha$ emission at different velocities.  To 
produce the channel maps we baseline subtracted each spatial resolution element by
fitting a straight line to the spectrum.  
This was then subtracted to remove the bulk of the 
continuum emission from the cube.  Channels at greater than $\pm$ 1000 km s$^{-1}$ 
relative to the redshift of the BCG were then used to determine the average continuum 
subtracted channel noise of the observation, these channels were then discarded. The 
remaining channels contained the H$\alpha$ emission from the cluster core. For channels 
where the continuum subtracted emission exceeds 7 times the average continuum subtracted 
noise a flux map was produced to show the location of the H$\alpha$ emission in that 
channel.  Each map gives 
the velocity of its channel with respect to  the median velocity of the 
H$\alpha$ emission for that object.  For reference the centre of the continuum emission
from the BCG for each object is marked on its respective maps with a blue cross. This 
position was determined by fitting 2--dimensional Gaussian profiles to a collapsed continuum image. 


\begin{figure*} 
\epsfig{figure=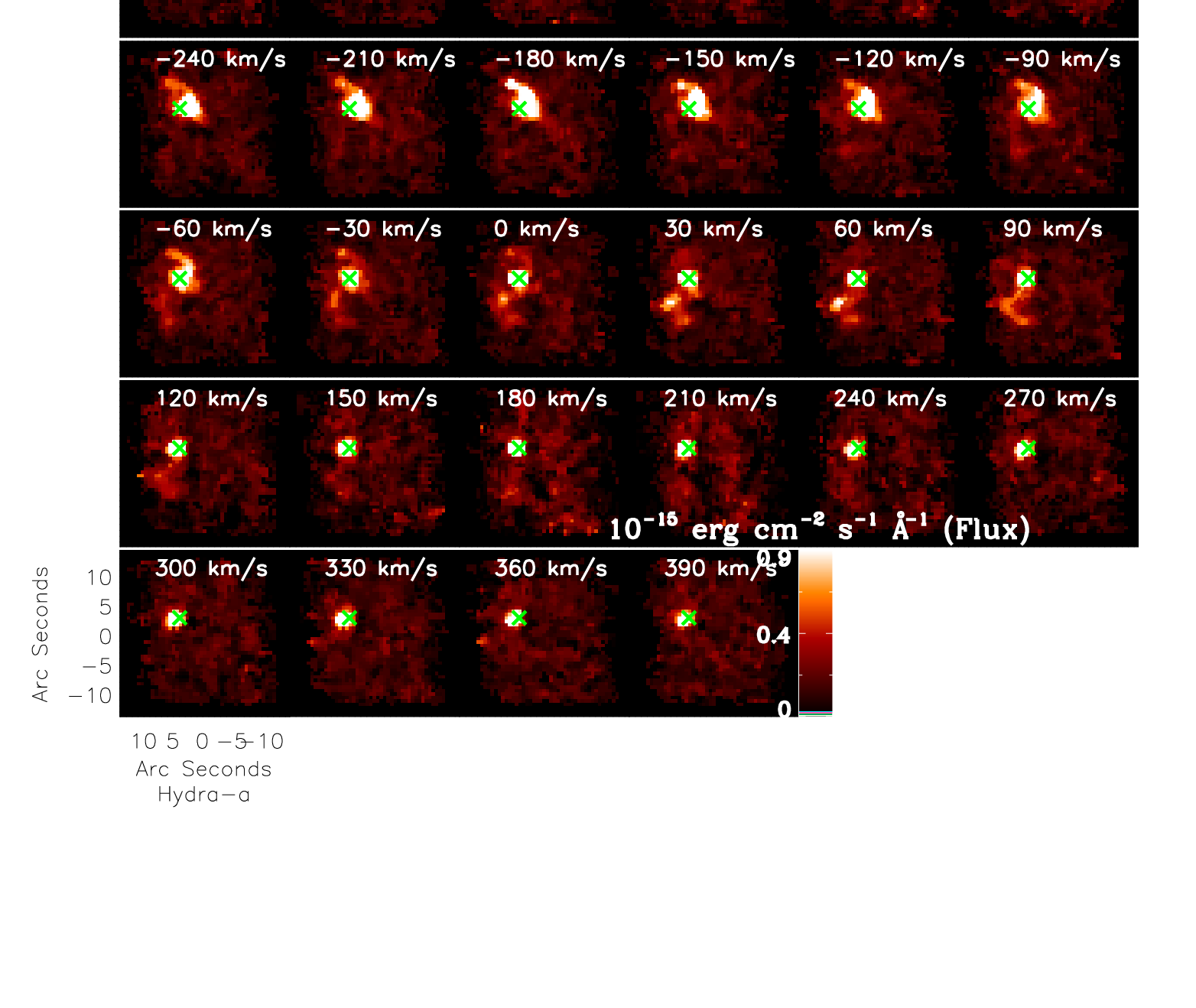,width=12.7cm}
\epsfig{figure=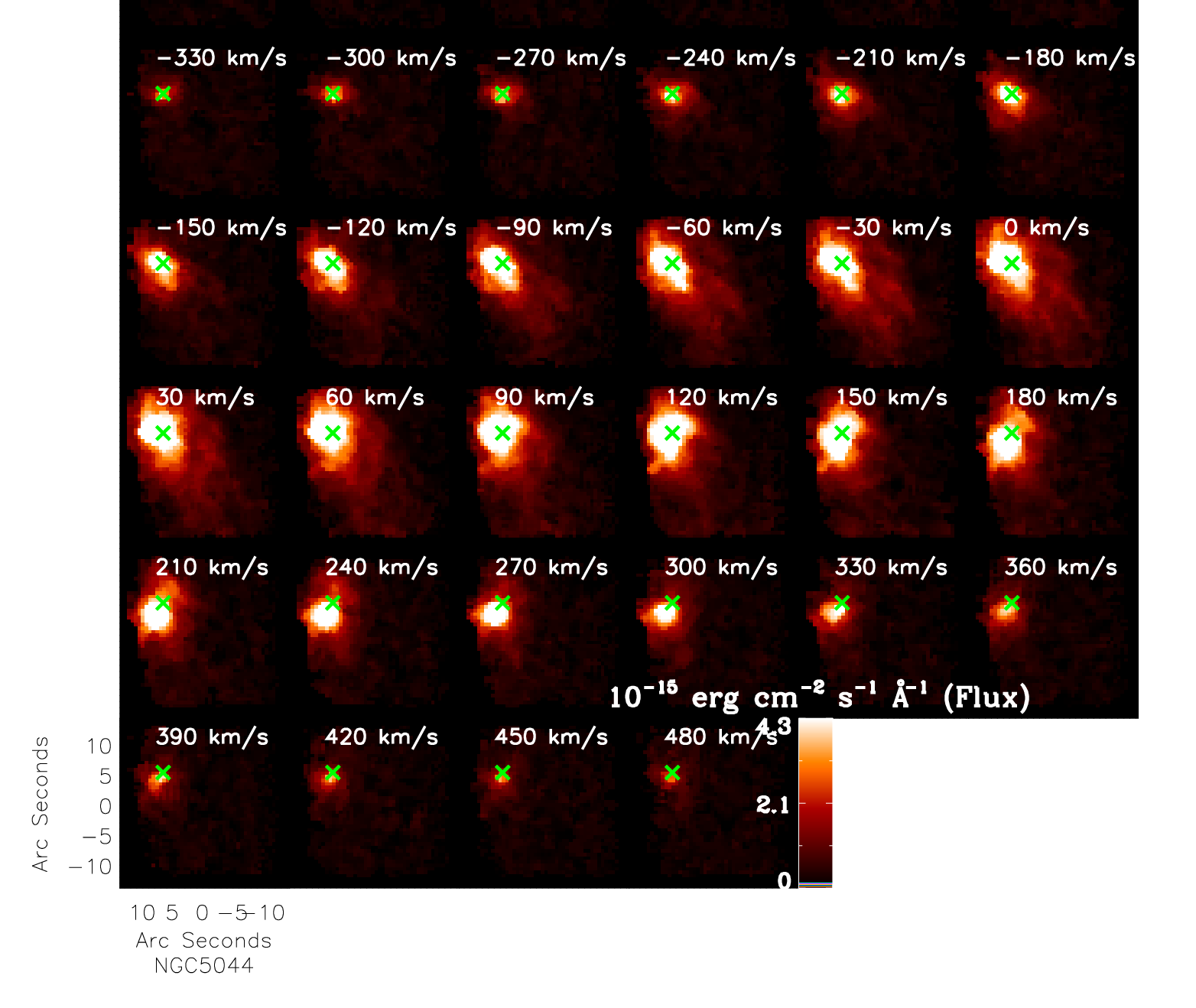,width=12.7cm}
\caption{{\em Top - }Here we display continuum subtracted maps of the H$\alpha$ emission in Hydra-A for each channel containing emission.  Each map gives the velocity of its channel with respect to  the median velocity of the H$\alpha$ emission.  The green cross marks the centre of the continuum emission from the BCG.  The maps clearly show a shift in the centroid of the H$\alpha$ emission as the velocity changes which follows the disc within this object.  {\em Bottom - }The channel maps of NGC 5044 showing the H$\alpha$ emission in velocity slices. The blue cross marks the centre of the continuum emission from the BCG.  These maps show a low surface brightness region of extended diffuse emission to the south where filaments are seen in narrow band images. The core region of the emission shows a chaotic variation in structure between velocity channels suggesting the kinematics in the core are quite complex and disturbed.  Channel maps for every object can be found in Appendix \ref{app:chan}}
\label{fig:hydra_chan}
\end{figure*}

We first show the channel map of Hydra-A (Figure \ref{fig:hydra_chan}) which shows 
evidence for a rapidly rotating disc
within its core \citep{ham14}. This is apparent from the channel maps which show a shift 
in the position of the bright component of the H$\alpha$ emission as the velocity changes.
Additionally we see that the filaments extending to the north and south also show a velocity 
shift, with the northern filament being slightly blueward of the core and the southern filament 
shifted towards the red.  This velocity shift is not clearly apparent in the velocity map displayed 
in \citet{ham14} as the shift is small relative to the total velocity shift in the system.  This kind of velocity structure, with a shift in the velocity 
of the bulk of the emission across the BCG, is common amongst the objects in the sample 
clearly visible in the channel maps of 31/73 clusters. 
The velocity shift in most is not as extreme as that seen in Hydra-A, however this is 
not surprising given we are viewing the disc in Hydra-A at close to edge on \citep{ham14}.

NGC 5044 has narrow band imaging that shows an extensive network or H$\alpha$ filaments 
\citep{wer14}
similar to those seen in the archetype cooling flow NGC 1275.  Our VIMOS observations do 
not resolve the individual filaments but do indicate diffuse extended emission to the 
south in the region where the filaments are found.  We see this emission extending to the 
south in channels close to the BCGs central velocity (Figure \ref{fig:hydra_chan}).
The kinematics correlate well 
with those of the molecular CO(2-1) emission seen by \citet{dav14s},  in the centre the 
line profile is broad (FWHM\,$\sim$\,400--500\,km\,s$^{-1}$) and double peaked 
suggesting 2 velocity components separated by $\sim$\,300\,km\,s$^{-1}$ are present. 
Multiple velocity components such as this are only apparent in the channel maps of 
NGC 5044.  However they have been observed in IFU observations of other systems 
\citep[e.g. Abell 1664][]{wil06,wil09} warranting a more detailed search for their 
presence.
The channel maps of NGC 5044 show that the brightest region of the emission can be 
found at different positions within the BCG at different velocities.  It is slightly 
extended towards the northwest at velocities blueward of the BCG 
($<$ $-$200\,km s$^{-1}$), runs from the northeast to southwest of the BCG at $\sim$ 
$-$120--0\,km s$^{-1}$ and 
extends slightly to the southeast when redward of the BCG ($>$ 150 km s$^{-1}$) 
suggesting that it has no ordered velocity structure.  This redshifted region matches 
the position and velocity relative to the centre of the system of GMA 18 identified 
in \citet{dav14s}.  Low surface brightness extended regions of H$\alpha$ emission 
similar to those seen in NGC 5044 are seen in 9/73 objects in our sample suggesting
that while other filamentary structures may be present in our sample, they are not common.
The channel maps for all objects can be seen in Appendix \ref{app:chan}.

It is clear from the channel maps that there is a wide variety of morphologies of 
the H$\alpha$ emitting gas between objects. Some objects appear very compact from their 
H$\alpha$ emission while others are very extended. Some of this results naturally from 
the variation in redshift of the objects within the sample however, there remains a 
large variation when we consider the a physical extent of the objects (See Table 
\ref{tab:obs}).  

In addition to their varied spatial morphology, many systems show a  
significant, coherent velocity gradient across the BCG (31/73).
This suggests the presence of an ordered 
velocity field within the gas which for one object (Hydra-A) we know 
traces a rotating disc of gas that aligns with the radio and
X-ray structure in the system \citep{ham14}.  The 
presence of such a velocity structure in a large fraction of BCGs would have 
important implications
for our understanding of cluster cores.

%% file: results_r2.tex
\section{The Structure and Kinematics of the Gas}
\label{chap:results}


In order to parameterise the data and make it easier to analyse we fit Gaussian emission 
line profiles to the H$\alpha$ and [NII] lines in each lenslet (see Section 
\ref{sec:fitting})
We now study the maps of these parameterised fits to the H$\alpha$ and [NII] triplet for 
all 73 objects in our sample.  
Five panels were produced for each object showing the continuum emission, the H$\alpha$ 
flux map, the ratio of the [NII] to H$\alpha$, the line of sight velocity field and the 
FWHM.  It should be 
noted that in some systems where the continuum is very high and H$\alpha$ flux is low
not all of the H$\alpha$ is recovered due to the underlying absorption feature caused 
by H$\alpha$ in stellar atmospheres.  In such cases the H$\alpha$ line often appears
very weak or even completely absent from some parts of the cube where it is otherwise 
expected to be (such as regions were the other ionised lines are strong).  Systems 
with this problem were typically of low redshift and thus filled much of the field 
of view making producing an empirical model of the absorption from a bright region 
of the galaxy devoid of line emission difficult or impossible. 

Rather than trying to 
model out the absorption as a component of the fitting routine (which would have reduced 
the degrees of freedom for these objects) we fitted the spectrum as normal. In these  
cases the spectral fit was primarily constrained by the bright [NII]($\lambda_{\rm rest}$ 
6583 \AA) line which could introduce some bias as the two lines may be tracing 
different excitation mechanisms (AGN processes typically produce more [NII] for a given 
H$\alpha$ than star formation for example).  
However, in all such cases the [NII]/H$\alpha$ ratio is high so [NII]
provides a higher significance tracer of the gas dynamics without the degeneracy of 
additional model parameters needed to model out the absorption.
In such cases the maps show the [NII] 
flux rather than the H$\alpha$.

\begin{figure*}
\psfig{figure=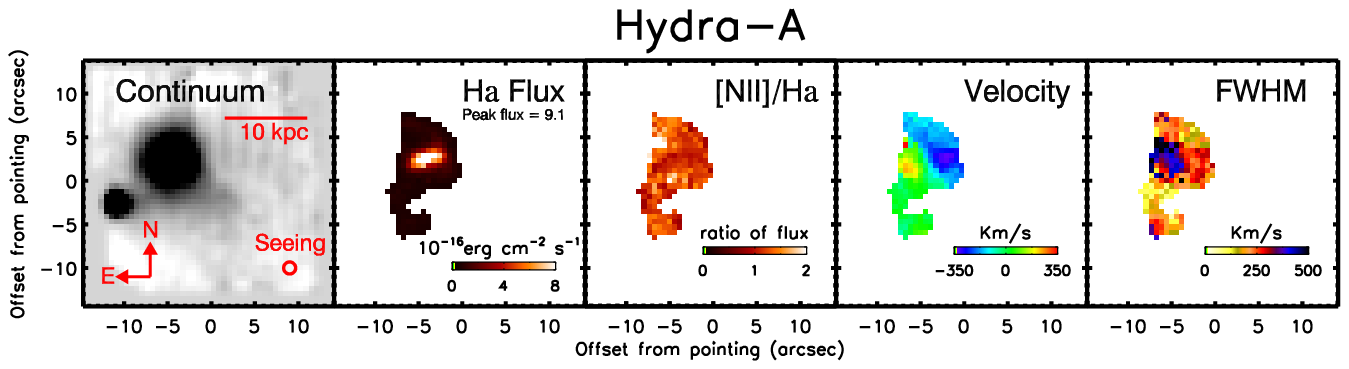,width=17cm,bbllx=74,bblly=367,bburx=459,bbury=470}
\psfig{figure=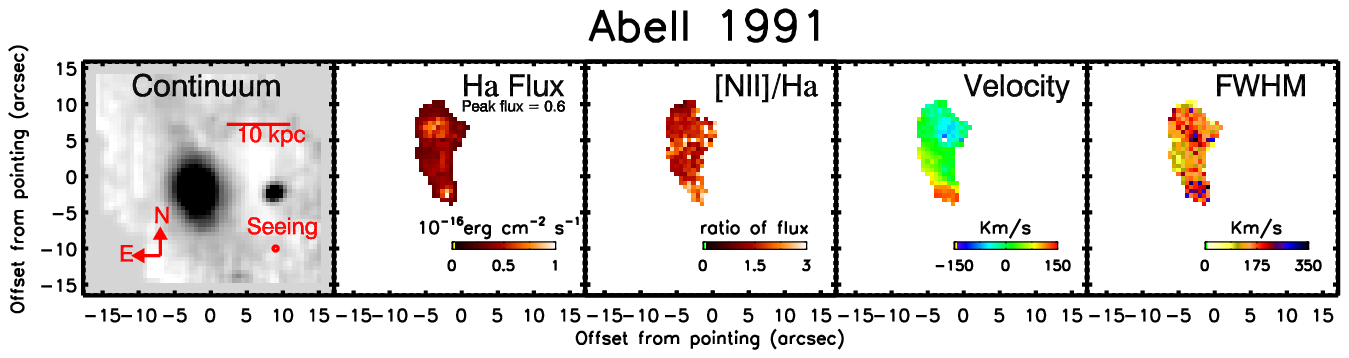,width=17cm,bbllx=74,bblly=367,bburx=459,bbury=470}
\psfig{figure=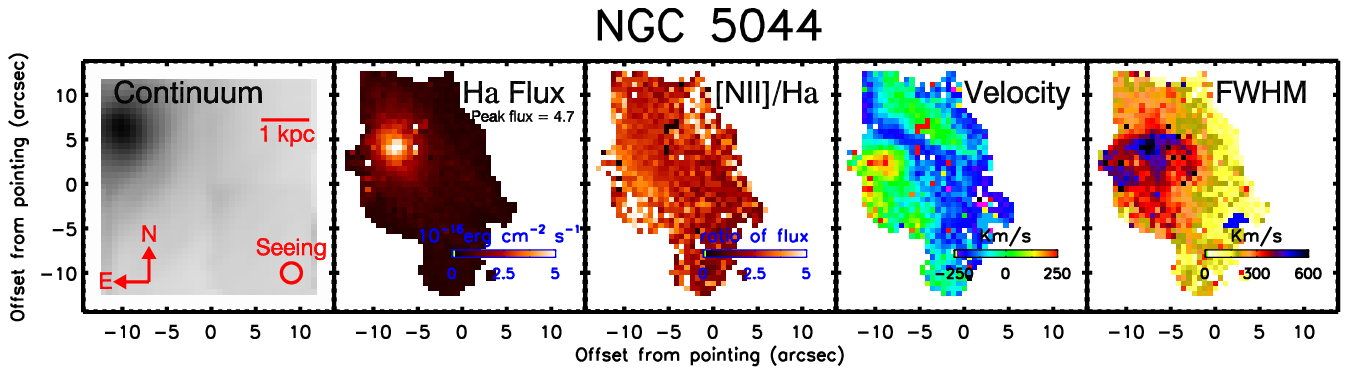,width=17cm,bbllx=74,bblly=367,bburx=459,bbury=470}
\psfig{figure=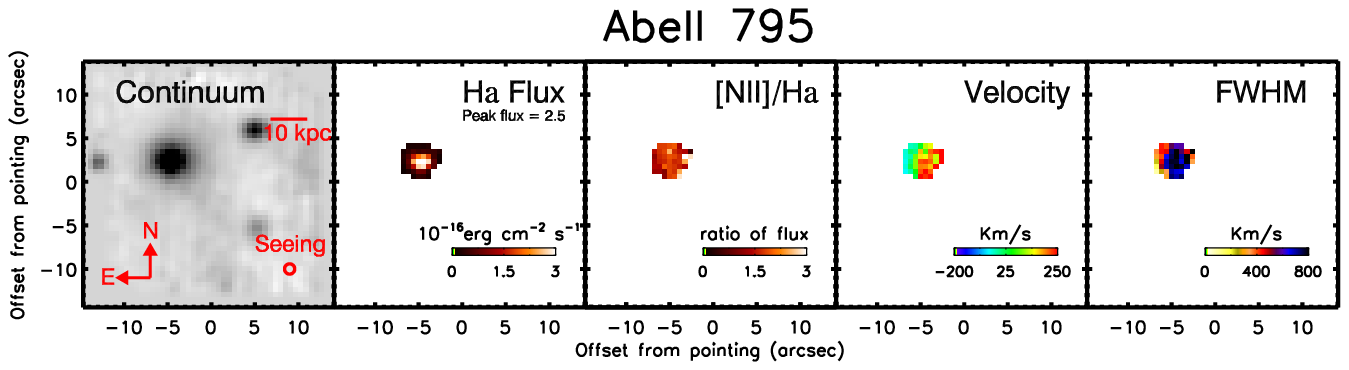,width=17cm,bbllx=74,bblly=367,bburx=459,bbury=470}
\psfig{figure=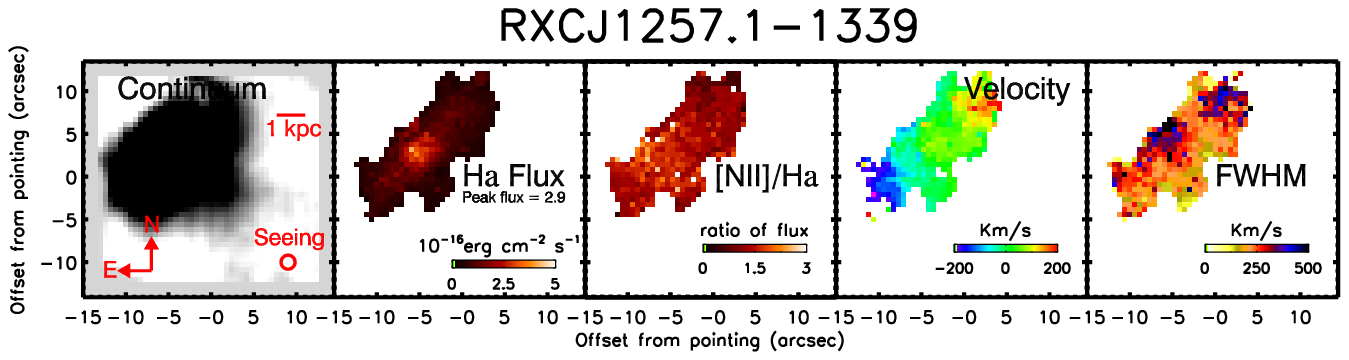,width=17cm,bbllx=74,bblly=367,bburx=459,bbury=470}
\caption[Maps of the spectral fits to the VIMOS data cubes]{Maps of the spectral fits to the VIMOS data cubes for the five objects from the sample.  From left to right the panels are: 1) A continuum image created by collapsing the cube over a wavelength range free of emission and sky lines, 2) The H$\alpha$ flux map, 3) A map showing the ratio of [NII] to H$\alpha$ flux, 4) the line of sight velocity profile of the H$\alpha$ and [NII] emission, 5) the Full Width at Half Maximum (FWHM) of the emission line deconvolved for instrumental resolution.}
\label{fig:maps}
\end{figure*}

To determine the effect of the bias introduced by this approach we produced 
maps constrained only by the [NII] for systems where H$\alpha$ and [NII] were equal
in strength and compared them to the H$\alpha$ constrained maps. We found no significant 
difference between the maps suggesting that the [NII] emission is tracing the same 
morphology and kinematics as the H$\alpha$, at least in the case of the galaxies we could 
perform this test on. We adopt the assumption that this concordance extends to objects 
for which H$\alpha$ emission was too weak and the stellar absorption component dominates  
to produce reliable maps constrained by H$\alpha$.
Figure~\ref{fig:maps} shows maps of the parameterised fits for a sub sample of objects, 
the maps of the full sample are available in Appendix \ref{app:maps}.

\subsection{H$\alpha$ morphology}
\label{sec:morph}
From the maps shown in Figure~\ref{fig:maps} we can calculate the extent of the  
line emitting region for each object.  We define the major axis to be the maximum 
length of continuous emission (at greater than 7$\sigma$) along a line through 
the peak of the H$\alpha$ emission  (throughout this paper we use extent and major axis
interchangeably).  The minor axis is then the length of 
continuous emission perpendicular to the major axis (also through the peak of the 
H$\alpha$).  This definition assumes that the distribution of H$\alpha$ emission is 
well approximated by an ellipse.  This is clearly not the case for all objects in the 
sample though it is a good approximation to regions of uniform H$\alpha$ distribution 
in most systems.  We note however, that it is also important to characterise the 
overall distribution of the emission for those objects which are extended.

As can be seen from Figure~\ref{fig:maps} the H$\alpha$ morphology varies greatly 
from system to system.  We categorise these morphologies into five distinct groups.
It is important to note that the definitions are not all mutually exclusive, as such 
it is possible that a single object may exhibit a morphology which is consistent with 
two or more of the definitions given below.  

\begin{description}
\item [{\bf Compact objects}] Compact objects are defined as objects in which the minor 
extent of the H$\alpha$ emission above a significance of 7$\sigma$ is less than 
twice the mean of the DIMM seeing during the observation of that object. The 
selection 
criteria of the VIMOS sample required that every object has an extent of more than 
2'' in the FORS spectra.  As such very few objects should fall into this definition however we include it 
to account for observations taken in very poor seeing conditions. It is important to note 
that this definition of compact makes no consideration of the physical scale of the 
objects. However, it is an important classification as any apparent spatial variations 
along the minor axis seen for such objects cannot be believed.  Objects that are given this 
classification are also classified according to the structure seen along their major 
axis unless this too is less than twice the seeing.

\item[{\bf Plumes}]  An object is classified as having a plume when the H$\alpha$ emission 
shows a clear extent in one preferential direction which is not shared by the continuum
emission.  Typically with this type of object the bright central region of the H$\alpha$ emission 
also shares the shape of the larger scale plume which results in the peak of the 
H$\alpha$ emission being slightly offset from the peak in the continuum.  
This offset is never 
a large physical separation ($\lesssim$ 7 kpc) but is an important parameter to 
distinguish plumes from quiescent objects in which H$\alpha$ emission is only detected 
to one side of the BCG, the latter showing no offset larger than the mean seeing.  

\item[{\bf Offset objects}] Objects classified as offset show a H$\alpha$ morphology in which
either all of the emission is offset from the peak of the continuum or that have a 
second H$\alpha$ peak which contains the majority ($>$ 50\%) of the H$\alpha$ flux that 
is substantially offset from the BCG ($>$ 8 kpc, the maximum physical extent of the 
seeing in any object from the sample).  Although we classify both of these 
definitions as offset they show distinctly different H$\alpha$ morphologies.  The 
objects which are completely offset from the centre of the BCG appear to be otherwise 
quite similar to quiescent objects in their morphology.  The latter kind however, appear 
morphologically more consistent with plumes with a clear extent of the low surface 
brightness H$\alpha$ emission extending in a single preferential direction.  However, 
there are two key differences to plumes. First, the extent seen in the low surface 
brightness emission is not shared by a similar structure in the bright emission as is 
the case with plumes. Secondly, those which are classified as offset have a second 
peak in the H$\alpha$ emission which is positioned at the end of the extended region 
of low surface brightness emission, away from the BCG.  For a more detailed analysis 
of the nature of objects with this kind of offset emission see \citet{ham12}.

\item[{\bf Quiescent objects}] Objects classified as quiescent have H$\alpha$ emission which 
is extended beyond twice the seeing and shows a simple elliptical and centrally 
concentrated morphology.  The peak of the H$\alpha$ emission must lie within the 
mean seeing of the peak in the continuum and the H$\alpha$ flux must fall off uniformly 
as the continuum does (though not necessarily at the same rate as the continuum).  
Objects with a H$\alpha$ morphology of this type are likely to be very relaxed and 
have had no major events occur recently enough to disturb the gas.

\item[{\bf Disturbed objects}] The final morphological classification we present are disturbed 
objects.  These are objects where the H$\alpha$ flux typically peaks close to the peak of the 
continuum but have lower surface brightness H$\alpha$ emission which is extended and forms 
non-uniform structures around the BCG.  This classification of object essentially 
includes all objects which are not classified under one of the previous four 
categories. The extended and non-uniform nature of the H$\alpha$ emission in this 
class of objects suggests that some event has disturbed the cold gas in the core of the 
cluster causing it to flow away from a uniform distribution.  Another possibility is 
that this emission is tracing the gas being accreted by the system.
Some of the objects which 
fall into this category are systems which are well known to have active AGN which can be 
seen to be disturbing the ICM \citep[for example Abell~2052][]{bla11}.  This suggests the 
possibility that objects which fall into this classification are currently undergoing 
feedback processes which has resulted in the disturbance of the cold gas reservoir.

\end{description}

\begin{figure}
\psfig{figure=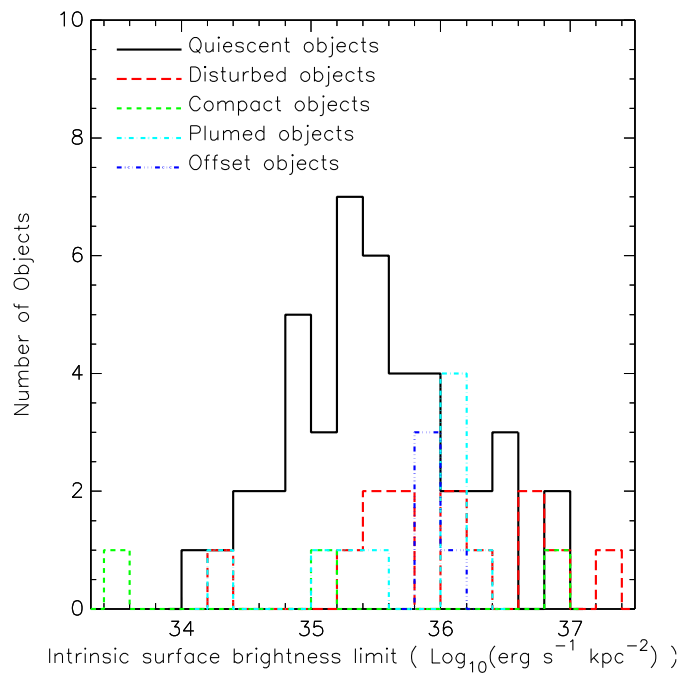,width=9cm}
\caption[Histogram of intrinsic surface brightness limit for each classification of object]
{A histogram showing the distribution of intrinsic surface brightness limits for each classification 
of object. We note that, with the exception of offset objects, all classifications sample the majority 
of the range of surface brightness limits.  Interestingly, the plot does show that 11 out of 14 objects 
with a surface brightness limit of below 10$^{35}$\,erg\,s$^{-1}$\,kpc$^{-2}$ are classified as quiescent 
suggesting that the higher surface brightness features associated with more complex morphologies
can be identified in all systems.}
\label{fig:sbhist}
\end{figure}

\begin{figure}
\psfig{figure=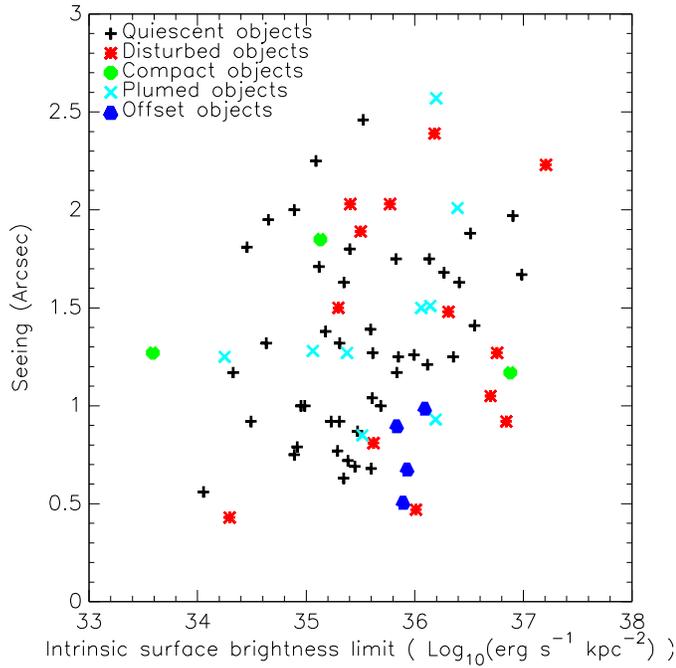,width=9cm}
\caption[Plot of the mean seeing for each object against the intrinsic surface brightness limit separated by classification of the object]{A plot showing the mean seeing of the observations for each object against that objects intrinsic surface brightness limit.  Each object is given a different symbol depending on its classification. If the seeing of our observations played a major role in our ability to distinguish the morphology of an object we would expect to see trends on this plot.  However, with the exception of offset objects, all others fill the parameter space with no obvious trends.  The offset objects are only seen at good seeing, $\sim1''$ or below.}
\label{fig:sbseeing}
\end{figure}

\noindent The classification assigned to each 
object in the sample can be found in Table \ref{tab:obs}. Some care must be taken when 
considering these classifications (in particular the distinction between quiescent and 
disturbed objects) as the surface brightness limits of our 
objects ranges considerably, from 0.38\,$\times$\,10$^{34}$\,erg\,s$^{-1}$\,kpc$^{-2}$ 
to 1.61\,$\times$\,10$^{37}$\,erg\,s$^{-1}$\,kpc$^{-2}$.  We note however, that these are 
extreme examples and the sample has a median surface brightness limit of 
3.97\,$\times$\,10$^{35}$\,erg\,s$^{-1}$\,kpc$^{-2}$ with a median absolute deviation 
of 3.19\,$\times$\,10$^{35}$\,erg\,s$^{-1}$\,kpc$^{-2}$.  A histogram of the surface 
brightness limits grouped by classification can be seen in Figure~\ref{fig:sbhist}.
The disturbed objects (listed in 
Table \ref{tab:obs}) sample the full range of surface brightness limits ranging from 
1.90\,$\times$\,10$^{34}$\,erg\,s$^{-1}$\,kpc$^{-2}$ to 
1.64\,$\times$\,10$^{37}$\,erg\,s$^{-1}$\,kpc$^{-2}$. However, the median surface 
brightness 
limit of the disturbed objects is 1.03\,$\times$\,10$^{36}$\,erg\,s$^{-1}$\,kpc$^{-2}$ 
with a median absolute deviation of 8.22\,$\times$\,10$^{35}$\,erg\,s$^{-1}$\,kpc$^{-2}$ 
which does suggest that objects classified as disturbed typically have higher surface 
brightness limits than average for the sample but is consistent with the median of the 
full sample within the errors. While the small sample size of disturbed 
objects (just 13) must be considered when interpreting these values the large range 
of surface brightness limits measured for the sample suggest that the surface brightness 
limit has only a small effect on our ability to distinguish an objects morphology.  
The seeing of our observations may also play a role in our ability to classify objects 
correctly, we test this in Figure~\ref{fig:sbseeing} by plotting the seeing of the 
observations against the surface brightness limit for the object.  No obvious trends are 
visible on this plot suggesting that seeing does not have a major impact on our ability 
to identify an objects morphology.  However, we do note that all offset objects where 
observed in relatively good seeing conditions. This is not unexpected as a poorer seeing will 
limit our ability to accurately determine the position of the H$\alpha$ peak and thus 
identify the object as offset if the separation is small.   

High resolution narrow-band imaging of H$\alpha$ in cluster cores has shown the 
presence of filamentary structures surrounding the BCG.  While a few objects in our 
sample do show narrow filaments of emission which extend out into the ICM we do not 
resolve these structures in the majority of cases.  Most likely this is a result of 
the poorer seeing of our observations smoothing the filamentary structures to such an 
extent that they appear blended as a continuous region of low surface brightness 
emission extended away from the BCG.  A good example of this effect is NGC 5044 which 
is known to have filaments of H$\alpha$ emission extended to the south-west of the BCG
\citep{dav11}.  Comparing to our H$\alpha$ flux map shown in Figure~\ref{fig:maps} 
it is possible to see low surface brightness emission extended to the 
south-west over similar distances to the filaments, however, the details of the 
filamentary structures are not visible in the VIMOS observation.  In the few cases where 
we do see structures which appear similar to filaments we note that their apparent thickness
is substantially higher than those observed with narrow-band imaging.  While this 
may be an effect of the seeing it is impossible to tell without high resolution 
H$\alpha$ imaging of these systems.

\begin{table*}
\begin{center}
\scriptsize
\centerline{\sc Table \ref{tab:obs}.}
\centerline{\sc The derived parameters listed for each object in the VIMOS sample}
\smallskip
\begin{tabular}{c c c c c c c}
\hline
\noalign{\smallskip} 
Cluster & Velocity & Extent & L$_{H\alpha}$  & FWHM & [NII]/H$\alpha$ & Morphology \\ 
 & km s$^{-1}$ & kpc(arcsec) & 10$^{40}$ erg s$^{-1}$ & km s$^{-1}$ & & \\ [0.5ex]
\hline
Abell 1060 & 68 $\pm$ 11 & 1.22(4.73) & 0.043 $\pm$ 0.0086 & 107 $\pm$ 17 & 0.7 $\pm$ 0.3 & Quiescent  \\
\noalign{\smallskip}  
Abell 1084 & 269 $\pm$ 30 & 11.23(4.75) & 11 $\pm$ 1.8 & 275 $\pm$ 108 & 0.9 $\pm$ 0.5 & Quiescent  \\
\noalign{\smallskip} 
Abell 11 & 247 $\pm$ 15 & 29.58(11.37) & 70 $\pm$ 6.5 & 340 $\pm$ 61 & 0.9 $\pm$ 0.5 & Quiescent  \\
\noalign{\smallskip} 
Abell 1111 & 478 $\pm$ 22 & 21.94(7.75) & 33 $\pm$ 3.9 & 297 $\pm$ 75 & 1.1 $\pm$ 0.5 & Plume   \\
\noalign{\smallskip} 
Abell 1204 & 490 $\pm$ 29 & 26.32(7.15) & 59 $\pm$ 5.4 & 514 $\pm$ 355 & 1.7 $\pm$ 0.6 & Disturbed   \\
\noalign{\smallskip} 
Abell 133 & 210 $\pm$ 32 & 7.18(6.53) & 1.2 $\pm$ 0.16 & 308 $\pm$ 14 & 1.7 $\pm$ 0.2 & Plume  \\
\noalign{\smallskip} 
Abell 1348 & 755 $\pm$ 40 & 18.05(8.35) & 46 $\pm$ 6.3 & 513 $\pm$ 153 & 1.6 $\pm$ 0.5 & Quiescent   \\
\noalign{\smallskip} 
Abell 1663 & 225 $\pm$ 45 & 13.71(8.66) & 1.7 $\pm$ 0.29 & 526 $\pm$ 120 & 3.4 $\pm$ 0.4 & Quiescent/Compact \\
\noalign{\smallskip} 
Abell 1668 & 302 $\pm$ 22 & 10.90(8.92) & 4.7 $\pm$ 0.45 & 339 $\pm$ 28 & 2.2 $\pm$ 0.7  & Quiescent  \\
\noalign{\smallskip} 
Abell 194 & 85 $\pm$ 7 & 1.49(4.03) & 0.50 $\pm$ 0.071 & 211 $\pm$ 122 & 0.8 $\pm$ 0.23 & Compact \\
\noalign{\smallskip} 
Abell 1991 & 124 $\pm$ 10 & 16.41(14.38) & 4.0 $\pm$ 0.35 & 209 $\pm$ 64 & 1.5 $\pm$ 0.3 & Offset  \\
\noalign{\smallskip} 		
Abell 2052 & 268 $\pm$ 18 & 6.44(9.41) & 1.8 $\pm$  0.18 & 454 $\pm$ 105 & 2.2 $\pm$ 1.1 & Disturbed  \\
\noalign{\smallskip} 
Abell 2390 & 490 $\pm$ 29 & 25.99(7.06) & 109 $\pm$ 15 & 422 $\pm$ 52 & 0.95 $\pm$ 0.48 & Quiescent  \\
\noalign{\smallskip} 
Abell 2415 & 186 $\pm$ 37 & 11.25(10.08) & 4.5 $\pm$ 0.42 & 275 $\pm$ 180 & 1.5 $\pm$ 0.55 & Plume  \\
\noalign{\smallskip} 
Abell 2495 &  201 $\pm$ 16 & 8.88(5.93) & 0.56 $\pm$ 0.090 & 235 $\pm$ 38 & 1.5 $\pm$ 0.074 & Quiescent  \\
\noalign{\smallskip} 
Abell 2566 & 325 $\pm$ 22 & 10.18(6.5) & 9.7 $\pm$ 1.2 & 172 $\pm$ 74 & 0.91 $\pm$ 0.56 & Offset  \\
\noalign{\smallskip} 
Abell 2580 & 203 $\pm$ 24 & 9.40(5.71) & 2.4 $\pm$ 0.48 & 340 $\pm$ 157 & 0.75 $\pm$ 0.55 & Compact/Plume \\
\noalign{\smallskip} 
Abell 2734 & 210 $\pm$ 20 & 6.09(5.09) & 1.1 $\pm$ 0.27 & 551 $\pm$ 105 & 2.5 $\pm$ 1.0 & Quiescent  \\
\noalign{\smallskip} 
Abell 291 & 294 $\pm$ 70 & 23.14(7.12) & 480 $\pm$ 31 & 626 $\pm$ 371 & 0.99 $\pm$ 0.55 & Disturbed \\
\noalign{\smallskip} 
Abell 3017 & 390 $\pm$ 31 & 25.03(7.06) & 110 $\pm$ 17 & 366 $\pm$ 106 & 0.62 $\pm$ 0.33 & Quiescent  \\
\noalign{\smallskip} 
Abell 3112 & 225 $\pm$ 19  & 8.94(6.20) & 7.1 $\pm$ 1.3 & 893 $\pm$ 256 & 1.9 $\pm$ 1.2 & Quiescent  \\
\noalign{\smallskip} 
Abell 3378 & 106 $\pm$ 15 & 20.63(8.30) & 6.1 $\pm$ 0.95 & 422 $\pm$ 170 & 1.1 $\pm$ 0.39 & Quiescent(Filament) \\	
\noalign{\smallskip} 	
Abell 3444 & 428 $\pm$ 41 & 35.70(8.99) & 66 $\pm$ 10 & 394 $\pm$ 140 & 0.67 $\pm$ 0.54 & Offset \\	
\noalign{\smallskip} 	
Abell 3574 &  161 $\pm$ 15 & 1.53(5.32) & 0.88 $\pm$ 0.061 & 82 $\pm$ 16 & 0.56 $\pm$ 0.37 & Disturbed  \\
\noalign{\smallskip} 
Abell 3581 & 464 $\pm$ 19 & 5.43(12.44) & 2.4 $\pm$ 0.14 & 379 $\pm$ 69 & 1.6 $\pm$ 1.0 & Disturbed  \\
\noalign{\smallskip} 
Abell 3605 & 269 $\pm$ 20 & 7.37(5.96) & 4.5 $\pm$ 0.72 & 484 $\pm$ 130 & 2.4 $\pm$ 1.0 & Quiescent  \\
\noalign{\smallskip} 
Abell 3638 & 111 $\pm$ 19 & 8.61(5.91) & 5.9 $\pm$ 0.90 & 403 $\pm$ 67 & 1.3 $\pm$ 0.97 & Quiescent \\
\noalign{\smallskip} 
Abell 3639 & 150 $\pm$  58 & 18.79(7.16) & 56 $\pm$ 4.5 & 367 $\pm$ 73 & 1.4 $\pm$ 0.56 & Quiescent  \\
\noalign{\smallskip} 
Abell 3806 & 136 $\pm$ 14 & 5.90(4.16) & 3.3 $\pm$ 0.69 & 231 $\pm$ 52 & 2.9 $\pm$ 0.88 & Quiescent  \\
\noalign{\smallskip} 
Abell 383 & 181 $\pm$ 52 & 24.25(7.69) & 40 $\pm$ 5.0 & 386 $\pm$ 73 & 1.5 $\pm$ 0.64 & Plume  \\
\noalign{\smallskip} 
Abell 3880 & 444 $\pm$ 20 & 13.24(11.72) & 12 $\pm$ 0.89 & 438 $\pm$ 50 & 1.8 $\pm$ 0.95 & Plume  \\
\noalign{\smallskip} 
Abell 3998 & 346 $\pm$ 66 & 17.85(10.61) & 7.5 $\pm$ 0.57 & 253 $\pm$ 137 & 1.4 $\pm$ 0.18 & Plume(Filament)  \\
\noalign{\smallskip} 
Abell 4059 & 396 $\pm$ 46 & 7.89(8.22) & 4.1 $\pm$ 0.48 & 536 $\pm$  97 & 2.5 $\pm$ 2.0 & Quiescent  \\

\hline
\end{tabular}  
\caption[Derived parameters listed for each object in the VIMOS sample]{Here we list the derived parameters for every object from the sample.  The velocity column refers to the peak to peak velocity measured across the mean line of sight velocity map. The extent is the length of the major axis. The luminosity is calculated only for flux that appears at greater than 7$\sigma$ significance. The FWHM is measured from the spectrum of the central $\sim$ 2 $\times$ 2 arcsec}
\label{tab:obs}
\end{center}
\end{table*}

\begin{table*}
\ContinuedFloat
\begin{center}
\scriptsize

\smallskip
\begin{tabular}{c c c c c c c}
\hline
\smallskip
Cluster & Velocity & Extent & L$_{H\alpha}$ & FWHM & [NII]/H$\alpha$ & Morphology \\ 
 & km s$^{-1}$ & kpc(arcsec) & 10$^{40}$ erg s$^{-1}$ & km s$^{-1}$ & & \\ [0.5ex] 
\hline 
Abell 478 & 176 $\pm$ 12 & 17.04(10.62) & 23 $\pm$ 2.3 & 261 $\pm$ 32 & 1.1 $\pm$ 0.92 & Quiescent  \\
\noalign{\smallskip} 
Abell 496 & 252 $\pm$ 8 & 8.19(12.46) & 3.1 $\pm$ 0.28 & 260 $\pm$ 46 & 2.1 $\pm$ 1.3 & Quiescent   \\
\noalign{\smallskip} 
Abell 795 & 315 $\pm$ 38 & 11.13(4.66) & 18 $\pm$ 3.7 & 755 $\pm$ 64 & 1.5 $\pm$ 0.85 & Quiescent  \\	
\noalign{\smallskip} 
Abell 85 & 217 $\pm$ 29  & 8.31(7.74) & 1.6 $\pm$ 0.20 & 339 $\pm$ 54 & 2.3 $\pm$ 0.47 & Quiescent(filament)   \\
\noalign{\smallskip} 
HCG62 & 168 $\pm$ 12 & 3.30(11.38) & 0.087 $\pm$ 0.011 & 204 $\pm$ 101 & 3.0 $\pm$ 0.53 & Quiescent    \\
\noalign{\smallskip} 
Hydra-A & 556 $\pm$ 6 & 7.95(7.68) & 13 $\pm$ 1.7 & 346 $\pm$ 58 & 0.86 $\pm$ 0.56 & Quiescent    \\
\noalign{\smallskip} 
NGC4325 & 255 $\pm$ 10 & 6.66(12.98) & 0.90 $\pm$ 0.089 & 236 $\pm$ 34 & 1.5 $\pm$ 0.59 & Disturbed    \\
\noalign{\smallskip} 
NGC5044 & 277 $\pm$ 18 & 5.60(30.51) & 0.54 $\pm$ 0.062 & 457 $\pm$ 99 & 2.5 $\pm$ 0.65 & Disturbed   \\
\noalign{\smallskip} 
NGC533 & 211 $\pm$ 19 & 2.77(7.14) & 0.19 $\pm$ 0.033 & 376 $\pm$ 218 & 3.1 $\pm$ 1.4 & Quiescent    \\
\noalign{\smallskip}
NGC5813 & 223 $\pm$ 24 & 1.48(10.61) & 0.044 $\pm$ 0.0059 & 503 $\pm$ 81 & 2.1 $\pm$ 0.63 & Disturbed   \\
\noalign{\smallskip} 
NGC5846 & 117 $\pm$ 12 & 1.27(10.12) & 0.076 $\pm$ 0.0056 & 255 $\pm$ 63 & 3.0 $\pm$ 0.88 & Quiescent   \\
\noalign{\smallskip} 		
RXCJ0120.9-1351 & 345 $\pm$ 8 & 4.66(4.71) & 1.5 $\pm$ 0.28 & 361 $\pm$ 40 & 2.2 $\pm$ 1.3 & Quiescent(filament)   \\
\noalign{\smallskip} 
RXCJ0132.6-0804 & 452 $\pm$ 16 & 24.78(9.57) & 43 $\pm$ 5.7 & 434 $\pm$ 66 & 1.5 $\pm$ 0.66& Quiescent    \\
\noalign{\smallskip} 
RXCJ0331.1-2100 & 125 $\pm$ 38 & 14.47(4.51) & 54 $\pm$ 7.6 & 592 $\pm$ 184 & 1.2 $\pm$ 1.0 & Quiescent   \\
\noalign{\smallskip} 
RXCJ0543.4-4430 & 245 $\pm$ 53 & 12.97(4.61) & 4.7 $\pm$ 0.98 & 259 $\pm$ 169 & 1.9 $\pm$ 0.35 & Quiescent    \\
\noalign{\smallskip} 
RXCJ0944.6-2633 & 289 $\pm$ 8 & 25.54(10.19) & 36 $\pm$ 4.4 & 277 $\pm$ 132 & 0.58 $\pm$ 0.40 & Disturbed    \\
\noalign{\smallskip} 
RXCJ1257.1-1339 & 468 $\pm$ 17 & 6.82(22.74) & 1.4 $\pm$ 0.087 & 274 $\pm$ 116 & 1.1 $\pm$ 1.0 & Quiescent    \\
\noalign{\smallskip} 
RXCJ1304.2-3030 & 343 $\pm$ 17 & 3.33(14.87) & 1.2 $\pm$ 0.072 & 375 $\pm$ 41 & 1.5 $\pm$ 0.9 & Quiescent(filament)    \\
\noalign{\smallskip} 
RXCJ1436.8-0900 & 199 $\pm$ 16 & 14.57(9.55) & 14 $\pm$ 1.7 & 273 $\pm$ 14 & 1.4 $\pm$ 1.0 & Quiescent    \\
\noalign{\smallskip} 
RXCJ1511.5+0145 & 138 $\pm$ 45 & 5.03(6.38) & 0.29 $\pm$ 0.062 & 553 $\pm$ 113 & 3.0 $\pm$ 0.40 & Quiescent    \\
\noalign{\smallskip} 
RXCJ1524.2-3154 & 310 $\pm$ 19 & 15.57(8.30) & 46 $\pm$ 3.9 & 492 $\pm$ 47 & 1.3 $\pm$ 1.2 & Plume    \\
\noalign{\smallskip} 
RXCJ1539.5-8335 & 333 $\pm$ 59 & 13.74(9.59) & 30 $\pm$ 2.0 & 572 $\pm$ 123 & 1.1 $\pm$ 0.56 & Disturbed    \\
\noalign{\smallskip} 
RXCJ1558.3-1410 & 275 $\pm$ 28 & 13.82(7.71) & 22 $\pm$ 2.6 & 465 $\pm$ 103 & 1.2 $\pm$ 0.71 & Quiescent    \\
\noalign{\smallskip} 
RXCJ2014.8-2430 & 539 $\pm$ 13 & 18.60(6.92) & 140 $\pm$ 12 & 462 $\pm$ 92 & 0.78 $\pm$ 0.55 & Quiescent    \\
\noalign{\smallskip} 
RXCJ2101.8-2802 & 169 $\pm$ 10 & 7.16(10.76) & 1.0 $\pm$ 0.11 & 106 $\pm$ 17 & 0.54 $\pm$ 0.32 & Offset    \\
\noalign{\smallskip} 
RXCJ2129.6+0005 & 157 $\pm$ 29 & 30.53(8.20) & 32 $\pm$ 4.6 & 314 $\pm$ 64 & 0.93 $\pm$ 0.32 & Quiescent    \\
\noalign{\smallskip} 
RXCJ2213.0-2753 & 185 $\pm$ 17 & 32.25(2.72) & 0.92 $\pm$ 0.32 & 660 $\pm$ 372 & 1.4 $\pm$ 1.1 & Compact/Quiescent   \\
\noalign{\smallskip} 
RXJ0000.1+0816 & 128 $\pm$ 3 & 3.55(4.57) & 1.9 $\pm$ 0.17 & 388 $\pm$ 74 & 1.7 $\pm$ 0.57 & Disturbed    \\
\noalign{\smallskip} 
RXJ0338+09 & 388 $\pm$ 6 & 8.60(12.54) & 8.3 $\pm$ 0.50 & 327 $\pm$ 77 & 1.1 $\pm$ 0.96 & Disturbed    \\
\noalign{\smallskip} 
RXJ0352.9+1941 & 244 $\pm$ 19 & 17.81(9.0) & 62 $\pm$ 5.3 & 551 $\pm$ 45 & 0.79 $\pm$ 0.64  & Quiescent     \\
\noalign{\smallskip} 
RXJ0439.0+0520 & 236 $\pm$ 54 & 26.15(7.70) & 62 $\pm$ 7.1 & 559 $\pm$ 92 &  & Quiescent    \\
\noalign{\smallskip} 
RXJ0747-19 & 354 $\pm$ 32 & 24.76(13.11) & 63 $\pm$ 6.1 & 422 $\pm$ 83 & 1.7 $\pm$ 0.77 & Disturbed    \\
\noalign{\smallskip} 
RXJ0821+07 & 270 $\pm$ 20 & 20.44(10.12) & 35 $\pm$ 3.7 & 180 $\pm$ 49 & 0.84 $\pm$ 0.49 & Plume    \\
\noalign{\smallskip} 
RXJ1651.1+0459 & 309 $\pm$ 25 &  17.50(6.52) & 29 $\pm$ 4.0 & 276 $\pm$ 152 & 0.54 $\pm$ 0.39 & Quiescent    \\
\noalign{\smallskip} 	
S555 & 287 $\pm$ 20 & 11.54(13.08)  & 4.5 $\pm$ 0.44 & 578 $\pm$ 234 & 1.8 $\pm$ 1.1 & Quiescent    \\
\noalign{\smallskip} 
S780 & 420 $\pm$ 17 & 33.04(8.87) & 240 $\pm$ 28 & 490 $\pm$ 49 & 1.1 $\pm$ 0.95 & Plume    \\
\noalign{\smallskip} 
S805 & 141 $\pm$ 22 & 2.75(8.91) & 0.18 $\pm$ 0.018 & 227 $\pm$ 59 & 3.7 $\pm$  0.95 & Quiescent(filament)    \\
\noalign{\smallskip} 
S851 & 523 $\pm$ 17 & 3.58(18.56) & 0.53 $\pm$ 0.039 & 425 $\pm$ 71 & 3.5 $\pm$ 3.1 & Quiescent    \\
\noalign{\smallskip} 
Z3179 & 278 $\pm$ 15 & 14.80(5.92) & 7.6 $\pm$ 1.5 & 482 $\pm$ 82 & 2.3 $\pm$ 0.97 & Quiescent    \\
\noalign{\smallskip} 
Z348 & 226 $\pm$ 12 & 42.02(10.66) & 390 $\pm$ 48 & 508 $\pm$ 71 & 0.57 $\pm$ 0.49 & Quiescent    \\
\hline
\end{tabular}  
\caption[]{continued.}
\end{center}
\end{table*}

\subsubsection{Offset Emission}
\label{sec:offset}

\begin{figure}
\psfig{figure=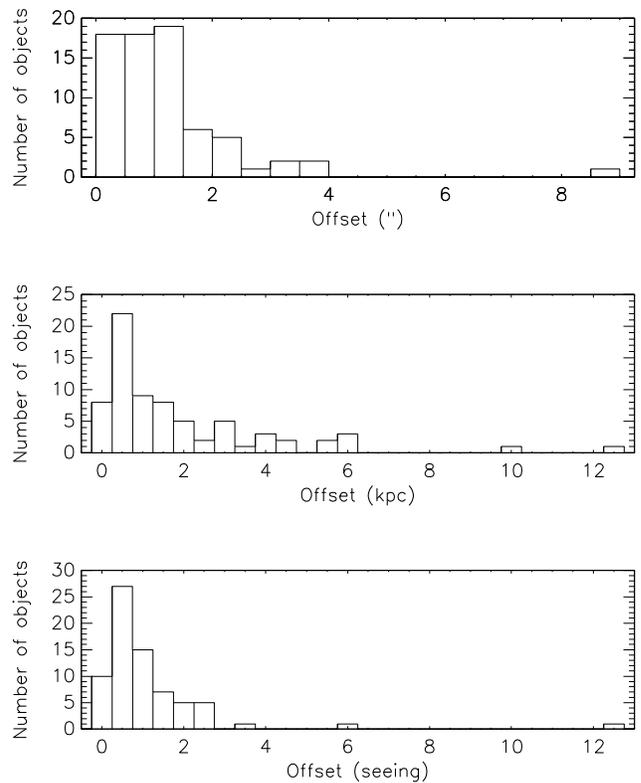,width=9cm}
\caption[Histogram of offsets between the continuum peak and the H$\alpha$ centroid]{The top plot (a) shows the apparent offset between the continuum emission of the BCG and the centroid peak of the H$\alpha$ emission 
for the full sample of objects.  Almost all objects show small apparent offsets of less than 4''.  Only one object shows a significant apparent 
offset, this is Abell 1991. The middle plot (b) shows the physical observed offset of the BCG and the majority of the line emission. Here we see a 
more spread out distribution of objects but with the majority still showing small offsets.The bottom plot (c) shows the offset as measured in terms of seeing units (offset in arcsec / mean seeing). We note that most objects are offset by less than 1 seeing unit confirming that there is no real offset in these objects. there are also two objects which show a very large offsets at 6 and 12.5 seeing units.  These are Abell 3444 and Abell 1991 respectively and clearly indicates that the offset in these objects is real.}
\label{fig:offsethist}
\end{figure}

In \citet{ham12} we report on the nature of three objects (two from this 
sample) which 
show a significant component ($>$ 50 per cent) of their line emission significantly 
offset from the centre of the BCG.  Figure~\ref{fig:offsethist} (top) shows a histogram plot of 
the apparent visible offset between the BCG and the majority of the clusters line 
emission.  It can be seen that almost all clusters show a small offset of less than 4'' 
and most well below the seeing limit ($\sim$ 1.5 arcsec). Only one object (Abell 1991) shows
an offset of greater than this which makes it clearly stand out amongst our sample.
In the middle panel of Figure~\ref{fig:offsethist} we present the physical offset of 
the BCG from the majority of the line emission calculated at the redshift of the host 
cluster.

We draw three conclusions from this plot, the first is that the majority of the objects 
show a small physical offset as is expected.  Second, we note that there 
are now two objects showing a significant offset, one at $\sim$ 10~kpc and one at 
$\sim$ 12.5~kpc.  These objects are Abell~1991 and Abell~3444 which were studied in 
\citet{ham12}.  
Additionally, there is a cluster of objects with an offset of $\sim$ 6 kpc in Figure~\ref{fig:offsethist}.
Studying the maps of these clusters, we note they are mostly plumes rather than separate 
offset emission.  However, one of these objects is Abell~2566 which is offset and shows 
no line emission at the location of the continuum peak.  In 
this respect it is similar to the Ophiuchus cluster identified by \citet{edw09}.

\citet{ham12} postulated that the apparent differences between Ophiuchus and 
Abell~1991/~3444 may have been the effect of viewing the offsets inclined at different 
angles to the plane of the sky.  Ophiuchus had a physical offset of just 2.2~kpc but a 
dynamical offset of $\sim$600~km~s$^{-1}$ while Abell~1991~/~3444 have much larger 
physical offsets but dynamical offsets on the order of $\sim$100~km~s$^{-1}$.  
Abell~2566 has an offset of 5.8~kpc putting it between the two extremes.  Comparing the 
redshift of the H$\alpha$ to that of the NaD from the BCG we find a dynamical offset of 
210~$\pm$~21~km~s$^{-1}$ which is consistent with this postulation.

\begin{figure*}
\psfig{figure=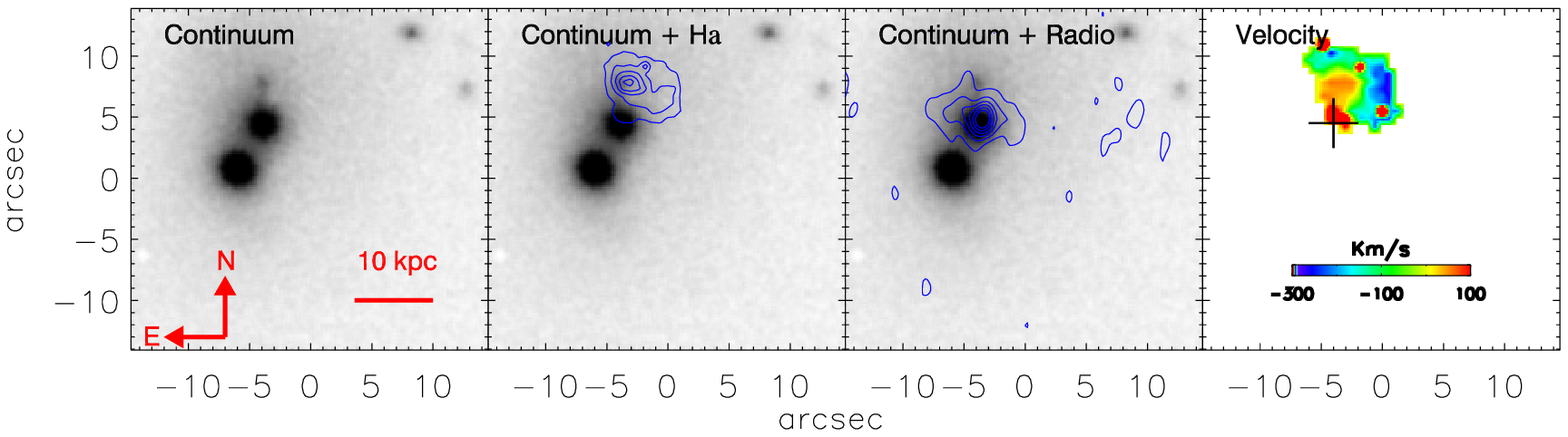,width=17cm}
\caption[Offset emission in Abell~2566]{{\em Left} - FORS1 R band image of Abell~2566, the BCG is the northern of the two objects.  A region of low surface brightness emission can be seen extending north from the BCG.  {\em Centre Left} - FORS1 R band image of Abell~2566 with the H$\alpha$ emission contoured in blue.  A clear offset between the BCG and the H$\alpha$ emission of $\sim$ 6kpc can be seen. We note that the peak of the H$\alpha$ emission is coincident with the position of the extended low surface brightness structure seen in the continuum image. {\em Centre Right} - FORS1 R band image of Abell~2566 with the radio map taken from \citet{ol97} contoured in blue.  {\em Right} - H$\alpha$ velocity field of Abell~2566 as shown in Appendix~\ref{app:maps}.  The cross on this image marks the centre of the BCG.}
\label{fig:A2566}
\end{figure*}

Abell~2566 differs from the other offset objects which are studied in \citet{ham12}
in that it shows a possible close companion in the VIMOS field of view 
(Figure \ref{fig:A2566}).  
Comparing the NaD of the second galaxy to that of the BCG we find a velocity difference 
of 279~$\pm$~29~km~s$^{-1}$ confirming it is a cluster member and not a foreground 
galaxy.
This suggests 
the possibility that the two are 
strongly interacting.  A FORS1 R band acquisition image shows an extent of the continuum emission 
in the direction of the H$\alpha$ similar to clumpy complex structures seen in other 
systems where the BCG and a companion are interacting.  The offset objects 
previously studied showed no evidence of an interaction with another cluster member, so 
while \citet{ham12} suggest an interaction as the likely cause of the offset it was not 
possible to confirm it unambiguously.  Unfortunately 
Abell~2566 lacks high resolution X-ray observations which are required to study the ICM 
in detail. So, while 
it may be tempting to speculate that the offset H$\alpha$ emission in this system is 
evidence of sloshing induced by an interaction more detailed 
observations are required before this can be confirmed.

\begin{figure}
\psfig{figure=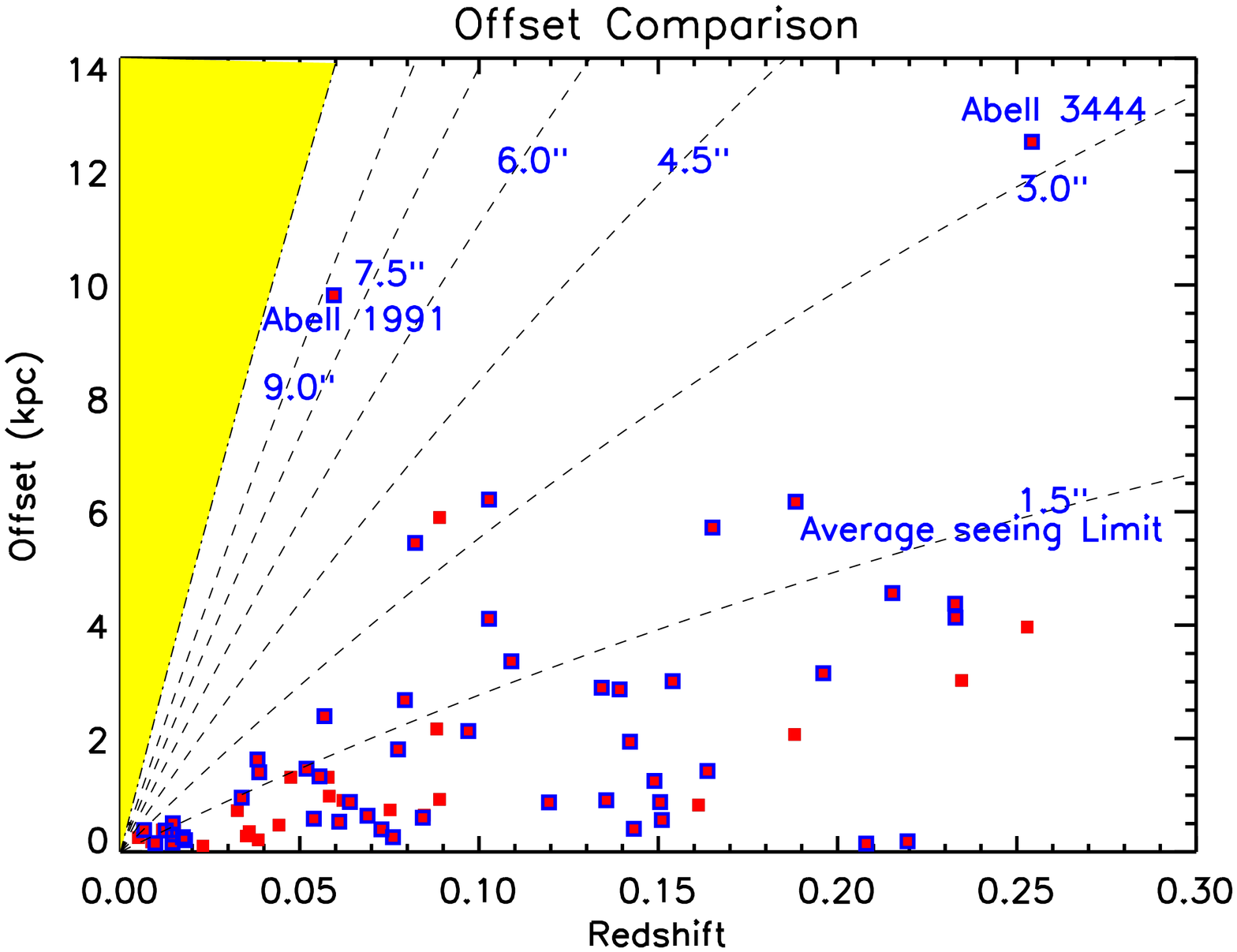,height=9cm,angle=90,bbllx=54,bblly=54,bburx=528,bbury=700,clip=}
\psfig{figure=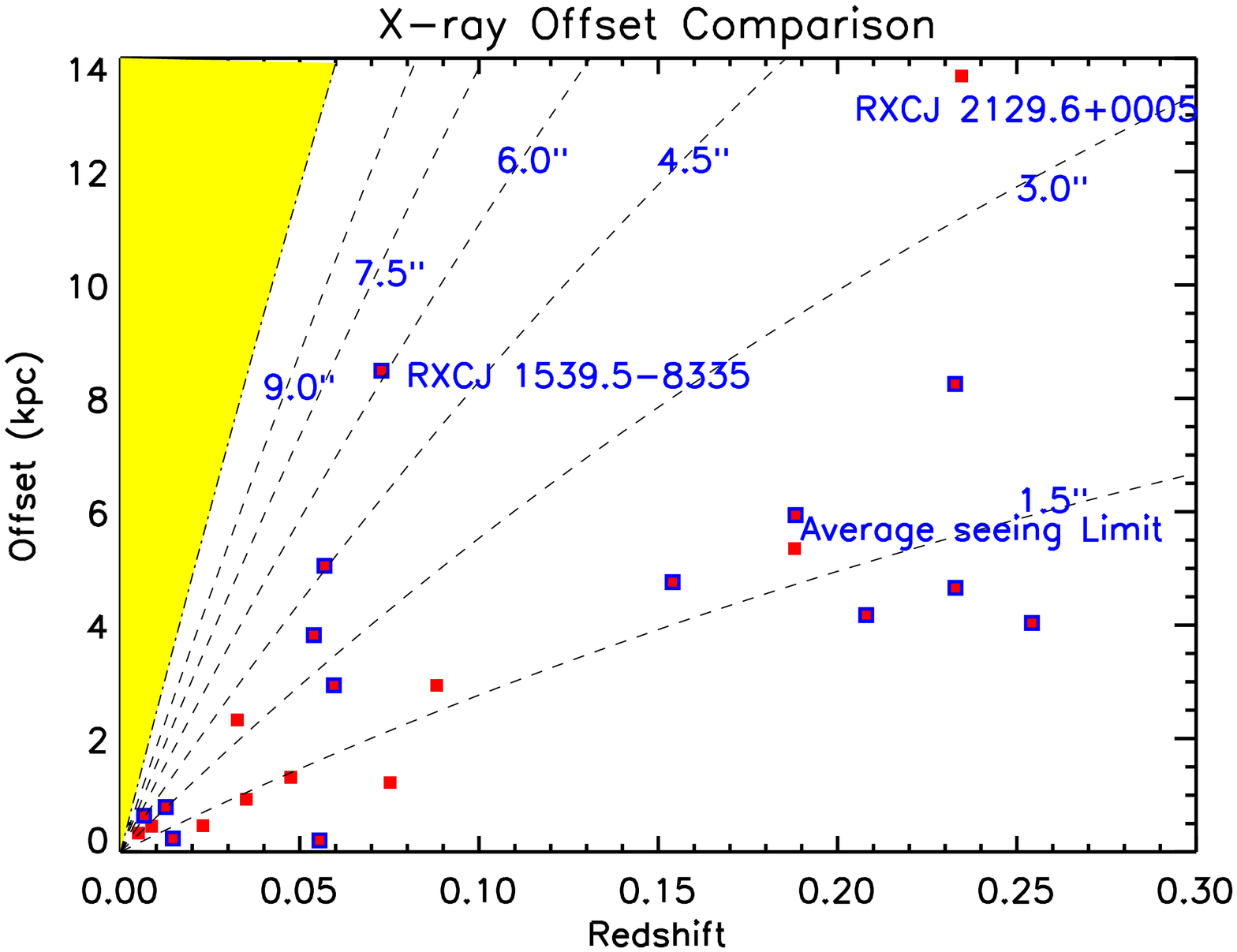,height=9cm,angle=90,bbllx=54,bblly=54,bburx=528,bbury=700,clip=}
\caption[Observed physical offset between the line emission and the BCG with cluster redshift]{{\em Top} - Observed physical offset between the BCG and the majority of the line emission at the cluster redshift. {\em Bottom} - Observed Physical offset between the centre of the clusters X-ray emission and the majority of the line emission at the clusters redshift. The dashed lines show constant visible offsets in 1.5'' steps.  We can see that the majority of objects fall well below the average seeing limit of 1.5''.  The dot-dashed line shows the extent of the VIMOS field of view ($\sim$ 12'' assuming the BCG was positioned at the centre of the field of view) with the shaded region marking the extent beyond the field of view. Objects whose seeing during the VIMOS observations was less than 1.5'' (the average seeing limit) have their points highlighted by a blue outline.}
\label{fig:offsetcomp1}
\end{figure}

Figure~\ref{fig:offsetcomp1} (top panel) shows the physical offset against the cluster redshift for 
the whole sample. It can clearly be seen here that the offsets for the majority of the 
sample fall below the seeing limit of the VIMOS observations. Abell 1991 and Abell 3444 again 
clearly stand out from the sample, and we see the group of plumed  
objects (along with Abell~2566) at an offset 
of $\sim$ 6 kpc.  The dot-dashed line in Figure~\ref{fig:offsetcomp1} marks the limit of 
the VIMOS field of view, assuming the BCG was at the centre of the pointing, objects 
with a separation placing them above this line would not be detected as the offset 
emission would fall outside the observed area of sky.  The bottom panel of 
Figure~\ref{fig:offsetcomp1} shows a similar comparison between the centre of the
clusters X-ray emission and the H$\alpha$ emission for objects with high resolution 
X-ray observations.  The X-ray centres are taken from the ACCEPT data base 
\citep{cav09} with the exception 
of Abell 3444 for which we use the position defined in \citet{ham12}.  Most of the 
objects have an offset close to the average seeing limit of VIMOS suggesting that 
there is no significant offset seen.  However, we note that 3 objects have an offset 
of more than 4~kpc that are significantly greater than the average seeing.  One is 
Abell 133 which shows an offset 5~kpc (4.5 arcsec). We note that the continuum image of 
this object shows a bright source 4.5 arcsec to the north of the BCG which matches the 
position of the X-ray centre suggesting that the true cluster centre may be obscured by a 
point source as was the case for Abell 3444 \citep{ham12}.  The other two objects are 
RXCJ 1539.5-8335 and RXCJ 2129.6+0005 which  show large offsets of 8.4~kpc (6 arcsec) 
and 13.7~kpc (3.6 arcsec) respectively.  RXCJ 1539.5-8335 may be erroneous as the VIMOS 
field of view only covers half of the BCG thus the position of the true peak of the 
H$\alpha$ emission is uncertain.  RXCJ 2129.6+0005 is quite high 
redshift for the sample (z\,$=$\,0.234) and we note that the seeing is above the average 
seeing limit suggesting the apparent offset may be a result of the seeing.  However, 
the average seeing during the observations of this object was 1.8 arcsec, roughly 
half the observed offset so it cannot be explained due to seeing alone. Additionally 
the position of the 
H$\alpha$ peak is well defined and no continuum source is present at the location of 
the X-ray centre suggesting that this offset may be real. Further dedicated study of 
these 3 objects is required to confirm if these offsets are real, however even if they 
all are they account for just 12\% (3/24) of the objects studied.  This would suggest 
that the optical line emission follows the X-ray peak in the majority of systems 
agreeing with the findings of \citet{ham12}.

Figures~\ref{fig:offsethist}$+$\ref{fig:offsetcomp1} highlight the difficulty in 
detecting offset emission 
with VIMOS observations.  Due to the limited field of view large offsets at low redshift 
would not be detected.  Likewise a smaller physical offset at the higher redshift end of 
our sample would have an apparent offset on the scale of the seeing (the offset in Abell 
1991 would be
$<$3 arcsec at the redshift of Abell 3444, for example) preventing its detection.  We also 
note that objects with an offset along the line of sight would not be easily detected  
when considering only the structure of the ionised gas.  The discovery of other offset 
objects does not significantly alter the statistics discovered in \citet{ham12} with 
just 4 of the 73 objects in this sample showing offset emission these objects account 
for less than 5\% of the line emitting BCG population.  If the sloshing of the ICM is 
responsible for these offsets then it suggests that events capable of causing sloshing 
(such as a strong interaction) are rare or do not cause a major disruption of the 
cold gas reservoir.


\subsection{Mean velocity structure}
\label{sec:velstruct}
The velocity structure of the gas can also indicate if the gas 
is disturbed and may be a more reliable measure in systems where the lowest surface 
brightness gas may not have been detected. Maps of the mean line of sight velocity (from 
here on the mean velocity) structure of each object were produced from the parameterised 
fits so the velocity structure could be studied. The 
ionised gas in many of the cluster cores studied show a clear velocity gradient from 
negative to positive velocities relative to the systemic velocity.  This velocity 
gradient is often seen to run across the bright peak of the line emission. Velocity 
profiles like these are often indicative of rotation but can also 
arise from uniformly entrained or out flowing material with an inclination to the 
plane of the sky.  
One caveat that must be considered here is the possibility that limited resolution in 
the higher redshift objects could smooth out variations in the velocity field and make 
the overall variation across the BCG appear as a much smoother gradient than is actually 
the case. To test this we take NGC~5044, a low redshift (z$\sim$0.009) object with a 
velocity field that shows a lot of variation, and degrade its resolution to that of a 
z=0.05 object before refitting the cube and remaking the velocity maps.  While this 
significantly 
reduces the small scale variations in the velocity field seen in the higher resolution 
map the overall velocity structure still does not show a smooth gradient like those seen 
in other objects suggesting that limited resolution 
has only a small impact on our ability to distinguish between objects with a smooth 
velocity gradient and those without. 

\begin{figure}
\psfig{figure=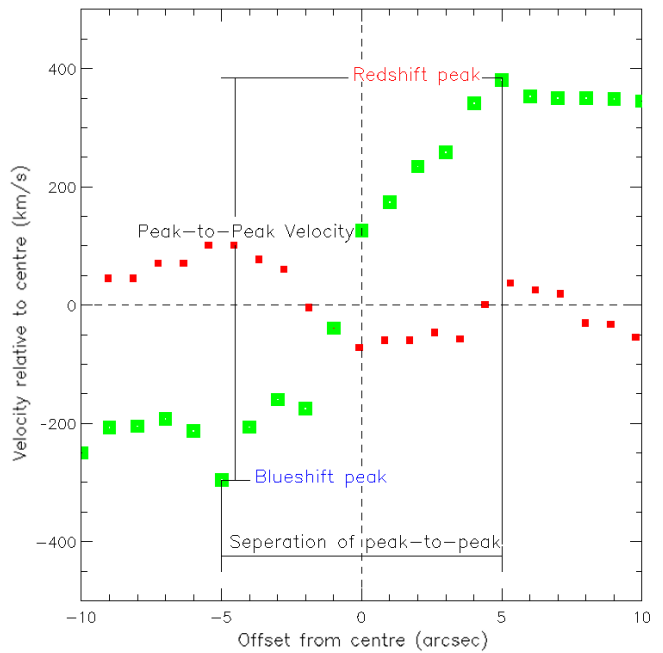,width=9cm}
\caption[Example profile for calculating peak--to--peak velocity]{Two examples of a velocity profiles taken from Hydra-A (large green points) and NGC\,5044 (small red points). These profiled were used to calculate the peak--to--peak velocity within each system. The contrast of these two objects shows how the peak--to--peak was clear in some objects (such as Hydra-A) while being much more difficult to constrain in others (such as NGC\,5044). The blueshift peak and redshift peak are labelled and the peak--to--peak velocity and peak--to--peak separation are indicated for Hydra-A.}
\label{fig:velpp}
\end{figure}

We define the peak--to--peak velocity to be the maximum velocity difference measured 
across the natural peak of the line emission. To calculate this the velocity maps were first 
median smoothed to eliminate any pixels with extreme velocities relative to their 
surrounding pixels.  A $\sim$ 2 arcsec wide slit was then placed on the map centred on 
the peak of the line emission, in the case of objects with significantly offset emission we use the 
peak of the continuum instead.  This slit was then rotated through 180$^o$ in 6$^o$ 
increments and at each position a profile was extracted 
by interpolating along the slit and smoothing 
to 2 arcsec to reduce the noise.
Figure \ref{fig:velpp} shows an example profile from Hydra-A with the blueshift peak, 
redshift peak and peak--to--peak velocity indicated. The maximum velocity 
difference between the blueshift peak and the redshift peak (relative to the redshift at 
the centre of the H$\alpha$ emission) from these profiles is then defined as the 
peak--to--peak velocity.  This is the maximum velocity difference across the line 
emission as given in Table \ref{tab:obs}. One caveat to note is that the blueshift and 
redshift peaks 
were required to be on opposite sides of the centre.  In some of the more disturbed 
velocity maps (NGC 5044 for example) the natural peaks did not occur on opposite sides 
of the centre, in such cases the peaks were defined as the points which gave the 
maximum velocity difference across the centre of the H$\alpha$ emission.

Within the more extended and disturbed objects the velocity maps show much more variation.
However, under closer scrutiny many of these objects do still often show smooth 
velocity gradients on smaller scales close to the centre of the system (defined by the 
location of the H$\alpha$ peak).  The bright central region of these systems show 
a coherent velocity structure with a strong gradient across the centre of the system.
Away from this central region however, in the lower surface brightness extended emission 
the velocity field becomes much more chaotic. Abell~3581 and RXJ~0338+09 are 
two good examples of such objects,  each show extended low surface brightness emission 
with no coherence to the velocity structure.  However, in the regions close (within 
$\sim$ 8 arcsec) to the peak of the emission the velocity field shows much less 
variation.  If it 
is the case that the extended emission is the result of some event which has disturbed 
the distribution of the cold gas then this small scale velocity structure may represent 
the overall velocity field of the object.  Events such as entrainment by radio jets 
or galaxy interactions which can disturbed the cold gas 
would initially affect the low density gas at the edges of the distribution.  The 
disturbance can impart a velocity on this gas which is not consistent with the overall 
velocity field of the system.  The densest gas however, is much more difficult to 
displace \citep{chu02} and, if it remains relatively undisturbed, should retain its previous 
velocity, which should be governed by the global velocity field within the cluster core.

If the velocity field of each object is roughly 
consistent, then any observed peak--to--peak velocity difference between them would simply
be the effect of the gas sampling different regions of the velocity distribution.  
This would present itself as a clear trend between the extent of the 
line emission in a given object and the magnitude of the velocity difference observed.

\begin{figure}
\epsfig{figure=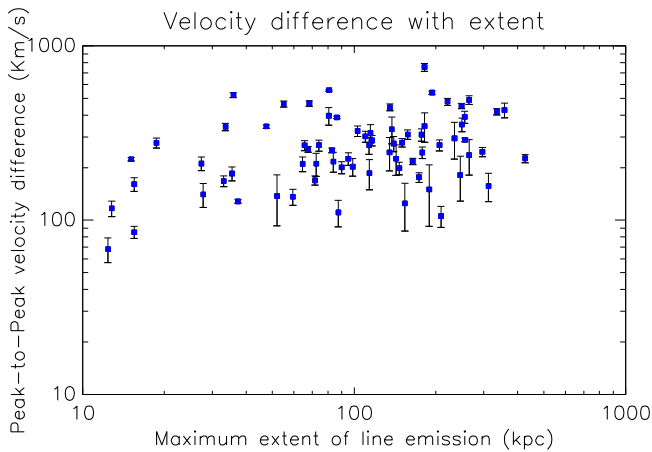,width=9cm}
\caption[Plot of peak--to--peak velocity difference against the measured extent of each object]{The peak--to--peak velocity difference against the measured extent of each object.  It is clear that there is no trend with the exception that very compact objects show a reduced velocity difference.  This suggests that the different peak--to--peak velocities seen between objects in the sample are not simply the effect of having gas sample a larger portion of the velocity field in some objects than in others.}
\label{fig:velkpc}
\end{figure}

In Figure~\ref{fig:velkpc} we show the peak--to--peak velocity
plotted against the maximum continuous extent of the line emission. A lower velocity 
difference is seen in very compact objects, however, as 
very compact objects are much more susceptible to blurring caused by the seeing this 
trend may not be real.  If the 
velocity difference is being measured over a region comparable to the seeing it is 
possible that gas at each peak may be blended with some emission from the other peak 
resulting in a shifting of the best fitting models closer together 
thus reducing the measured velocity difference. For objects with
 an extent of $>$~2~kpc the median velocity difference of the objects 
remains constant at $\sim$ 250 km s$^{-1}$ with a very large scatter.  It is important 
to note however, that any inclination of the velocity structure towards the plane of 
the sky will reduce the observed peak--to--peak velocity.

The average velocity gradient of each system can be 
calculated from the peak--to--peak velocity difference and the separation of the 
peak--to--peak (thus V$_{grad}$ = $\Delta$V/$\Delta$r). 



\begin{figure*}
\psfig{figure=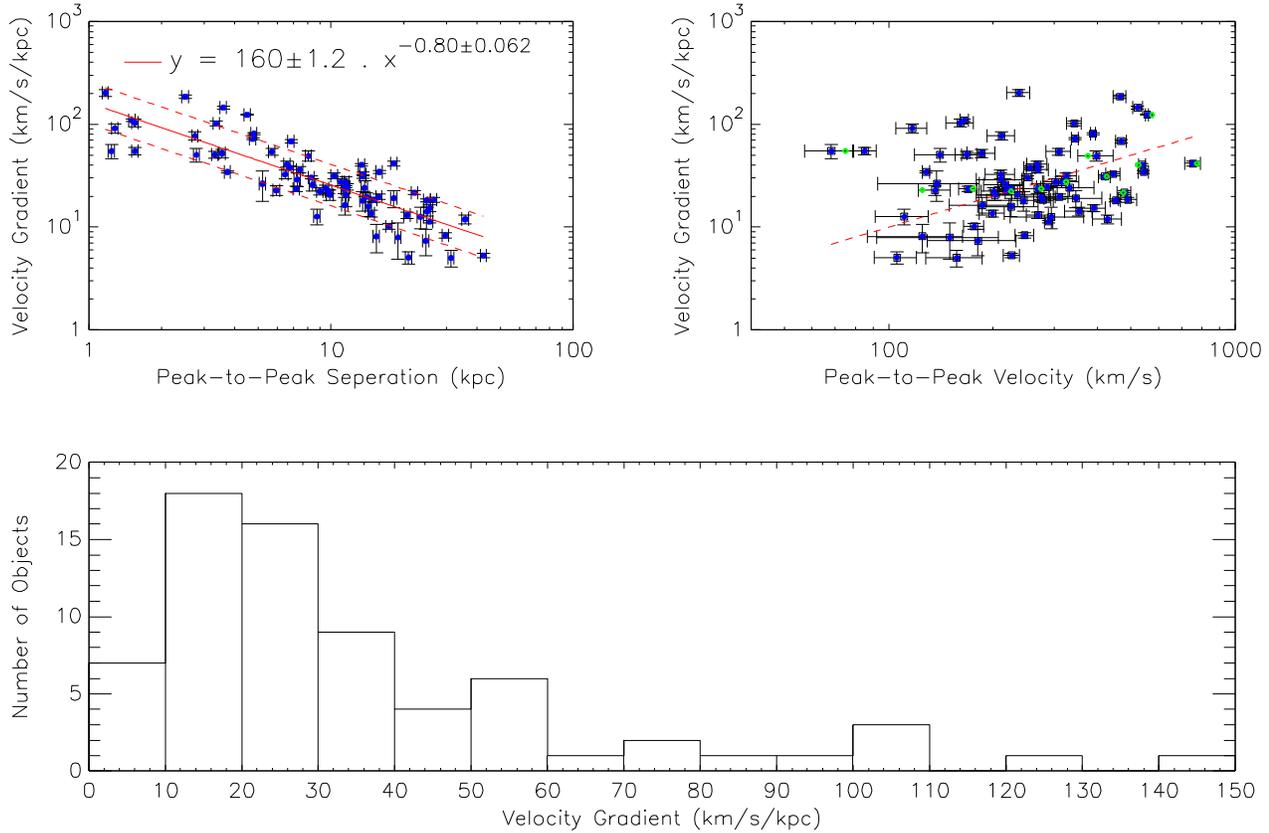,width=17cm}
\caption[Analysis of the velocity gradients present in the VIMOS sample]{
{\em Top Left} - Plot of the velocity gradient of every object in the VIMOS sample against the separation of the peak--to--peak velocity.  A clear trend can be seen in this plot indicated by the red line and the equation of the best fit is given. {\em Top Right} - In this plot we compare the velocity gradient of our objects to the measured peak--to--peak velocity difference.  No clear trend is seen, the dashed line shows a linear increase with velocity and the green diamonds show the moving average of the data. {\em Bottom} - This shows a histogram of the velocity gradients calculated for the objects in the VIMOS sample. A clear peak can be seen at $\sim$ 15 km s$^{-1}$ kpc$^{-1}$ with a steady decrease towards higher velocity gradients.  This suggests that the velocity gradients seen in the sample are sufficiently small that their effect on the velocity dispersion should be negligible. The combined interpretation from the top two plots is that the velocity gradient responds to the separation of the peak--to--peak in a non linear way while showing little correlation to the peak--to--peak velocity itself.  
}
\label{fig:velgrad}
\end{figure*}

In Figure~\ref{fig:velgrad} we plot the calculated values of velocity gradient for each 
object, the top two panels of this figure show the velocity gradient plotted against 
the two parameters used to calculate it (the peak--to--peak velocity and the separation 
of the velocity peaks).  It can be seen from these plots that the measured
velocity gradient of each object is well correlated with the separation 
with a smaller gradient in objects with a larger $\Delta$r.  By contrast the 
plot showing the velocity gradient against the velocity difference shows no obvious 
correlation.  
It is not unexpected that a correlation would be seen when plotting $\Delta$V/$\Delta$r 
against $\Delta$r.  The simple expectation if $\Delta$V and $\Delta$r are uncorrelated 
would be that V$_{grad}$ $\propto$ $\Delta$r$^{-1}$ and V$_{grad}$ $\propto$ $\Delta$V.  
However, our best fit trend line has the equation 
V$_{grad}$ = 160$\pm$1.2 $\times$ $\Delta$r$^{-0.8\pm0.06}$ which is not consistent 
with this expectation.  No significant trend is found in the plot of 
V$_{grad}$ against $\Delta$V, however we note that higher values of the peak--to--peak 
velocity have (on average) higher velocity gradients.  This is shown by the green 
diamonds which show the median velocity gradient within 50\,km\,s$^{-1}$ bins.
It can be seen that at higher peak--to--peak velocities the median velocity gradient 
is higher, as would be expected. The red dashed line shows the expected trend 
V$_{grad}$ $\propto$ $\Delta$V with the form V$_{grad} = \Delta$V\,/\,10, the points of 
median velocity follow this trend (except at extreme velocities where few points are 
sampled) suggesting that the velocity gradient is proportional to the peak--to--peak 
velocity as is expected, albeit with considerable scatter. 

In order to test the expected correlations we replace the peak--to--peak velocity 
($\Delta$V) in our analysis with the total H$\alpha$ flux of an object.  Since the 
H$\alpha$ flux comes from the full extent of each object we expect it to have no 
correlation to the separation of the peak--to--peak velocity.  We thus define 
H$\alpha_{grad}$ = F$_{H\alpha}$/$\Delta$r and plot H$\alpha_{grad}$ against 
F$_{H\alpha}$ and $\Delta$r.  We find that H$\alpha$$_{grad}$ = 62$\pm$1.5 $\times$ 
$\Delta$r$^{-0.94\pm0.15}$ and H$\alpha$$_{grad}$ = 0.11$\pm$1.4 $\times$ F$_{H\alpha}$$^{0.97\pm0.09}$.
Since the exponents of these equations are consistent with -1 and 1 respectively the 
equations are consistent with our expectation that the total H$\alpha$ flux has no 
dependence of the separation of the peak--to--peak velocity.  Thus for two unrelated 
values a and b, a/b $\propto$ a and a/b $\propto$ b$^{-1}$.  Since the correlation 
of velocity gradient ($\Delta$V/$\Delta$r) is proportional to $\Delta$r$^{-0.8}$ not 
$\Delta$r$^{-1}$ (Figure~\ref{fig:velgrad}) this suggests that $\Delta$V already has some 
dependence on $\Delta$r that is a physical phenomenon, rather than a mathematical 
artefact. As such the velocity fields have some structure as a function of radius. 

One possible explanation is that the velocity field traces a rotation. 
On large scales gas rotating under gravity should 
form a Keplerian disc about the centre of mass of the system.  The rotation curves of 
discs are well understood and can be approximated to follow the curve of an arctan 
function \citep{cou97} of the form

\begin{equation}
\rm V(r) = V_0 + \frac{2}{\pi}V_c arctan(R)
\label{eqn:arctan}  
\end{equation}

where R = (r -- r$_0$)/r$_t$, V$_0$ is the velocity at the centre of rotation, r$_0$ is 
the spatial centre of the galaxy, V$_c$ is the asymptotic velocity corrected for 
inclination (V$_c$ = V$_{asym}$ $\times$ sin~i) and r$_t$ is the 
radius at which the rotation curve transitions from rising to flat (where the rising 
part of the rotation curve has a gradient of 1). Thus 
for a rotating disc the velocity at a given distance from the centre of rotation can be 
assumed to be a function of that distance. 

If the velocity field followed such a rotation then the expected peak--to--peak velocity 
($\Delta$V) would be 2 $\times$ the rotational velocity (V(r) from 
Equation~\ref{eqn:arctan}) and the separation ($\Delta$r) would be twice the maximum 
radius (r from Equation~\ref{eqn:arctan}).  
V$_0$ can be set to 0 as we are only interested in the velocity difference not the 
absolute velocity. We used Equation~\ref{eqn:arctan} to produce 500 velocity 
profiles using randomly generated variables from within the parameter limits of our 
data (V$_{asym}$ of 200--700 km~s$^{-1}$, sin(i) of 0.0--1 and r$_t$ of 1--3 kpc). For each 
we calculated the velocity gradient (V$_{grad}$) and reproduced Figure~\ref{fig:velgrad}
for these models.  We find a correlation between V$_{grad}$ and the peak--to--peak 
velocity with the form V$_{grad}$ = 130\,$\pm$\,1.1 $\times$ $\Delta$r$^{-0.78\pm0.03}$ 
and no strong correlation of V$_{grad}$ with $\Delta$V which matches what is 
seen in the data.  This suggests that the velocity structure in the majority of our 
objects may be consistent with disc like rotation.  
However, we note that the velocity maps for some of our objects are very chaotic (e.g. 
NGC~5044) and clearly not dominated by rotation, thus a more detailed analysis is needed 
to identify rotation in these systems on a case by case basis. We will further explore 
the idea that the kinematics are related to a rotating disc in the next paper in this 
series but have already shown this to be true for one object \citep{ham14}.

\subsubsection{Velocity dispersion}
\label{sec:veldisp}
The FWHM of an emission line is a measure of the velocity dispersion of the gas 
from which it was emitted.  In observations with ordered motions (such as a difference 
in the mean velocity) the velocity dispersion is a measure of the random motions of the 
particles which make up the gas.  There is of course some care required in this 
interpretation as it requires the assumption that the variation of any ordered velocity 
field, within the region the spectrum is extracted from, is negligible compared to the random 
velocities within that same region.  
Figure~\ref{fig:velgrad} also shows a histogram of the mean velocity gradients for each 
object in the sample.  From this it can be seen that the velocity gradient for 
most objects is quite low with the distribution peaking at $\sim$ 15 km s$^{-1}$ kpc$^{-1}$. This suggests that the velocity gradients seen in the sample are sufficiently small 
that their effect on the measured linewidths should be negligible.

We give the average line width measured within the 
central $\sim$ 2 $\times$ 2 arcsec region of the system in Table \ref{tab:obs}.  
We note that the gradient of the clusters velocity field should be highest 
in the centre where the mass concentration naturally peaks and it is within this region 
that the velocity dispersion is most likely to be broadened by the velocity gradient.  
The region used to extract the FWHM is comparable to the mean 
seeing ($\sim$ 1.5 arcsec) and is thus the scale on which we would expect the mean 
velocities to be blended.  

\begin{figure}
\psfig{figure=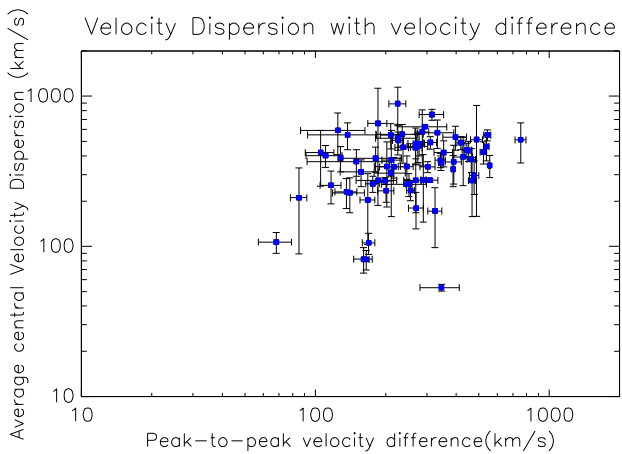,width=9cm}
\caption[Plot of central velocity dispersion (FWHM) against the peak--to--peak velocity difference of each object]{Plot of the central velocity dispersion (FWHM) against the peak--to--peak velocity difference of each object.  If the outliers with very low velocity dispersion ($<$ 100 km s$^{-1}$) and/or low velocity difference ($<$ 100 km s$^{-1}$) are ignored then there is no obvious trend suggesting the velocity dispersion is unrelated to the velocity across the object.  This confirms the assumption that the velocity dispersion measured is real and not an effect of blending gas clouds with different mean velocities. }
\label{fig:sigvel}
\end{figure}

In Figure~\ref{fig:sigvel} we show the average velocity 
dispersion within the central 2 $\times$ 2 arcsec of each objects plotted against 
the peak--to--peak of the mean velocity field.  If we ignore the points with a peak--to--peak velocity below 100 km 
s$^{-1}$ (i.e. below the spectral resolution of the {\tt HR\_Orange} and {\tt HR\_Red} VIMOS gratings)
this plot would suggest that there is no obvious correlation between the 
average central velocity dispersion and the peak--to--peak velocity difference for our 
sample. We performed a Spearmans test to confirm this lack of correlation and find that 
$\rho$ = -0.016 with a p-value of 0.9 confirming the lack of correlation.  This lack of 
correlation confirms the assumption that the velocity dispersion measured is real and 
not an effect of blending gas clouds with different mean velocities.  Higher spectral and 
spatial resolution data would be required to study the broad line regions in more detail.

The maps of most objects show a FWHM profile 
which peaks towards the centre of the system, where the flux map peaks, and falls 
away as the surface brightness of the line emission does.  This line width profile is 
exactly what is expected as in the central regions there should be a higher
projected mass density. Within most objects the extended low surface brightness gas has a 
consistently low FWHM on the order of $\sim$100--200\,km\,s$^{-1}$.
Interestingly some objects show additional peaks in the 
line width away from the centre suggesting the gas is more kinematically disturbed 
in these regions.  Indeed the fact that most of these offset FWHM peaks appear in 
systems with spatially disturbed morphologies (Abell~3574, RXJ~0338+09) would seem to support this 
assumption.


\subsection{Statistics of the sample}
\label{sec:stats}
\begin{figure}
\psfig{figure=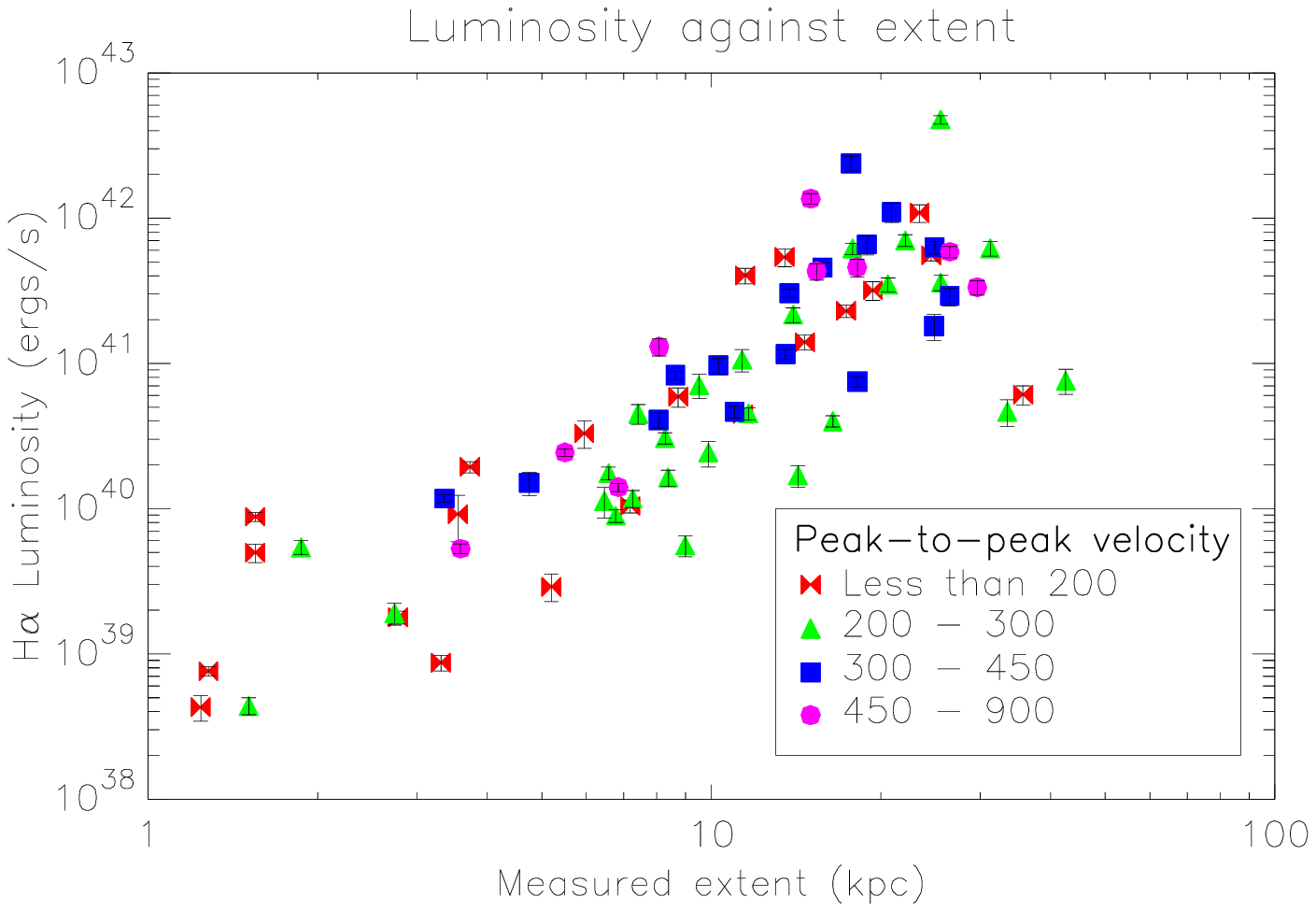,width=9cm}\\
\psfig{figure=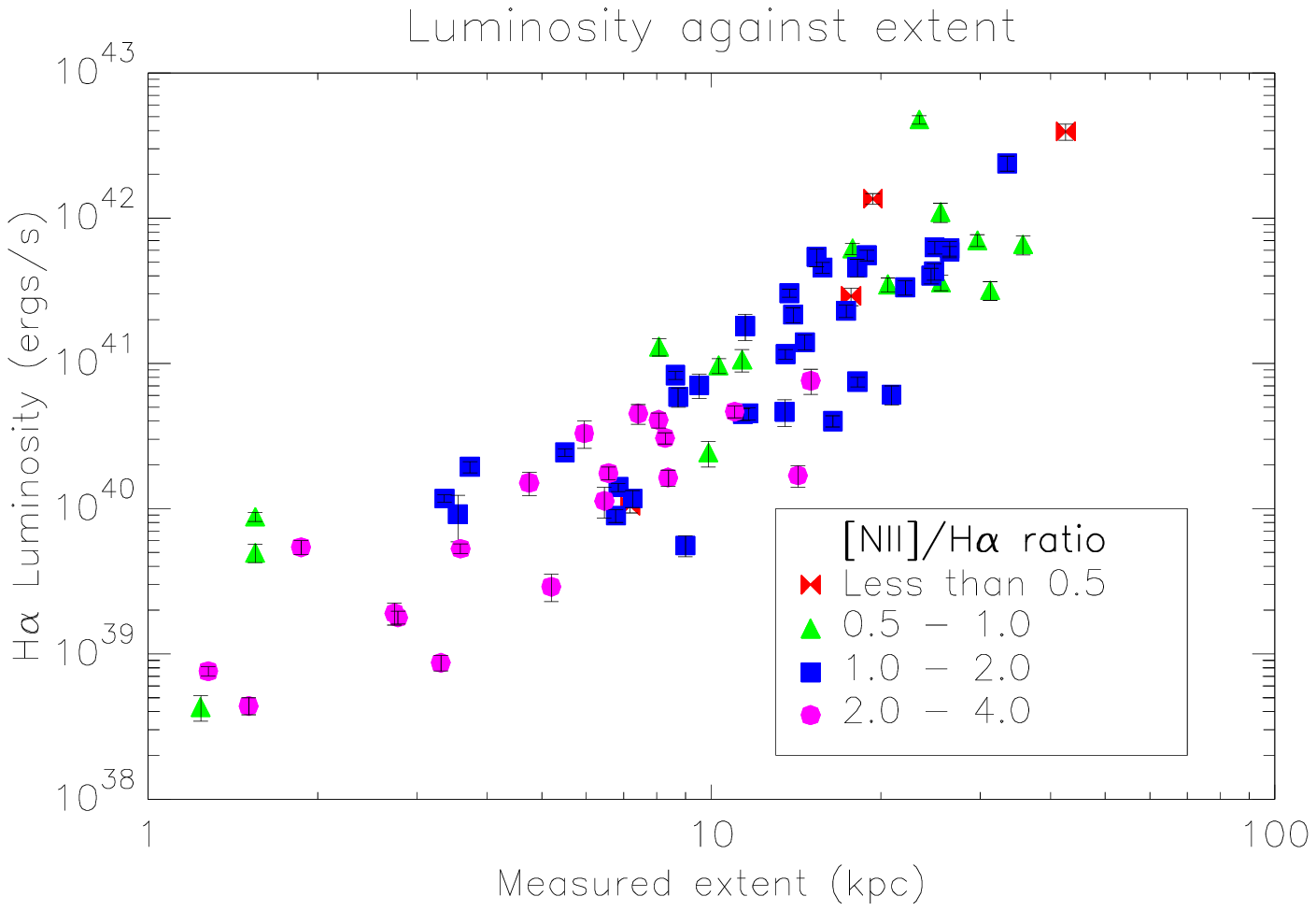,width=9cm}\\
\psfig{figure=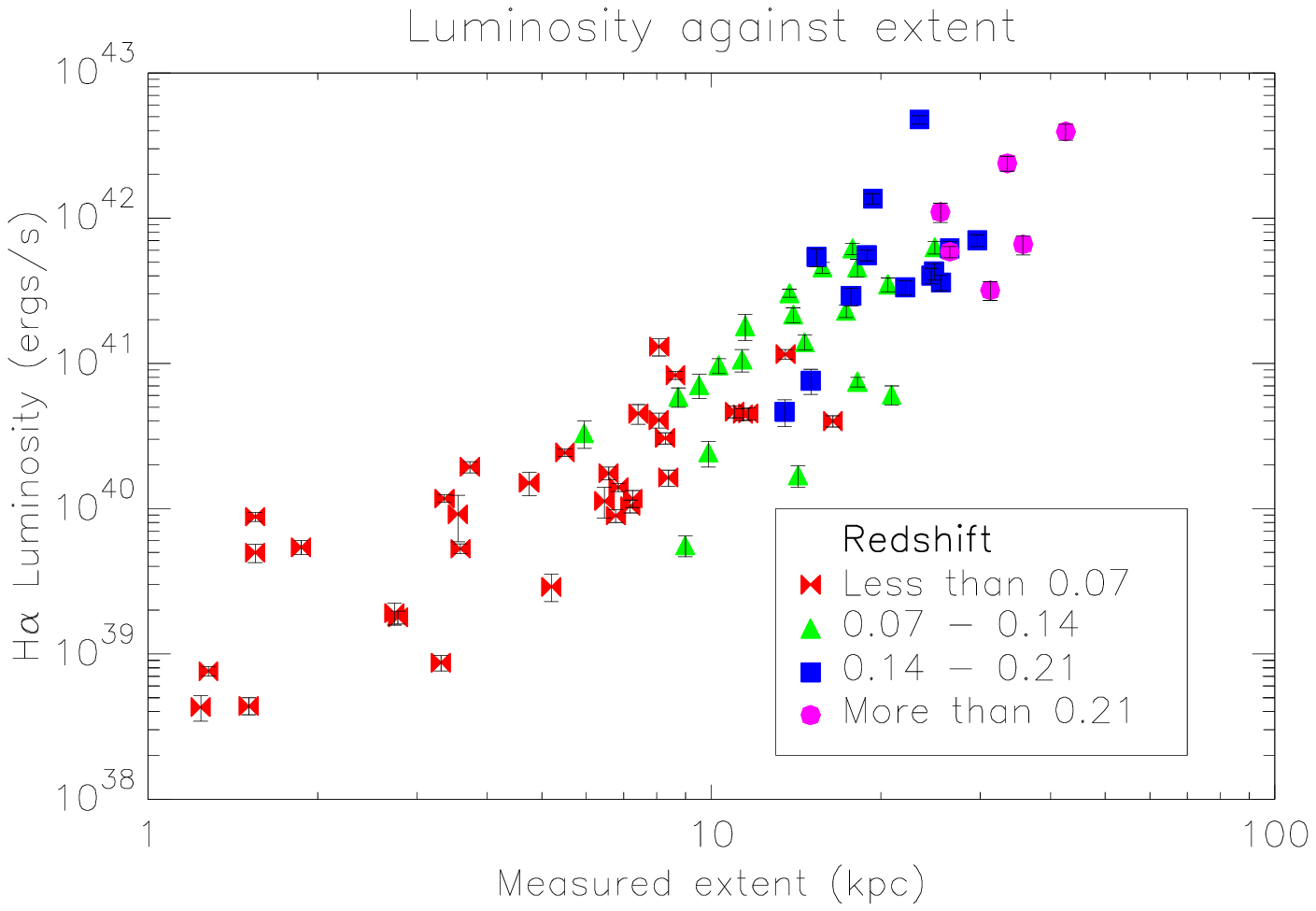,width=9cm}
\caption[H$\alpha$ luminosity plotted against H$\alpha$ extent, with points colour coded by the peak--to--peak velocity, {[}NII{]}/H$\alpha$ ratio and redshift of the systems]{The H$\alpha$ luminosity plotted against the extent of the line emitting region. A clear trend can be seen between the luminosity and extent as would be expected.  {\em Top} - The points are colour coded into bins of peak--to--peak velocity.  The velocity shows no clear trend on this plot however, we do note that the most compact (with an extent of less than 3 kpc) and least luminous objects have low velocities with all but 2 falling into the lowest velocity bin.  {\em Middle} - The points are colour coded into bins of overall [NII]/H$\alpha$ ratio.  The trend of [NII]/H$\alpha$ ratio with extent can be seen in this plot.  Again we see that the low luminosity compact objects typically have high [NII]/H$\alpha$ ratios. {\em Bottom} - The points are colour coded into redshift bins.  The higher redshift sources typically show higher luminosities as would be expected for a flux limited sample.} 
\label{fig:lumkpc}
\end{figure}

In Figure~\ref{fig:lumkpc} we plot the H$\alpha$ luminosity against the extent for each 
object.  As would be expected we see a clear trend in this plot with the more luminous 
objects being more extended due to the 
increased surface brightness.  The points on the plots in Figure~\ref{fig:lumkpc} are 
colour coded, in the top plot by the peak--to--peak velocity, and in the middle by the 
[NII]/H$\alpha$ ratio (see Section \ref{sec:spec} for a discussion of the line ratio 
analysis).  
The top plot of Figure~\ref{fig:lumkpc} shows no general trend 
of the velocity, as might be expected from the lack of a trend in 
Figure~\ref{fig:velkpc}. We do note however that the lowest luminosity ($<$ 10$^{40}$ ergs 
s$^{-1}$),  most compact objects (extent $<$ 3 kpc) have low peak--to--peak velocities 
with all but two falling into the lowest velocity bin.  The middle plot of 
Figure~\ref{fig:lumkpc} shows how the [NII]/H$\alpha$ ratio varies with the extent and 
luminosity.  Again we note that the objects to the lower left (compact and low luminosity)
have quite consistent [NII]/H$\alpha$ ratios with all but two falling into the highest 
bin.  So collectively from Figure~\ref{fig:lumkpc} we can identify a collection of 
objects with low extent and luminosity, little velocity structure and high ionisation 
states.  We would therefore suggest that the majority of the objects in Figure~\ref{fig:lumkpc} 
below an H$\alpha$ luminosity of $\sim$ 4 $\times$ 10$^{39}$ ergs s$^{-1}$ with an 
extent less than 4~kpc are most likely to be LINERS.  The bottom plot of 
Figure~\ref{fig:lumkpc} colour codes the points by redshift bin.  The higher redshift 
sources are typically brighter and more extended as would be expected for a flux 
limited sample.

\begin{figure}
\psfig{figure=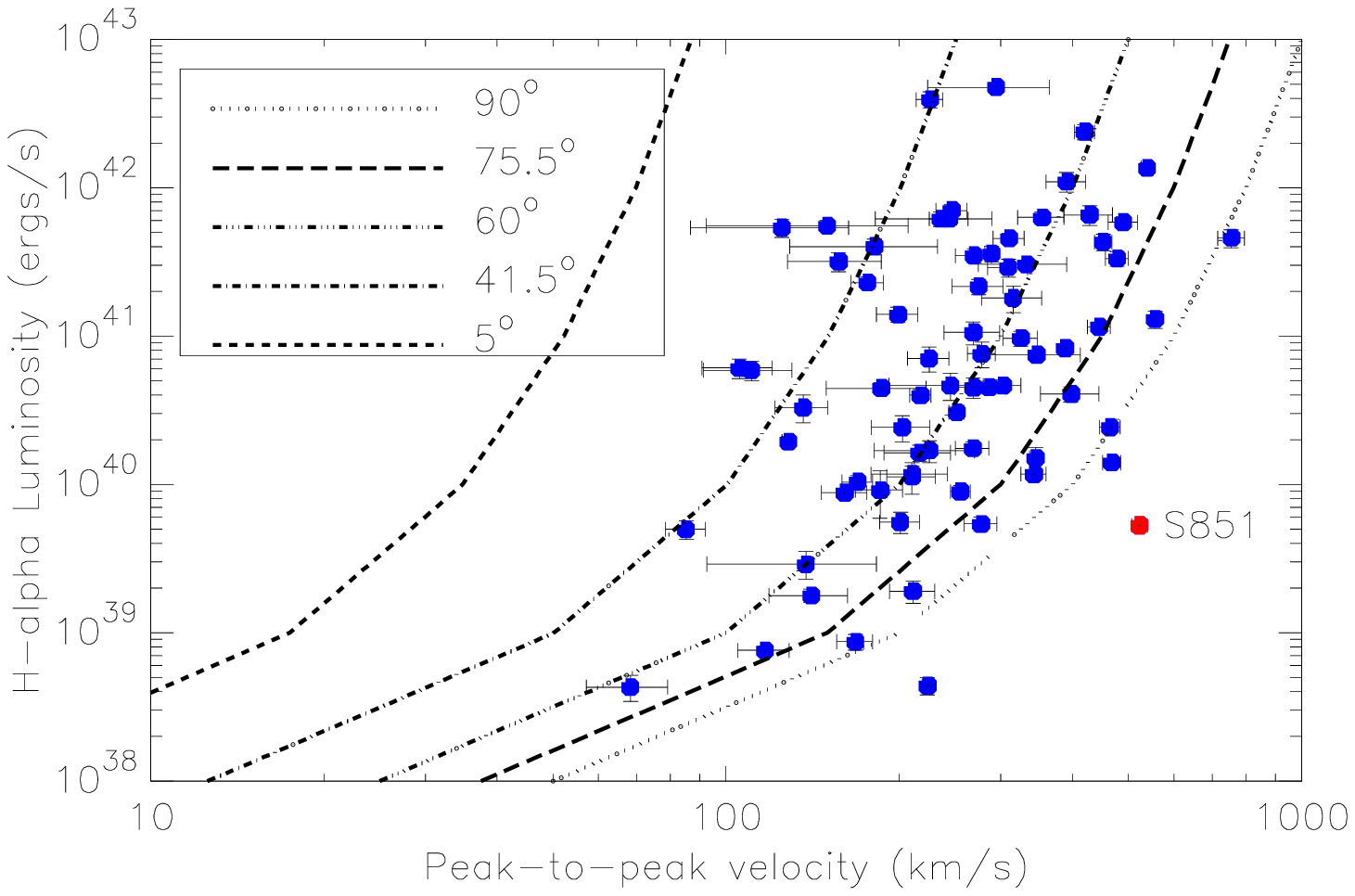,width=9cm}
\caption[H$\alpha$ luminosity plotted against the peak--to--peak of the velocity field]{The H$\alpha$ luminosity plotted against the peak--to--peak velocity. It is apparent from this plot that the mean velocity of the sample is dependent on the luminosity.  This may be an effect of the more H$\alpha$ luminous systems being more massive or that we can detect H$\alpha$ sampling more of the velocity field in the more luminous systems.  The lines on this plot show what the velocity would be at a given inclination if we assumed this trend was true, took the median velocity in each luminosity bin and assumed this came from an object inclined at 60 degrees to the plane of the sky (the most likely inclination of an unbiased sample).  In such a case the velocity of only one object in the sample, S851, cannot be explained by an inclination effect alone.} 
\label{fig:vellum}
\end{figure}

In Figure~\ref{fig:vellum} we compare each objects H$\alpha$ luminosity with its 
peak--to--peak velocity.  There is no trend immediately apparent from this plot as might 
be expected given the trend between luminosity and extent, and apparent lack of trend 
between velocity and extent.  
When considering an observed velocity structure it is 
always important to consider the effects of inclination which can reduce the observed 
peak--to--peak velocity regardless of the origin of the velocity structure.  The most 
likely inclination to view a velocity structure along, given a heterogeneous population 
and unbiased sample (a fair assumption for our objects given the minimal selection 
criteria), is 60 degrees to the plane of the sky.  
If we assume that the average velocity 
for each luminosity bin is from an object at this inclination, then we can show how this 
trend would be affected by the inclination.  We selected a number of inclinations which
would split the velocity range approximately into quarters, these inclinations are plotted
as a series of lines in Figure~\ref{fig:vellum}. We can see that the same trend shifted 
to 90$^o$ encloses all of the objects in the sample bar one, S851.  If we ignore the 
single outlier then it is possible to account for the different velocities seen in objects 
with similar luminosities as purely an effect of inclination.

\begin{figure}
\psfig{figure=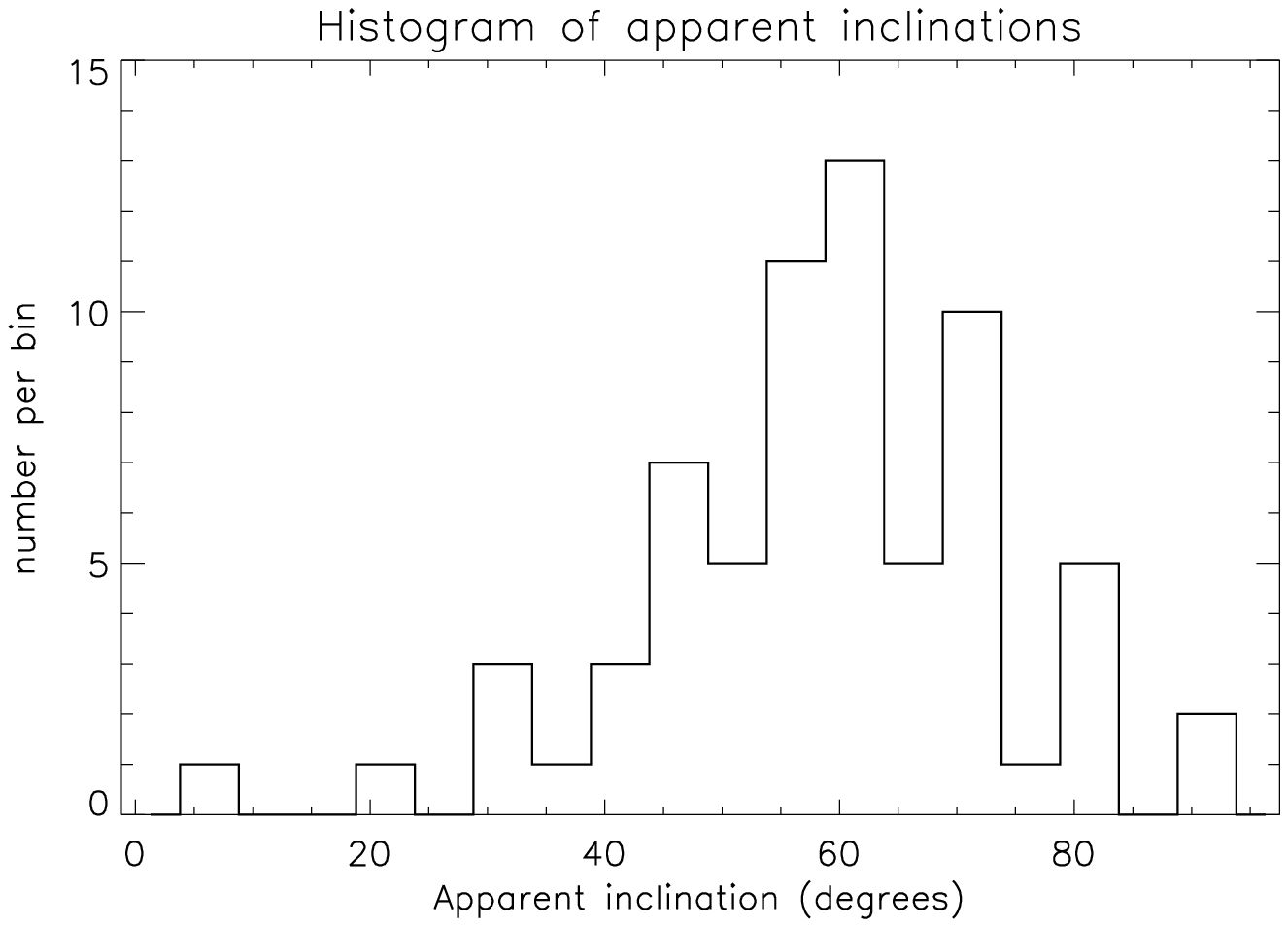,width=9cm}
\caption[Histogram of apparent orientations assuming velocity difference at a given Luminosity is a factor only of inclination]{Here we show the apparent inclination of the objects in the sample if it is assumed that the variation in their velocity at a given luminosity is a function only of inclination.  The apparent inclination of the velocity is relative to the plane of the sky such that a velocity in the plane of the sky would have an inclination of 0 degrees.  There is a peak at $\sim$ 60 degrees around which the apparent inclinations are distributed likely caused by noise obscuring face on systems and degeneracy between the effect of inclination and natural variations of the peak--to--peak velocity.  
} 
\label{fig:orien}
\end{figure}

If we assume this distribution in velocities to be a result of the inclination of the 
system then we can calculate the apparent inclination angle of each system using its 
luminosity and peak--to--peak velocity.  In Figure~\ref{fig:orien} we show a histogram 
of this apparent inclination for the systems in the sample.  This plot shows that 
the velocity structure of the systems are distributed about an apparent inclination of 
60$^o$ to the plane of the sky.  There are very few objects with an apparent inclination 
around 0$^{o}$ and few with an inclination of greater than 80$^{o}$.  This distribution 
is consistent with what is typically seen when calculating the inclination of galaxies 
from their axial ratios. It is important to note however, that using such a method has 
degeneracy with the thickness and ellipticity of the galaxy \citep{tmp13}. 
Thus edge on and face on disc galaxies have their inclination incorrectly calculated 
resulting in a distribution similar 
to that in Figure~\ref{fig:orien}. However, the method used to determine the inclination 
in Figure~\ref{fig:orien} is not subject to these same degeneracies and thus the 
distribution should be constant with sin($\theta$). The lack of face on systems can be 
explained by the fact that the measurable true peak--to--peak velocity difference would 
be very small.  In such a situation the velocity field will be dominated by noise 
obscuring face on systems.  We also note that some scatter in velocity at a given 
luminosity is expected, especially given the size of the luminosity bins used (an 
order of magnitude).  So the highest velocity systems in a given bin may skew the 
mean thus making naturally lower velocity systems appear more inclined resulting in an 
under-abundance of edge on systems.



\subsection{Comparison to the Stellar Kinematics}
\label{sec:starkin}
The sodium D stellar absorption features (NaD$_{\lambda5895.9}$
and NaD$_{\lambda5889.9}$) allows for 
the extraction of stellar kinematics but it requires 
the continuum emission to be bright in order for the absorption feature to be visible 
in the spectrum.  It is also intrinsically a broad feature which makes contamination 
by sky lines a serious problem.  
We performed the fitting using Voigt absorption profiles following the same procedure 
used to fit the Gaussian emission line profiles.  
From the sample only 12 objects had spectra 
in which the NaD absorption was clear enough to be fitted at the 
7$\sigma$ significance level (so as to match the emission lines).  

While these objects all show a strong continuum and strong NaD absorption, they are 
otherwise quite dissimilar in their X-ray and optical properties.  They have X-ray 
luminosities in the range of 0.1\,--\,42.10\,$\times$\,10$^{43}$\,erg\,s$^{-1}$ and 
H$\alpha$ luminosities  in the range of 0.087\,--\,46\,$\times$\,10$^{40}$\,erg\,s$^{-1}$ 
which is representative of most of the objects in our sample.  We note that none of the 
most luminous systems (either in X-ray or H$\alpha$) appear in this subsample. However, 
such objects make up only a small fraction of our sample.  The 12 objects show no common 
features in their other optical properties (velocity, extent, FWHM, [NII]/H$\alpha$ 
ratio and morphology) and were observed in a wide range of seeing conditions 
(0.63\,--\,2.39\,arcsec) suggesting they are representative of the sample.  We do note
that the objects are all typically low redshift sources (11 out of 12 have z$<$0.06) 
but this is the result of the requirement for bright continua to 
detect the NaD absorption feature.

Recent studies have separated the NaD absorption in 
star forming galaxies into two components (one from the stars and one from the ISM) by 
fitting stellar population models to isolate the stellar component \citep{hec00,jim07}.  
This may be possible for the brighter objects in our sample through binning to increase 
the signal to 
noise. However, we expect the stellar component to be dominant in our systems as 
NaD is strong in the spectrum of cool stars \citep{jac84} common in BCGs and is one of the 
strongest optical stellar absorption lines in the spectra of early type galaxies 
\citep{hec80,bic91}. As such, given
the focus of this paper is on the ionised gas emission we assume that all the NaD 
absorption is stellar in origin.

We produce maps from the fits to the NaD absorption
feature for these 12 objects and compare them to the line 
emission maps (see Appendix \ref{app:NaD}).  Determining the velocity dispersion of the 
NaD absorption from these line fits is not a trivial calculation and the fits themselves 
constantly overestimated the FWHM.  To account for this we used the fitting method to 
fit the total spectrum of each object and compared the FWHM derived by this method with 
the FWHM derived by fitting stellar templates to the same spectrum using 
penalised pixel-fitting \citep[][which could not be used on individual pixels due to insufficient signal to noise]{cap04}.  In this way we calculated a normalisation 
factor for the FWHM in each object and applied this to the velocity dispersion maps. 

\begin{figure}
\psfig{figure=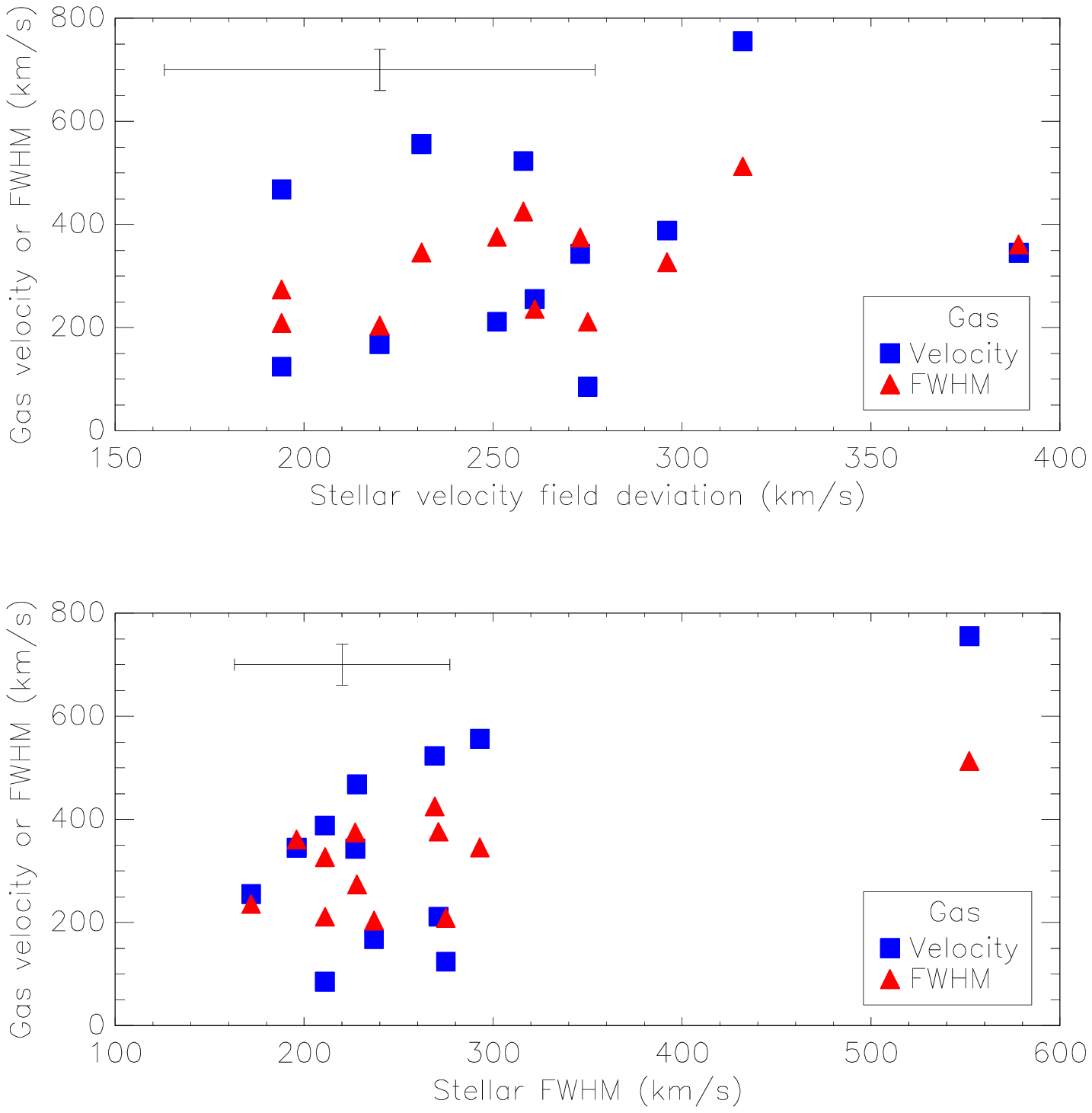,width=9cm}
\caption[Stellar kinematics compared to those of the ionised gas]{{\em Top} - Ionised gas peak--to--peak velocity and FWHM compared to the standard deviation of the stellar velocity field. {\em Bottom} - Ionised gas peak--to--peak velocity and FWHM compared to the overall FWHM of the stellar absorption.  No trend can be seen in either of these plots suggesting that the ionised gas kinematics are decoupled from the stellar component of the BCG. Representative error bars are shown in the top left of each plot.} 
\label{fig:kincomp}
\end{figure}

One thing that is immediately apparent from studying the NaD absorption maps is that 
the stellar component of the BCGs shows no evidence of an ordered velocity structure.
Indeed the mean velocity field appears to be completely random which is expected
for a massive elliptical galaxy such as a BCG.  This is in stark contrast to the 
ordered velocity maps produced from the H$\alpha$ emitting gas.
We also note that although the line width maps do generally show a peak towards the 
centre of the object they are overall substantially more uniform than their counterparts 
from the line emission.  This implies that the stellar component of the BCGs has a 
velocity field which is dominated primarily by random motion suggesting that the movement 
of the stars and gas in the galaxy are decoupled.  In Figure \ref{fig:kincomp} we 
compare the kinematics of the gas to those of the stars.  No trends can be seen for 
any of the comparisons in this figure however we note the large errors on the parameters 
from the stellar fits make this ambiguous.  

To test this we performed a Monte Carlo 
procedure to reproduce the plots with parameters sampled from within the Gaussian errors 
shown.  We produced 10,000 of each of the four comparisons: stellar velocity against 
gas velocity, stellar velocity against gas FWHM, stellar FWHM against gas velocity and 
stellar FWHM against gas FWHM. For each we performed a Spearman 
correlation test and found no correlation in 90.0\%, 87.6\%, 88.9\% and 89.2\% of the 
realisations respectively. This lack of correlation confirms that the kinematics of the 
ionised gas are decoupled from those of the BCGs stellar component.
Hydra-A is a clear example of this, 
it is one of the most ordered velocity fields in our sample and also has one of the 
largest peak--to--peak velocities.  Moreover it is known that the gas forms a rotating 
disc \citep{sim79,ham14}, despite this however the stellar motions appear to remain 
random, producing a velocity field with no obvious order.

In Section \ref{sec:offset} we identify Abell 1991 as an object with offset line emission. 
The extent of this offset can be seen in the maps comparing the stellar absorption to the 
ionised gas emission which can be found in Appendix \ref{app:NaD}.
While both have a peak in roughly the same position, the line emission map shows little 
emission around this point with most of it located $\sim$ 10 arcsec to the north where it 
shows a second peak.  By contrast the majority of the NaD absorption is centred almost
uniformly around the southern peak of the line emission with possibly a slight extent to 
the south. The comparison of the line emission maps with the stellar absorption maps 
clearly highlights this offset

The line emission map of RXJ0338+09 shows a very disturbed morphology suggesting that 
some major event is disrupting the line emitting gas within this system.  Interestingly 
the stellar absorption map shows a secondary peak of absorption at $\sim$ 7 arcsec 
($\sim$ 4.7 kpc) to the north-west of the BCG.  This suggests the presence of a second 
galaxy very close in projection to the BCG.  Consulting the stellar velocity map 
we find that this galaxy also has a velocity which is close to that of the BCG, 
redshifted by just a few hundred km s$^{-1}$ and matching the velocity of the line 
emission at this point.  It is tempting to interpret these findings as evidence 
of an interaction which is disrupting the cold gas in the cluster core. However, we 
note that the companion galaxy has a bright X-ray point source \citep{sar92} suggesting 
AGN activity so it is possible that the H$\alpha$ emission seen at the redshift 
of the companion is associated to its nucleus and superimposed on that of the cluster 
core.  Additionally the ICM shows evidence of cavities from multiple AGN outbursts 
\citep{snd09} which may be responsible for disrupting the gas on large scales. Finally 
the ro-vibrationally excited molecular gas appears to form a rotating structure close 
to the BCG \citep{wil11}.  Thus to fully characterise the nature of the ``interaction'' 
in this system requires a dedicated study of the complex processes at work using
high sensitivity multiwavelength observations and is beyond the scope of this paper.

%% file: spectra_r2.tex


\section{The Physical properties of the gas}
\label{sec:spec}

In order to investigate the spectral differences between the objects
we now present spectra extracted from two regions for each object in our sample.  We 
initially extracted spectra from the full region of line emission for each object. We 
defined this extent as being the region where the H$\alpha$ flux exceeded the noise 
by a factor of 7.  To determine this we first collapsed the cube (see Section 
\ref{chap:meth} for a description of this procedure) over the region of the spectra 
containing the H$\alpha$ emission.  Additionally, we collapsed the cube over two other 
regions of equal spectral length which contained no emission lines or sky lines to 
produce two continuum images. A continuum subtracted H$\alpha$ map was then produced 
by subtracting the first of these continuum maps from the H$\alpha$ map. The first 
continuum image was also subtracted from the second continuum image to produce a map 
from which the noise could be determined.  

A mask was then produced to exclude spatial pixels where the H$\alpha$ emission did not 
exceed a Signal to Noise (S/N) of 7.  The flux in the remaining pixels was then summed 
for each channel to produce a spectra of the full H$\alpha$ emitting region. A second 
spectra for each object was extracted from the nuclear region of the BCG, defined as a 
3 $\times$ 3 pixel ($\sim$ 2 $\times$ 2 arcsec$^2$) region (roughly the extent of the 
mean seeing) centred on 
the peak of the continuum emission.  All spectra were sky and continuum subtracted, and 
have been filtered to eliminate contamination by cosmic rays. The two spectra from each 
object can be found in Appendix \ref{app:specs} and the line fluxes and ratios are given 
in Appendix \ref{app:lines}.

Our IFU observations of these clusters allow us to study the physical properties of the 
gas within the ionised nebulae using the ratios of the various optical emission lines 
visible in these spectra.  Many of the lines commonly used in such analysis (H$\alpha$, 
[NII], [OI] and [SII]) are covered by our observations for all objects and are detected 
in most (see Table \ref{tab:sam1}).  Importantly as our observations are resolved they 
allow us to determine how the properties of the gas vary throughout the nebula.  However,
some care must be taken with this analysis, especially as we lack the [OIII]/H$\beta$ 
ratio (commonly used in conjunction with the other line ratios for this kind of 
analysis), as there are many factors which can all affect the observed line ratios.

\subsection{Metallicity}

The line ratios are subject to change due to variations in the metallicity of the gas 
with the [NII]/H$\alpha$ being the most strongly affected, log$_{10}$([NII]/H$\alpha$) varies 
linearly with metallicity until it saturates at log$_{10}$([NII]/H$\alpha$)\,$\sim$\,-0.5.
For cool core clusters metallicity profiles of the ICM are consistent  
between objects \citep{bw09}, falling from close to solar metallicity in the BCG to 
$\sim$ 1/3 solar metallicity \citep{loe04} in the ICM (beyond $\sim$ 0.25\,r$_{200}$).
We thus expect that the metallicity of the gas is approximately the same in all of the 
objects and, as our observations only cover at most a few 10s of kpc, that any change in 
metallicity within a given object will be small. Thus the metallicity should have a 
negligible affect on the variation of line ratios both between objects and within a 
single object when compared to other effects.

As all objects in our sample have H$\alpha$ and [NII] emission we can test this 
assumption by using the $N2$ index to estimate the metallicity of the gas in each of our 
objects as measured by the 
oxygen abundance (O/H).  We use equation 1 from \citet{pp04} to estimate the range of 
oxygen abundances in our sources from the [NII]/H$\alpha$ ratio. From Table 
\ref{tab:obs} we see that the maximum and minimum mean [NII]/H$\alpha$ ratios in our 
sample are 3.7 and 0.54 respectively which corresponds to oxygen abundances 
(12\,$+$\,log$_{10}$(O/H)) of 9.22 and 8.75.  \citet{pp04} find that 95\% of their measured 
oxygen abundances lie within $\pm$\,0.41 of the value derived for a given [NII]/H$\alpha$
 ratio.  Our two extremes fall well within these limits suggesting that the metallicities 
within all our objects are consistent.  For objects where we also have the [OIII] and 
H$\beta$ lines we can further test this consistency by using the $O3N2$ index as an 
additional measure of the oxygen abundance.  Using equation 3 from \citet{pp04} we 
estimate the oxygen abundances for the seven sources for which we could obtain the 
[OIII]/H$\beta$ ratios. We find oxygen abundances (12\,$+$\,log$_{10}$(O/H)) of 8.67 to 9.16 
which are consistent with the values derived from the [NII]/H$\alpha$ ratios.
As such we assume that the metallicity is the 
same in all objects and has no effect on the measured line ratios.

\subsection{Electron Density}
\label{sec:ED}

\begin{figure}
\psfig{figure=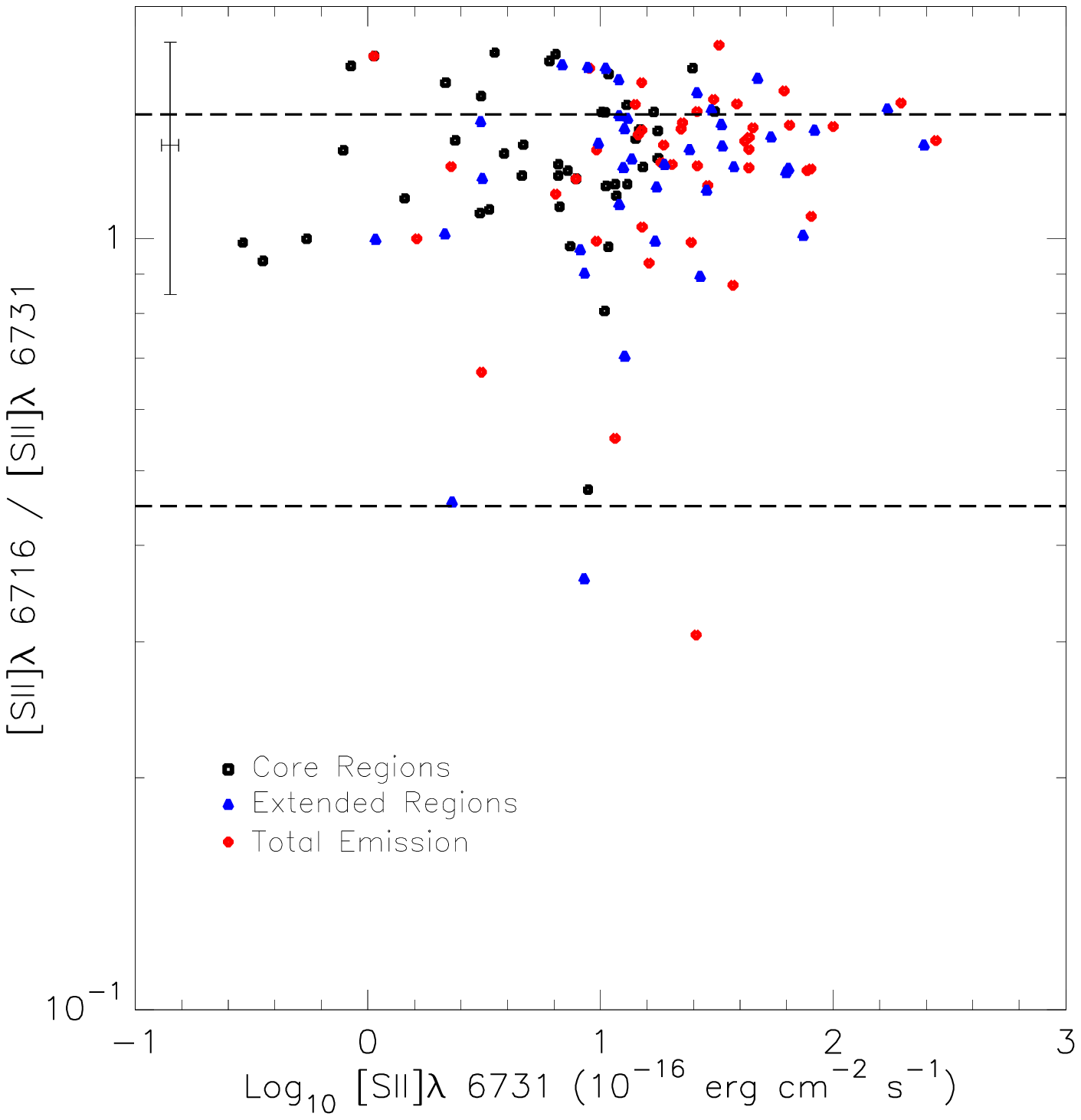, width=9cm}
\caption{A comparison of the [SII]$_{\lambda 6716}$/[SII]$_{\lambda 6731}$ ratio and [SII]$_{\lambda 6731}$ flux for all objects in the sample which had detections of both lines.  The points show the core regions (black squares), extended regions (blue triangles) and total emission (red diamonds) for each object. The dashed lines indicate upper and lower bounds where the [SII]$_{\lambda 6716}$/[SII]$_{\lambda 6731}$ ratio is sensitive to changes in electron density.  The upper boundary is set to a value of 1.45 corresponding to a density of $\sim$ 100 cm$^{-3}$ and the lower boundary to a value of 0.45 corresponding to a density of $\sim$ 10$^4$ cm$^{-3}$. Most of the points fall towards the upper boundary suggesting densities commonly in the region of a few 100 cm$^{-3}$.}
\label{fig:SIISII}
\end{figure}

The density of the gas can also have an effect on the line ratios measured for a given 
excitation mechanism.  Importantly models of star formation excitation often 
used to distinguish between star forming galaxies and those excited by more energetic 
phenomena typically assume electron densities of a few hundred or less 
\citep[for example][]{dop00,kew01}.  The strengths of several lines in 
our observations are sensitive to the electron density but in particular the [SII] 
doublet line 
fluxes, available for all but a few objects in our sample, are sensitive to the effects of 
collisional de--excitation.  Since both lines have very similar excitation energies their 
relative excitation rates are almost insensitive to variations electron temperature.
As such the effect of temperature is negligible and the relative excitation rates 
of the two lines depends only on the the ratio of their collision 
strengths and thus the ratio of intensities depends only on the density of the gas 
\citep{ost89}.  
This analysis is only sensitive to densities over $\sim$ 1--2 orders of magnitude 
between 100 cm$^{-3}$ and 10$^4$ cm$^{-3}$, fortunately this is consistent with the 
densities assumed in models of star formation excitation and we expect the electron 
density in the cores of clusters to be towards the lower end of this density range.  
In Figure 
\ref{fig:SIISII} we plot the [SII]$_{\lambda 6716}$/[SII]$_{\lambda 6731}$ ratio against the 
 [SII]$_{\lambda 6731}$ flux for all objects in the sample.  The dashed horizontal lines on this plot 
contain the region that is sensitive to density changes.  The majority of the points 
are found towards the upper boundary of this region suggesting that most objects in our 
sample have electron densities on the order of a few 100 cm$^{-3}$. 
We also note however that a small number of points (just four) fall within the region below a value of $\sim$ 0.7 suggesting these objects have densities in excess of 10$^3$ cm$^{-3}$.     

The fact that some points fall outside of the region sensitive to density must also be 
considered.  Some of these can be accounted for by the fact that 
in some cases some or all of the [SII]$_{\lambda 6716}$ line 
fell into an atmospheric absorption feature and was lost. An example is Abell 795 which has a 
ratio in its core of $\sim$ 0.2 despite relatively strong lines being present as the
[SII]$_{\lambda 6716}$ line is redshifted into the atmospheric A-band.
A similar 
situation also affects some of the objects which have a ratio above 1.45.  In this case 
the [SII]$_{\lambda 6731}$ line falls into the absorption feature causing the 
[SII]$_{\lambda 6716}$ to appear unusually strong. Some of these points
can also be attributed to objects whose redshifts put the [SII] lines right at the 
edge of the spectral coverage of our observations.  In some cases this resulted in some 
of the [SII]$_{\lambda 6731}$ line falling off the edge of the spectrum in part or all of 
the field of view. Objects with such obvious factors causing errors in the [SII] flux 
measurements have been removed from the analysis in Figure \ref{fig:SIISII}.
There remain however, many objects with ratios falling above a threshold of 1.45 
and still two that fall below a ratio of 0.45.  We note however, that the errors 
on the ratio are considerable, in Figure \ref{fig:SIISII} we show representative errors
positioned at the median ratio of [SII]$_{\lambda 6716}$/[SII]$_{\lambda 6731}$.  We note that 
the upper error limit encompasses most of the points above a ratio of 1.45, as the 
majority of our objects tend to have high [SII]$_{\lambda 6716}$/[SII]$_{\lambda 6731}$ ratios 
(above $\sim$0.8) these errors can explain the large number of points falling above our 
expected upper limit.  Indeed the majority of ratios fall within these errors suggesting 
that they are consistent with the median ratio of 1.32 which corresponds to densities on 
the order of 100--200\,cm$^{-3}$. As such the measured gas densities are consistent with 
those used in models of star formation so we can use the expected line ratios from such 
models to determine if star formation plays a major role in the excitation of the 
ionised gas in cluster cores.

\subsection{Excitation state}
\label{sec:excit}
The [NII] to H$\alpha$ ratio is also a good measure of the excitation state of the gas
and in particular is a good indicator of even small contributions to the excitation from 
an AGN thanks to its sensitivity to metallicity. The 
log$_{10}$([NII]/H$\alpha$) ratio saturates at high metallicity peaking at $\sim$ -0.5 
\citep{den02,kd02,pp04,kew06}, as such [NII]/H$\alpha$ ratios of above $\sim$ 0.6 are a
strong indicator of an AGN (or other non star formation) contribution to the excitation.
Table \ref{tab:obs} lists the mean [NII]/H$\alpha$ ratio for all objects in our sample as 
calculated from the fits to each pixel.  
 The ratio varies substantially between objects
suggesting that  some objects are substantially more highly ionised than others. 
As noted earlier some of the objects showed 
H$\alpha$ emission which was absorbed, this will have the effect of artificially 
rising the [NII]/H$\alpha$ ratio calculated from fits to the total spectrum. We clearly 
indicate which objects show H$\alpha$ absorption in Table \ref{tab:sam1}. 
For these objects the [NII]/H$\alpha$ ratio 
calculated and presented in Table \ref{tab:obs} can at best be considered an upper limit.

For the majority of objects the [NII]/H$\alpha$ ratio can be seen to vary within 
the maps presented in Figure \ref{fig:maps} suggesting that the ionisation state is not 
constant throughout the objects.
This is not surprising as sources of ionisation (AGN, star formation, shocks, etc.) are
often seen to be localised to specific parts of a galaxy \citep[AGN are localised to the 
centre of the galaxy, star formation is often seen to be clumpy rather than uniformly 
distributed][]{ode10}.  The maps also indicate the 
regions of objects which have substantial H$\alpha$ absorption as the 
[NII]/H$\alpha$ ratios are extremely high appearing white on the maps.



While the maps show a lot of variation in the ionisation state within 
different regions of each object, the line ratios do not show any 
obvious structures.  Studying the overall ionisation state allows for a direct 
comparison between objects and with other global parameters.  The ionising radiation 
within galaxies is believed to come from two primary sources, AGN and star 
formation \citep{bpt81}. AGN typically produce a harder radiation field which results in a higher 
ionisation state while star formation has a softer radiation field.  Thus one would 
expect gas ionised by AGN to have a high [NII]/H$\alpha$ ratio while that ionised by 
star formation should have a lower value.  However, this distinction is more complicated 
for gas in BCGs where both AGN and star formation are likely to be occurring.
Shocks, produced from either internal (e.g. AGN outbursts) or external (e.g. galaxy 
interactions) sources, can also produce optical line emission.  Generally, faster 
shocks produce higher line ratios \citep[though there is also a dependence on 
metallicity, gas density and magnetic fields,][]{all08} comparable to those produced by 
AGN.  However, if such fast shocks were present they should be visible in Figure 
\ref{fig:maps} as broad regions in the FWHM panel. Post-AGB stars have been shown to be 
a possible source for LINER like emission \citep{sin13} which can produce line ratios 
inconsistent with star formation models. Other possible sources of the 
ionisation include photoionisation by emission from the hot gas in the ICM or collisional 
heating \citep{fer09,fab11} which can also produce high [NII]/H$\alpha$ ratios. 

The majority of the extracted spectra show little variation in appearance between 
the central region and the total emission suggesting that an AGN is not 
significantly affecting the distribution of ionising radiation in most objects. 
However, some objects do show a variation in the [NII] to H$\alpha$ line ratio, which might 
suggest the influence of an AGN.  Indeed for some objects the [NII] to H$\alpha$ 
ratio is sufficiently high (above [NII]/H$\alpha$\,$\sim$\,0.6) to rule out star formation 
as the dominant component of the ionisation. One striking 
observation from the spectra is the fact that some objects, which show strong 
lines in their total spectrum, have little or no emission present in their central 
regions.  These are objects in which the peak of the line emitting gas is offset from 
the BCG which results in weak or absent lines at the peak of the continuum emission.
Such objects are discussed in section \ref{sec:offset} and studied in detail 
in \citet{ham12} and while interesting, make up only a small fraction of BCGs (just 4 
of the objects in our sample are offset in this way) and are thus not representative of 
the population.

\begin{figure}
\psfig{figure=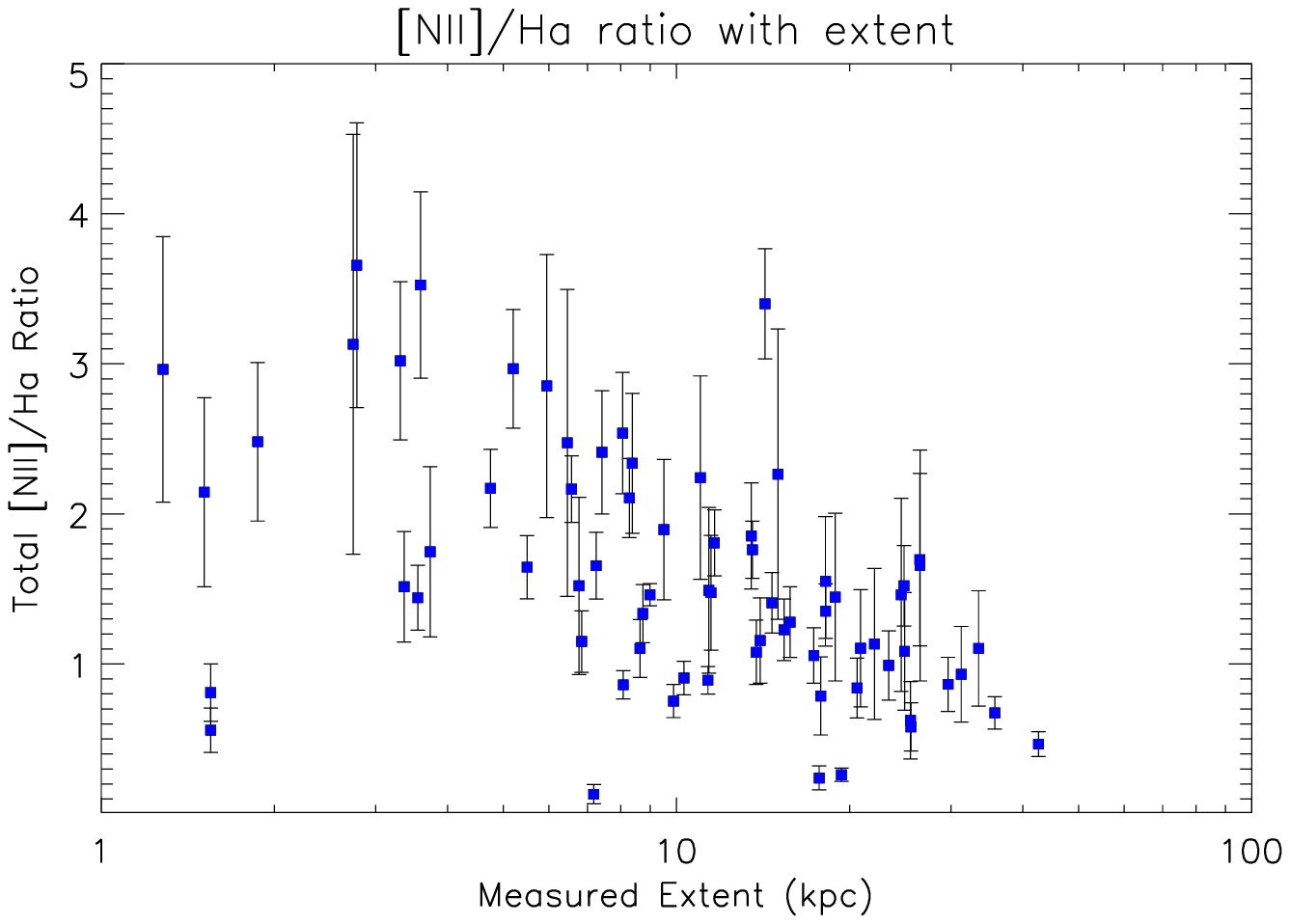,width=9.1cm,bbllx=89,bblly=360,bburx=469,bbury=626,clip=}
\caption[Average {[}NII{]}/H$\alpha$ ratio for each object in the VIMOS sample plotted against the extent of the emission lines.]{The overall [NII]/H$\alpha$ ratio for every object in the VIMOS sample plotted against the extent of the line emitting region.  A loose trend can be seen in this plot which suggests that the overall [NII]/H$\alpha$ ratio decreases as the extent of the emission line region increases.  This suggests that the dominant source of the ionisation is more energetic for compact objects than for extended ones.  Likely the compact objects feel more influence from a central AGN while the extended regions are ionised by other mechanisms. 
}
\label{fig:N2Hkpc}
\end{figure}

In Figure~\ref{fig:N2Hkpc} we show the global measure of the [NII]/H$\alpha$ ratio of 
each system as a function of the extent of its line emitting region.  With the exception 
of a few outliers this plot shows a trend with the [NII]/H$\alpha$ ratio decreasing
as the extent increases.  Since the ionising radiation in BCGs likely to be generated 
from numerous sources (AGN, star formation, soft X-rays from the ICM, cosmic rays, 
shocks, collisional excitation etc.) 
this trend can be interpreted as an effect of the relative contributions of each. 
Radio observations 
of all BCGs in this sample have already been taken with a substantial number detected.
\citet{hog15} find a flat spectrum radio source at 10\,GHz in 36 of the 73 objects in 
our sample, and a steep spectrum component at 1\,GHz in 53/73 of our objects. This 
suggests that AGN activity is currently active, or has recently been active within 
the cores of many of the BCGs in our sample. As such in the very compact objects the gas 
is likely to be concentrated near an AGN so the ionisation 
will thus be dominated by fast shocks and photoionisation producing a 
high [NII]/H$\alpha$ ratio.  The more  extended objects will have a higher relative 
contribution from other less energetic sources of ionisation producing a lower overall 
ratio explaining the global trend we see in Figure~\ref{fig:N2Hkpc}.
In the next paper in this series we will use resolved radio observations to compare the 
radio structure to the kinematics of the ionised gas to determine how the gas and AGN 
interact.

\subsection{The source of the ionisation}
\label{sec:bpt}

We have already discussed the ionisation state of the gas in the previous section, 
however it is important to note that while the [NII]/H$\alpha$ ratio can provide 
a measure of how highly ionised any gas is it cannot on its own be used to determine the 
source of the ionising radiation.  For this, other line ratios are also required, 
such as [OIII]/H$\beta$ for a BPT analysis \citep{bpt81}. Unfortunately for 
most of the objects in our sample the observations did not cover the wavelength 
range required to observe these lines.  However, as can be seen from Table 
\ref{tab:sam1}, only seven of the objects in our sample were at a redshift high enough to 
allow H$\beta$ and [OIII] to be present at the blue end of the HR\_orange grism.

We produced BPT diagrams for the 7 objects (Figure \ref{fig:bpt}) which had both 
H$\beta$ and [OIII] line present in the spectrum.  From Figure \ref{fig:bpt} it is 
apparent that all 7 of these objects appear to have ionisation states which are not
consistent with being the result of star formation  activity based on the empirical 
separation from \citet{bpt81} (dotted line). It is interesting to note that one of 
the objects for which the 
BPT analysis was possible was Abell 2566.  Abell 2566 was discussed in Section 
\ref{sec:offset} because the emission is completely offset from the BCG by $\sim$ 
9 kpc.  \citet{kew01} calculate the maximum theoretical boundary for star formation which 
is shown as a dashed line in Figure \ref{fig:bpt}.   We note that Abell 2566 falls 
below this line, within the region enclosed by the two dividing lines which 
\citet{kew06} define as a region of composites between AGN and SF ionisation. 
Most of the points sampled in the other 6 of our 7 objects fall above or on this line 
suggesting that star formation alone cannot explain the ionisation. These objects 
show line ratios typically attributed to AGN. While this is not unexpected as BCGs 
do often host AGN \citep{rus13,hla13} there are also other processes which could 
produce the line ratios
seen, shocks \citep{all08} or particle heating \citep{fer09,fab11} for example.  
Additionally we note that the log$_{10}$([OIII]/H$\beta$) values typically fall below 
$\sim$ 0.5 putting these objects in the region typically occupied by LINER-like 
objects which \citet{sin13} suggest do not require an AGN as the ionising source. 
If AGN are responsible for ionising the gas in all these systems then this would
suggest that the influence of the AGN is able to reach far beyond the core of 
the galaxy. 

\begin{figure*}
\psfig{figure=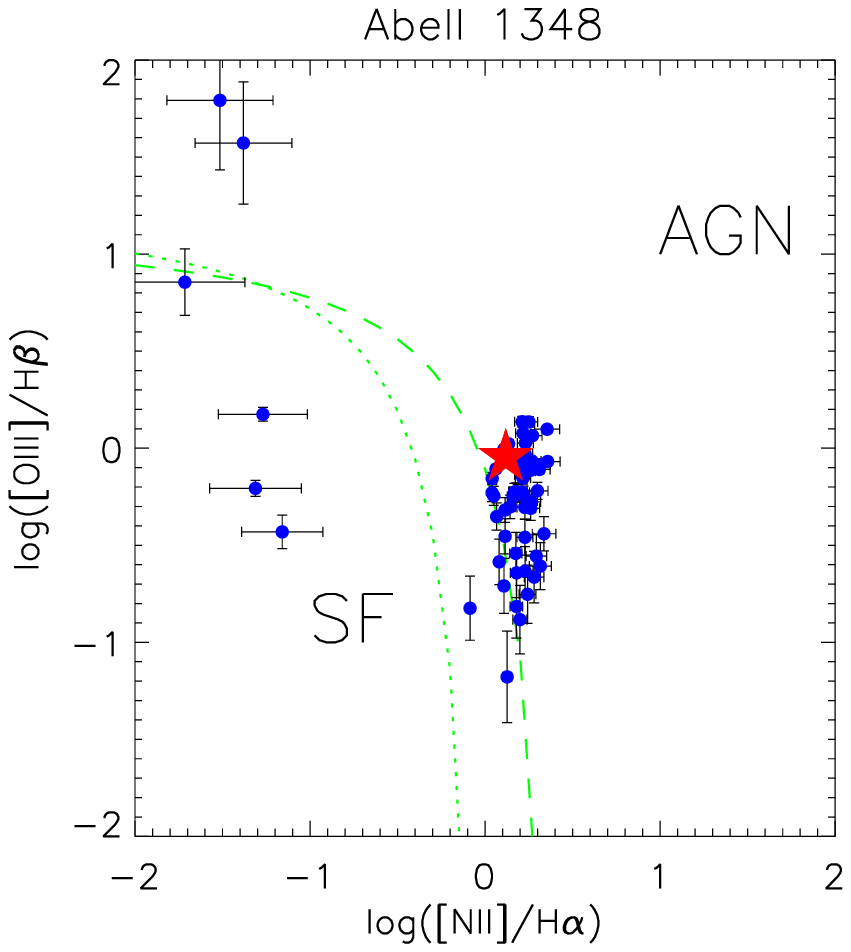,width=7cm}
\hspace{-1cm}
\psfig{figure=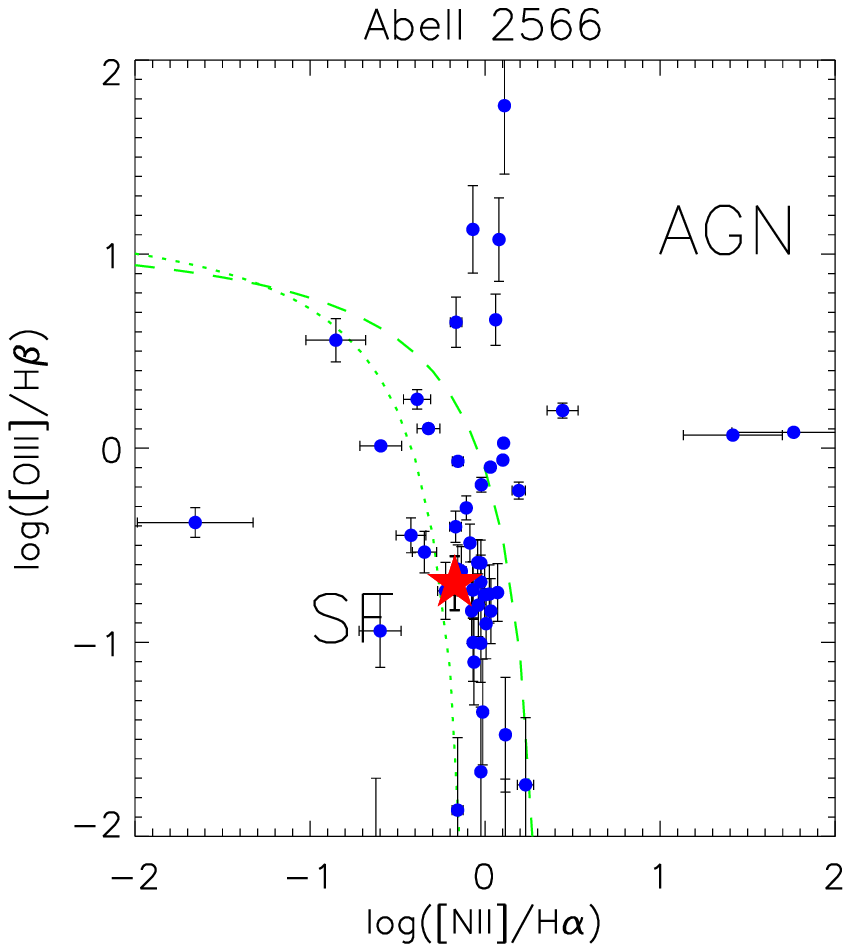,width=7cm}
\psfig{figure=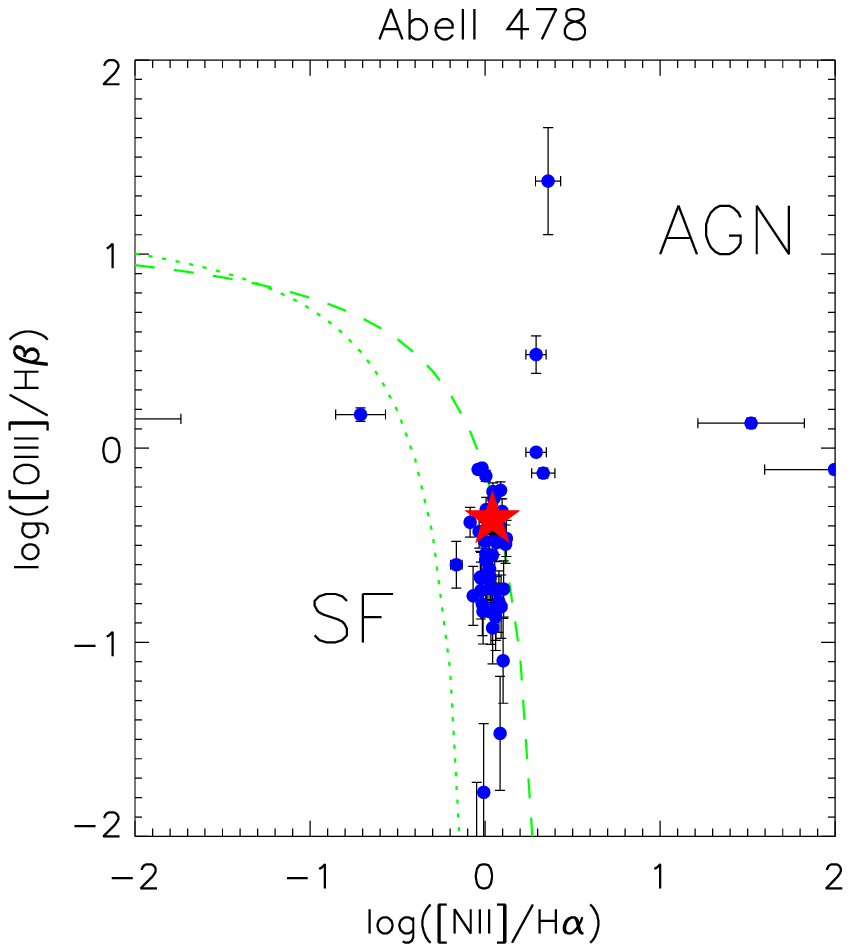,width=7cm}
\psfig{figure=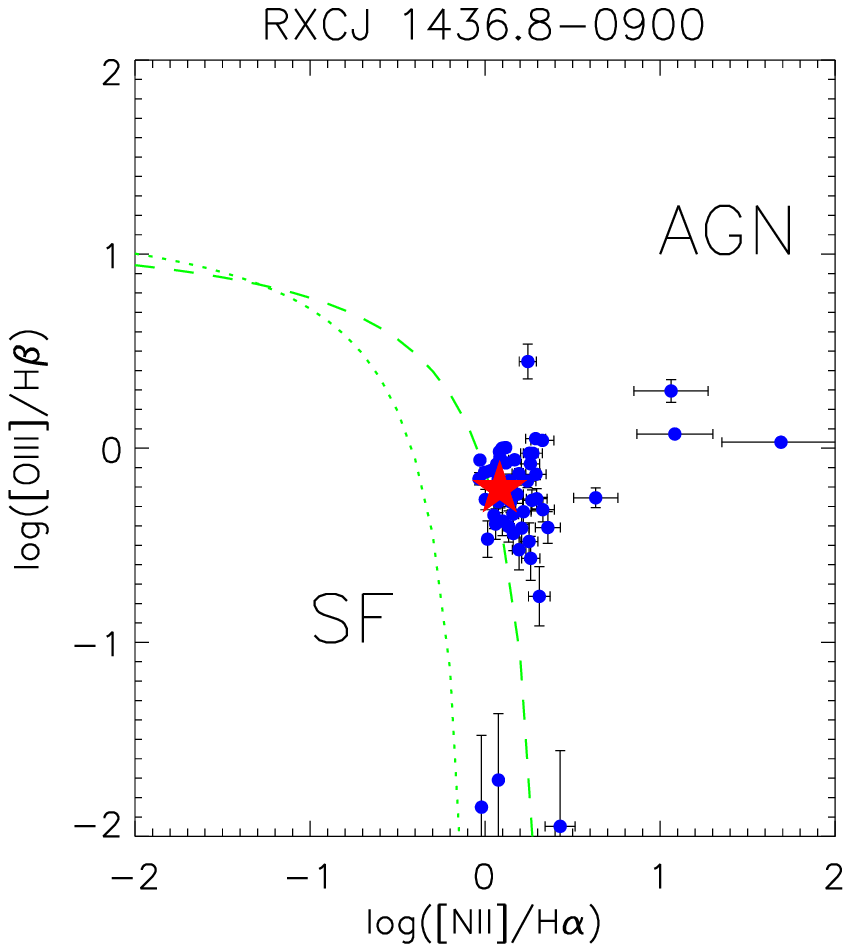,width=7cm}
\caption[BPT diagrams of objects in the VIMOS sample]{Here we show the BPT analysis of the objects from our sample which have redshifts that allow us access to the [OIII] and H$\beta$ lines in our VIMOS observations.  The blue points show the locations of every pixel of detected emission on the diagram, the red stars then show the location of the central 2 $\times$ 2 arcsec$^2$ which should be the location of the AGN.  The green dotted line marks the empirical boundary between star formation and AGN dominated ionisation while the green dashed line marks the theoretical upper limit of star formation dominated ionisation.  The region between these two lines likely contains objects which are a composite of star formation and AGN ionisation.   Note that for Abell 2566 the emission was offset so the central region was determined from the H$\alpha$ emission and not the continuum.} 
\label{fig:bpt}
\end{figure*}

\begin{figure*}
\ContinuedFloat
\psfig{figure=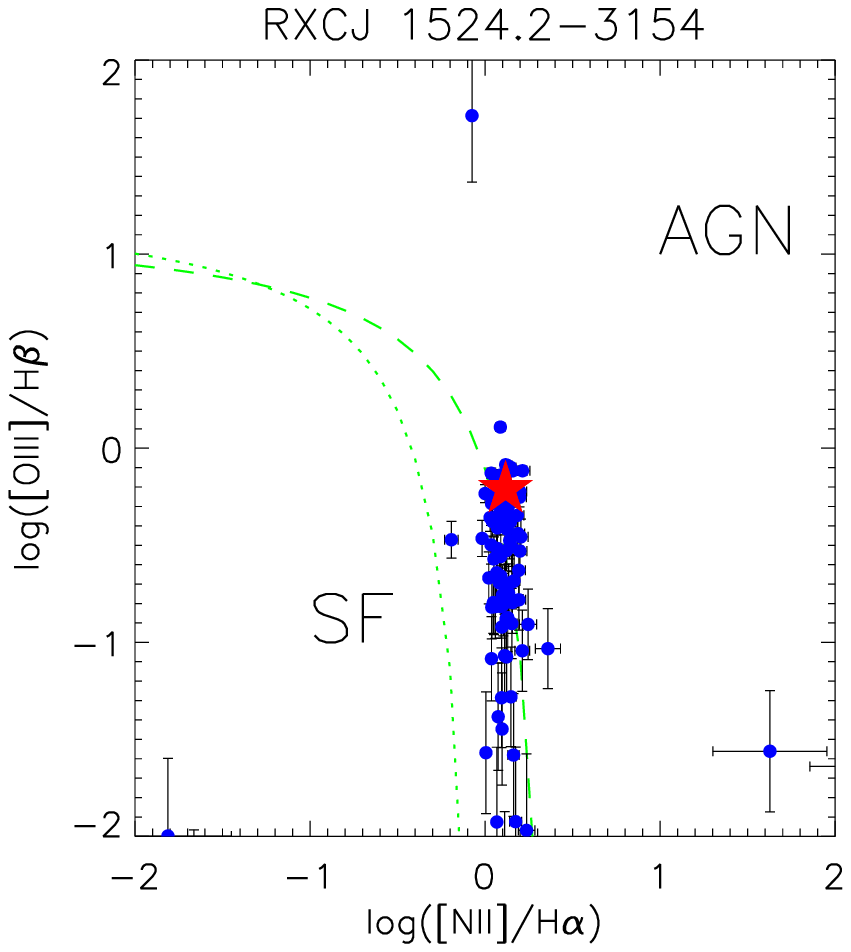,width=6.3cm}
\psfig{figure=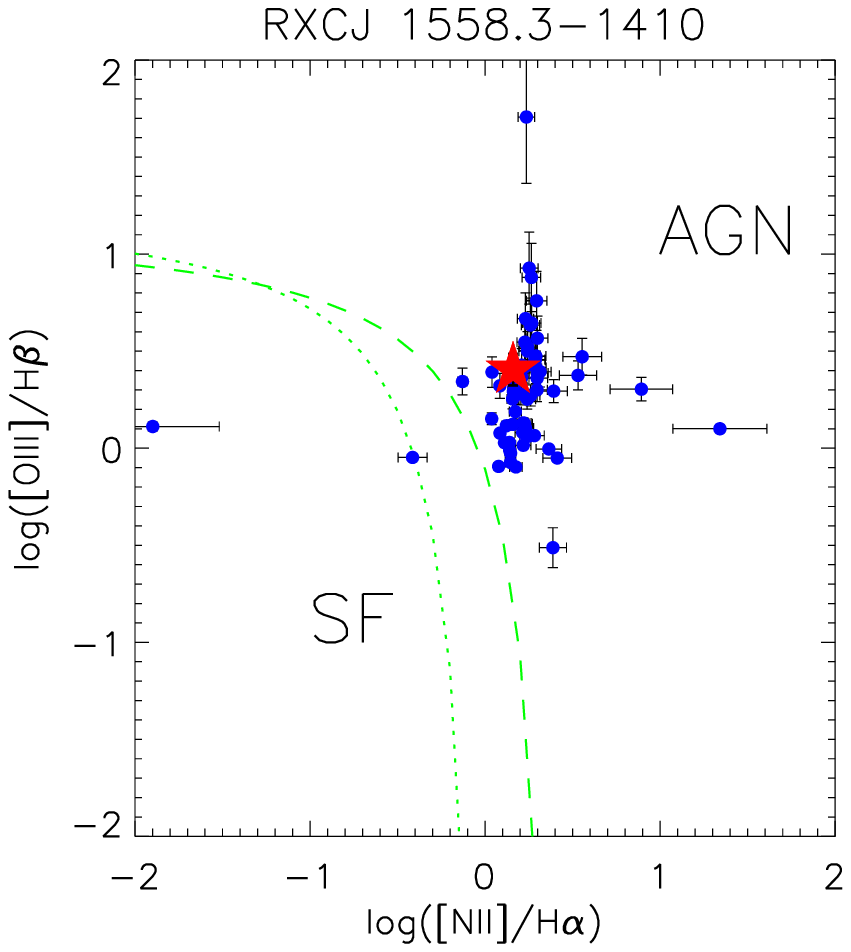,width=6.3cm}
\psfig{figure=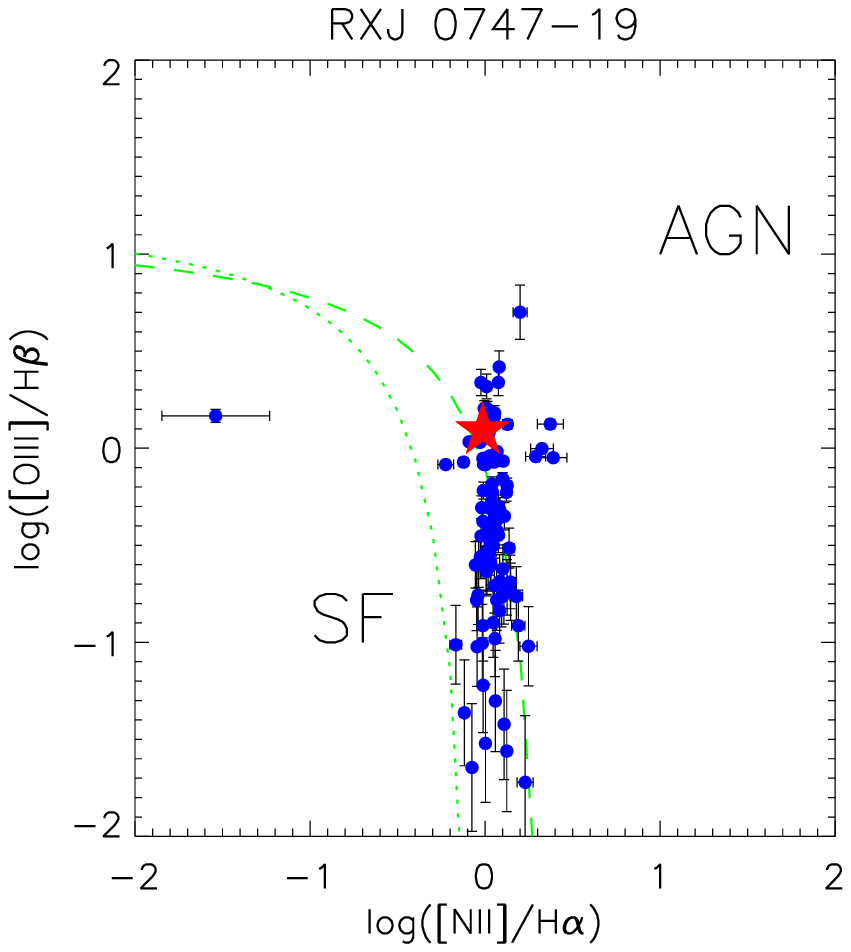,width=6.3cm}
\caption[]{continued.} 
\end{figure*}
    
\begin{figure*}
\psfig{figure=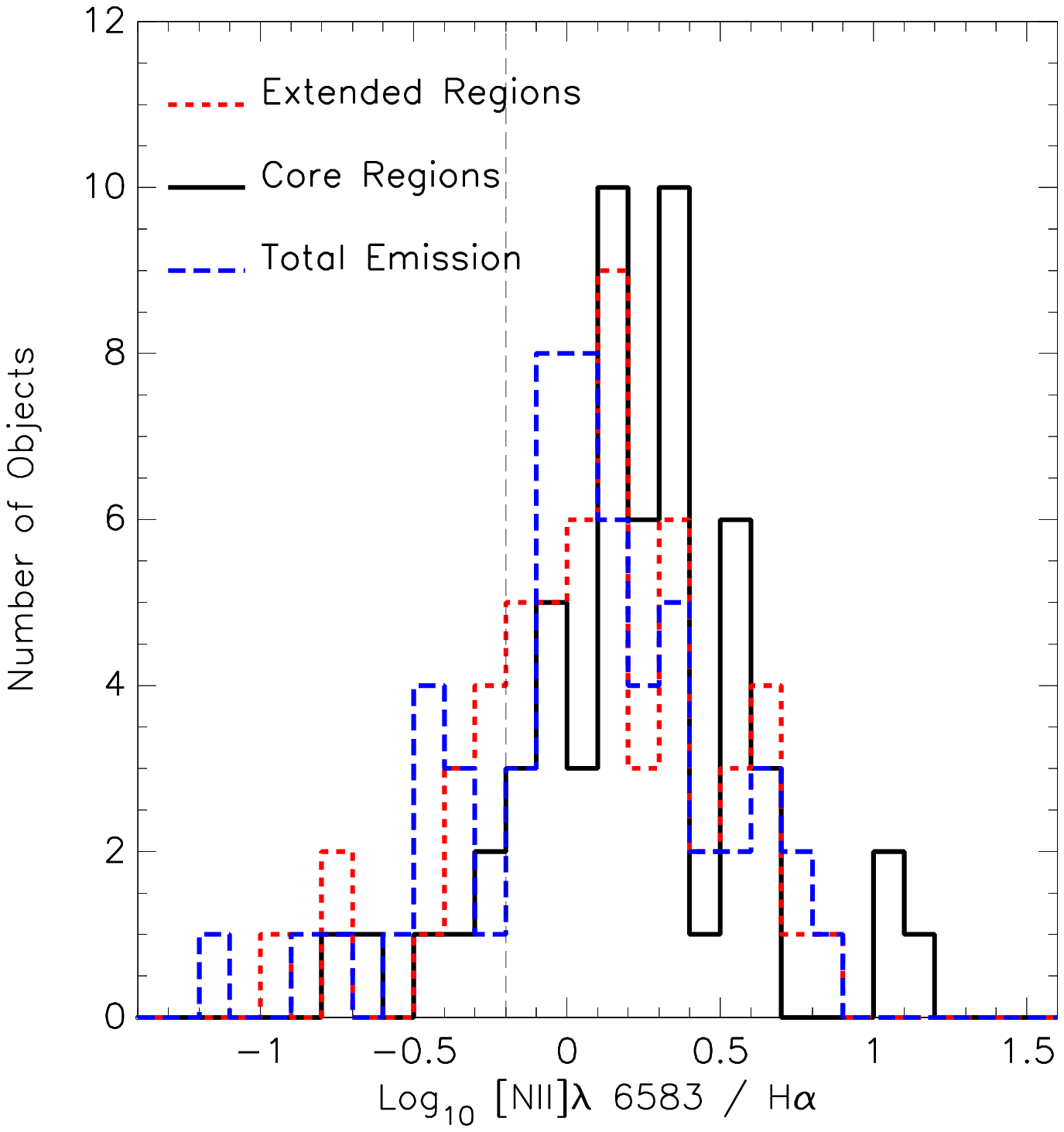, width=8cm}
\psfig{figure=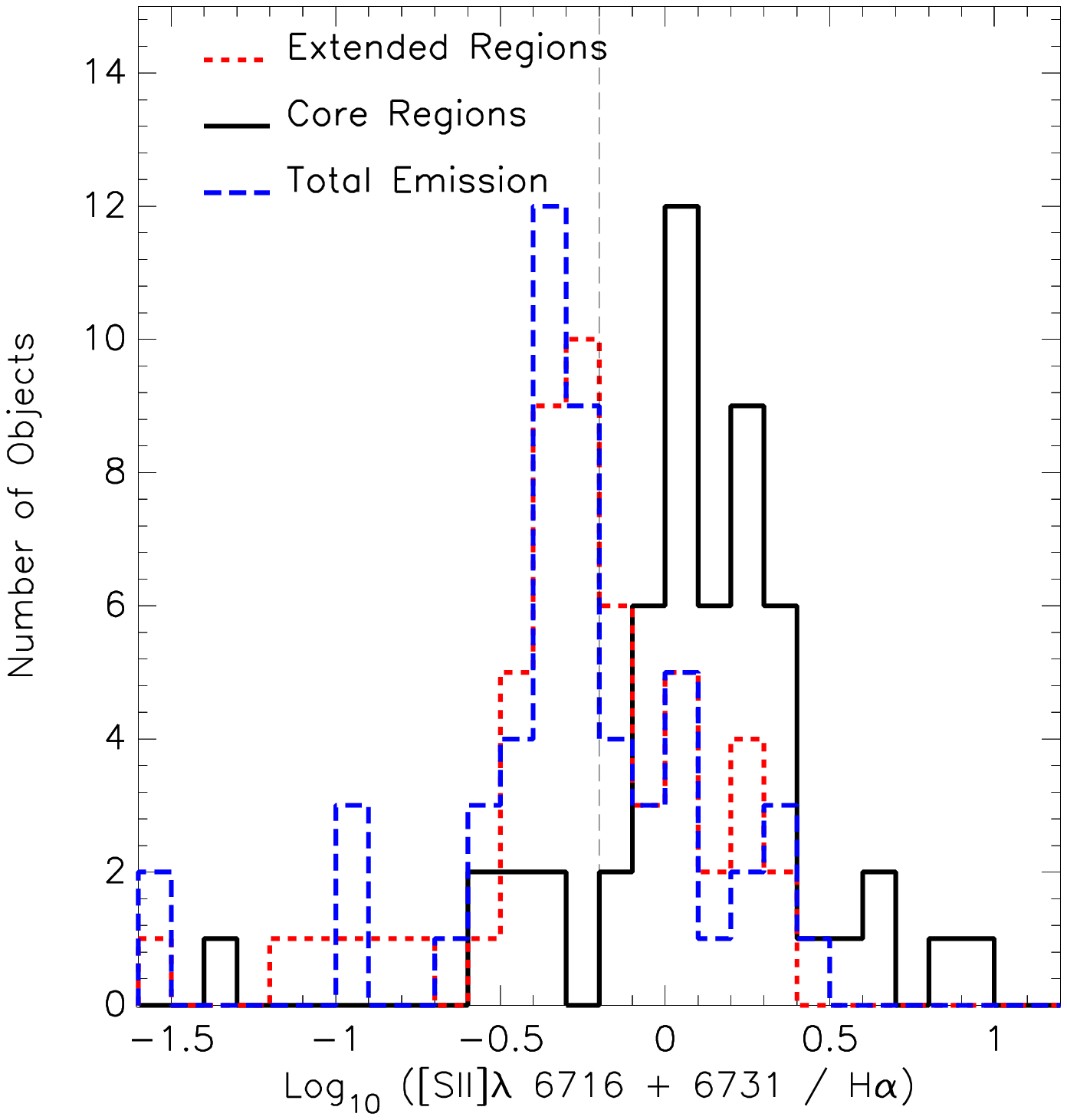, width=8cm}
\psfig{figure=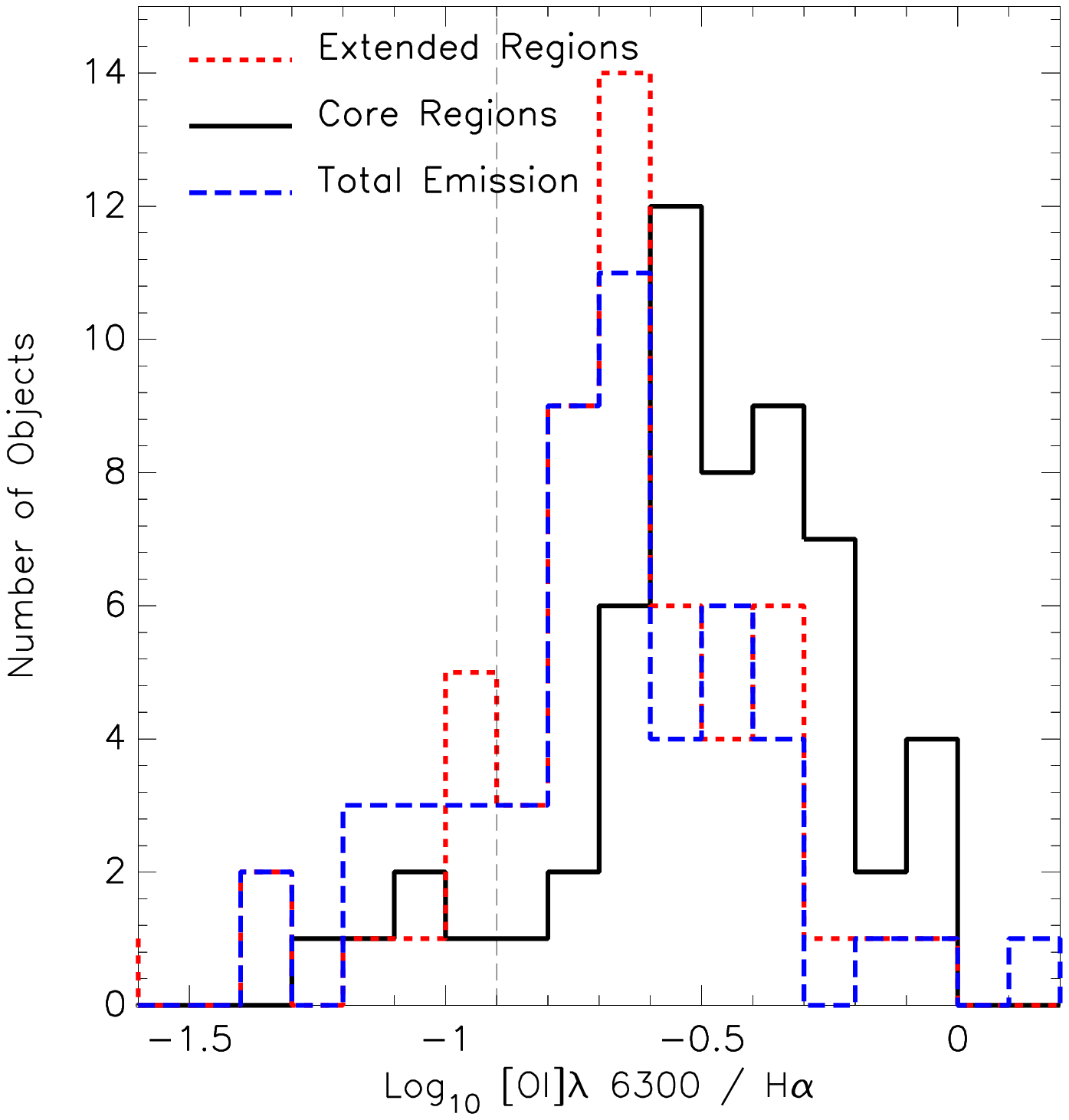, width=8cm}
\caption{Histograms of the [NII] (left), [SII] (right) and [OI] (bottom) to H$\alpha$ ratios for all object in the sample.  The ratios from the total emission as well as from the extended and core regions are shown separately to allow for easy comparison.  The vertical long dashed line shows the division between low and high energy ionisation for each line ratio. Note that core regions typically fall at higher ratios suggesting they are ionised by more energetic sources such as AGN or shocks implying an AGN is at least partially responsible for the elevated ratios.}
\label{fig:linerats}
\end{figure*}

While the [OIII] and H$\beta$ lines were not covered by all of our observations, the [OI] 
and [SII] lines appear in many of the observations by virtue of being close in wavelength to 
the principal H$\alpha$ line.  While these lines are not as good an indicator of the mean 
level of ionisation and temperature as [OIII]/H$\beta$, their ratio with H$\alpha$ are 
indicators of the relative importance of high--energy ionisation in large partially 
ionised regions \citep{ost89}. These lines were detected in the extracted spectrum 
in the majority of our sample (see Appendix \ref{app:lines} for a list of line fluxes) 
allowing us to make some measure of the importance of 
high--energy ionisation sources within the cores of clusters.  In Figure 
\ref{fig:linerats} we show histograms of the ratio of [NII], [SII] and [OI] with 
H$\alpha$ and note that most ratios fall above the division between low and high 
energy ionisation \citep{ost89,kew06}.

\begin{figure}
\psfig{figure=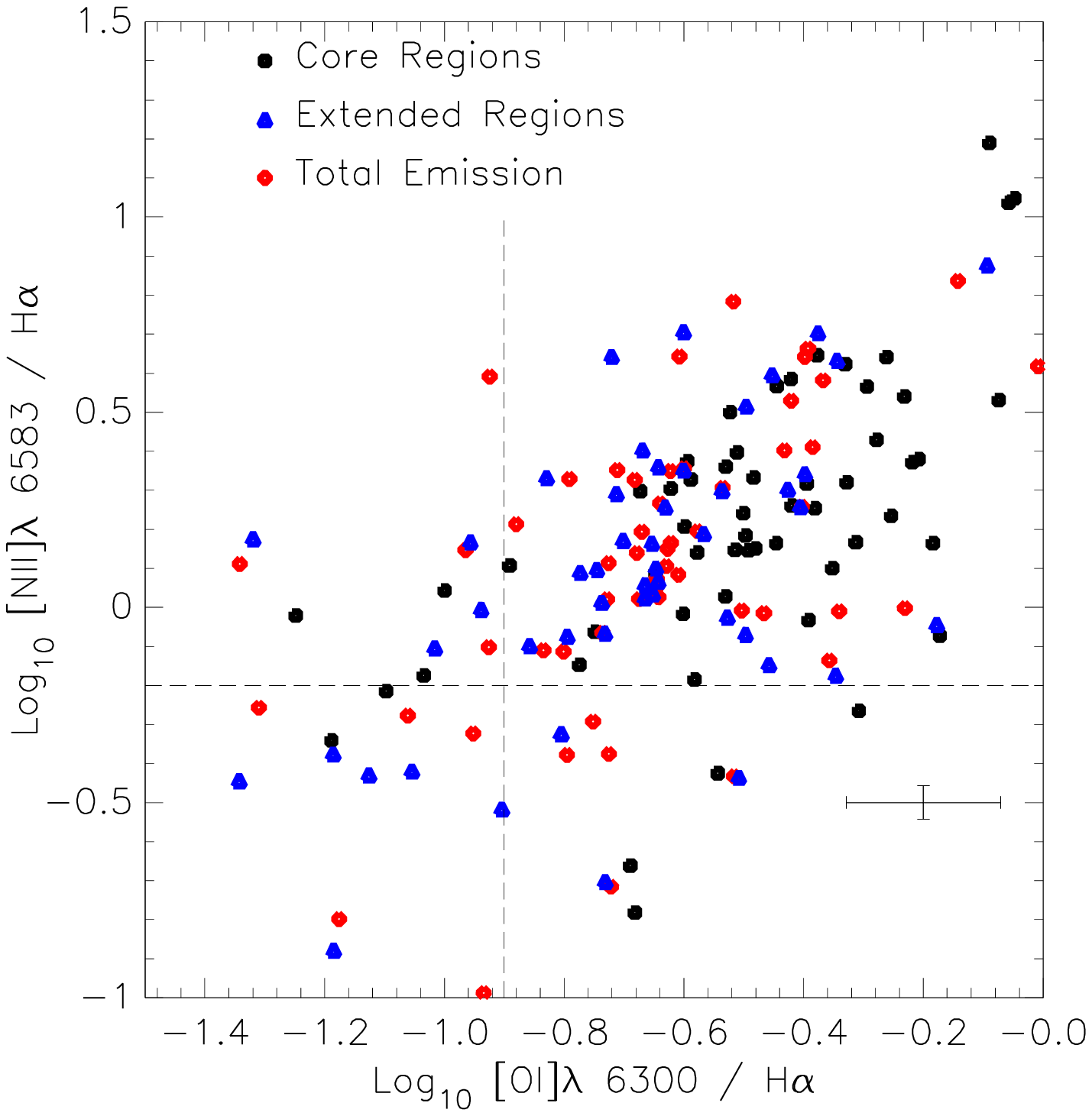, width=9cm}
\caption{A comparison of the [NII]/H$\alpha$ and [OI]/H$\alpha$ ratios for all objects in the sample which had detections of all three lines.  The points show the core regions (black squares), extended regions (blue triangles) and total emission (red diamonds) for each object. The dashed lines indicate the level above which empirical evidence suggests ionisation is caused by non stellar processes such as AGN or shocks.  The majority of sources fall within the upper right hand quadrant, this is especially true for the core regions which show only 9 objects falling below one of these lines and only 2 falling beneath both.}
\label{fig:OINII}
\end{figure}

In Figure \ref{fig:OINII} we plot the [NII]$_{\lambda 6583}$/H$\alpha$ ratio against the 
[OI]$_{\lambda 6300}$/H$\alpha$ ratio.  The [OI]$_{\lambda 6300}$/H$\alpha$ ratio shows a 
clear separation between sources ionised by low energy sources (such as UV emission 
from star formation) and those ionised by other sources (such as AGN or shocks).   
\citet{kew06} shows this divide occurs at 
approximately Log$_{10}$([OI]$_{\lambda 6300}$/H$\alpha$) = -0.9  with all their sources 
above this value being from non stellar processes. This is shown as a vertical line on 
Figure \ref{fig:OINII}.  
Similarly for the [NII]$_{\lambda 6583}$/H$\alpha$ ratio, \citet{ost89} shows a divide at 
approximately  Log$_{10}$([NII]$_{\lambda 6583}$/H$\alpha$) = -0.2.  While this separation 
is less obvious since at higher values of Log$_{10}$([OIII]$_{\lambda 5007}$/H$\beta$) 
lower values of Log$_{10}$([NII]$_{\lambda 6583}$/H$\alpha$) are seen it is still true that 
all objects above this value in the \citet{ost89} sample are from AGN.  Thus without access 
to [OIII] or H$\beta$ we use it as a first order estimate of the separation and show it 
in Figure \ref{fig:OINII} as a horizontal dashed line.  

It is immediately apparent 
from  Figure \ref{fig:OINII} that the majority of the points fall in the upper right 
quadrant suggesting they are high ionisation regions dominated by excitation from non 
star formation processes.  It is not surprising then that a lower fraction 
of the core regions fall below these cut offs than the regions containing extended or 
total emission (9/73 compared to 13/73 and 14/73 respectively). 

\begin{figure}
\psfig{figure=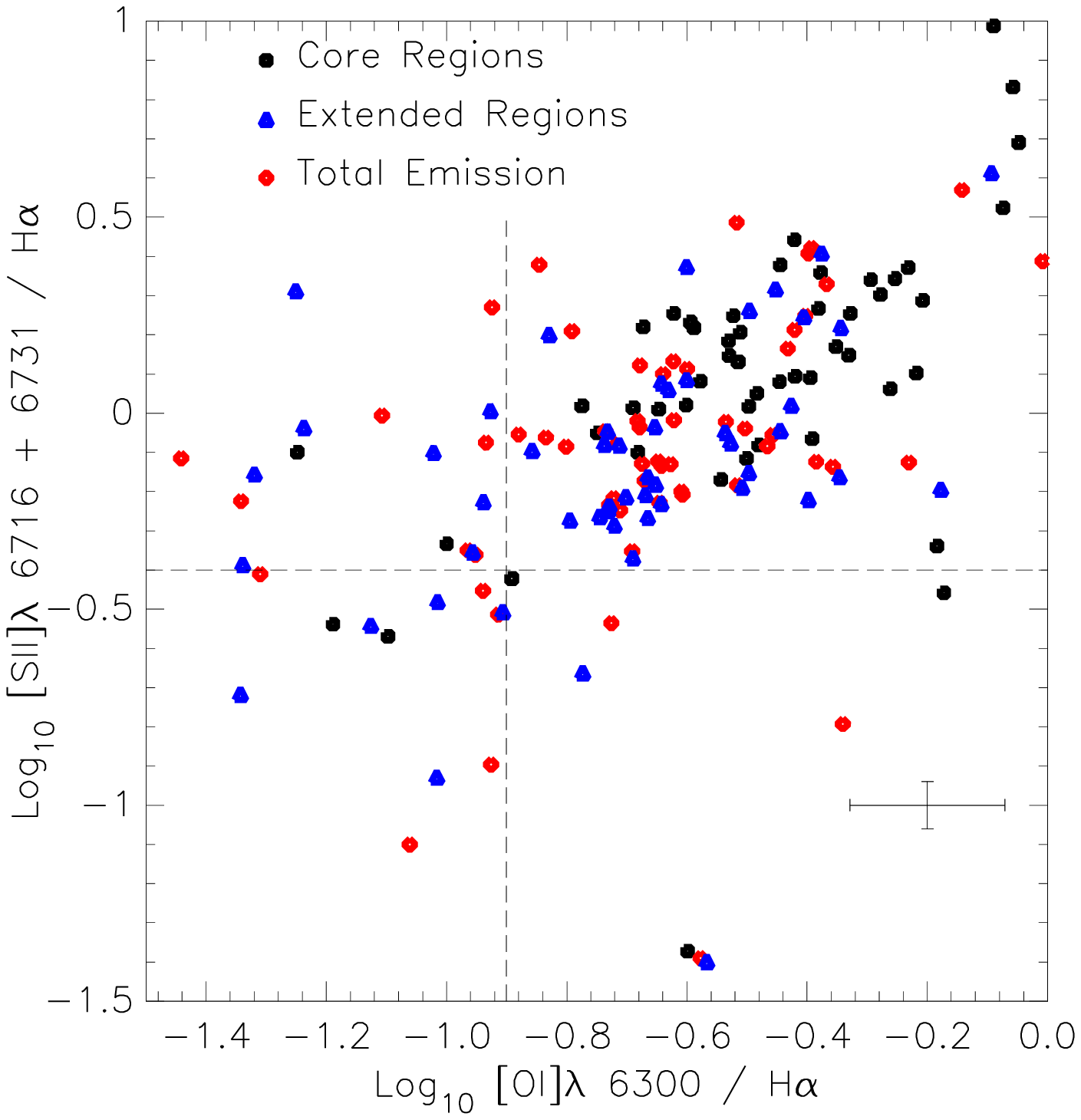, width=9cm}
\caption{A comparison of the [SII]/H$\alpha$ and [OI]/H$\alpha$ ratios for all objects in the sample which had detections of all three lines.  The points show the core regions (black squares), extended regions (blue triangles) and total emission (red diamonds) for each object. The dashed lines indicate the level above which empirical evidence suggests ionisation is caused by non star formation processes.  The majority of sources fall within the upper right hand quadrant, this is especially true for the core regions which show only 7 objects falling below this for one of these ratios and only 2 falling beneath both.}
\label{fig:OISII}
\end{figure}

We show a comparison of the [SII]$_{\lambda 6716 + 6731}$/H$\alpha$ ratio against the 
[OI]$_{\lambda 6300}$/H$\alpha$ ratio in Figure \ref{fig:OISII}. Like the 
[OI]$_{\lambda 6300}$/H$\alpha$ ratio,  [SII]$_{\lambda 6716 + 6731}$/H$\alpha$ shows a 
clear separation between sources ionised by high energy photons (such as from narrow line 
AGN) and those ionised by other sources \citep{ost89}.  Like the 
[NII]$_{\lambda 6583}$/H$\alpha$ the separation point shows some variation at high values 
of [OIII]$_{\lambda 5007}$/H$\beta$ but as the BPT diagrams in Figure \ref{fig:bpt} 
suggest this is generally not the case in our sample we adopt the value 
Log$_{10}$([SII]$_{\lambda 6716 + 6731}$/H$\alpha$) = -0.4 as the separation between 
high--energy ionisation regions and low--energy ionisation regions.  This is 
the maximum value that a HII region had in the sample of \citet{ost89}, above this value 
all sources were ionised by non stellar processes.  We plot this separation 
as a horizontal dashed line on Figure \ref{fig:OISII} along with the same separation 
for the [OI]$_{\lambda 6300}$/H$\alpha$ ratio as used earlier.  Once again the objects 
show a clear trend towards higher values of both ratios suggesting 
the ionisation is dominated by high energy processes such as AGN or shocks. Like for 
[NII]$_{\lambda 6583}$/H$\alpha$ we see a lower fraction of core regions (7/73) fall below
these cut offs than the regions containing extended (15/73) or total (15/73) emission.


\begin{figure}
\psfig{figure=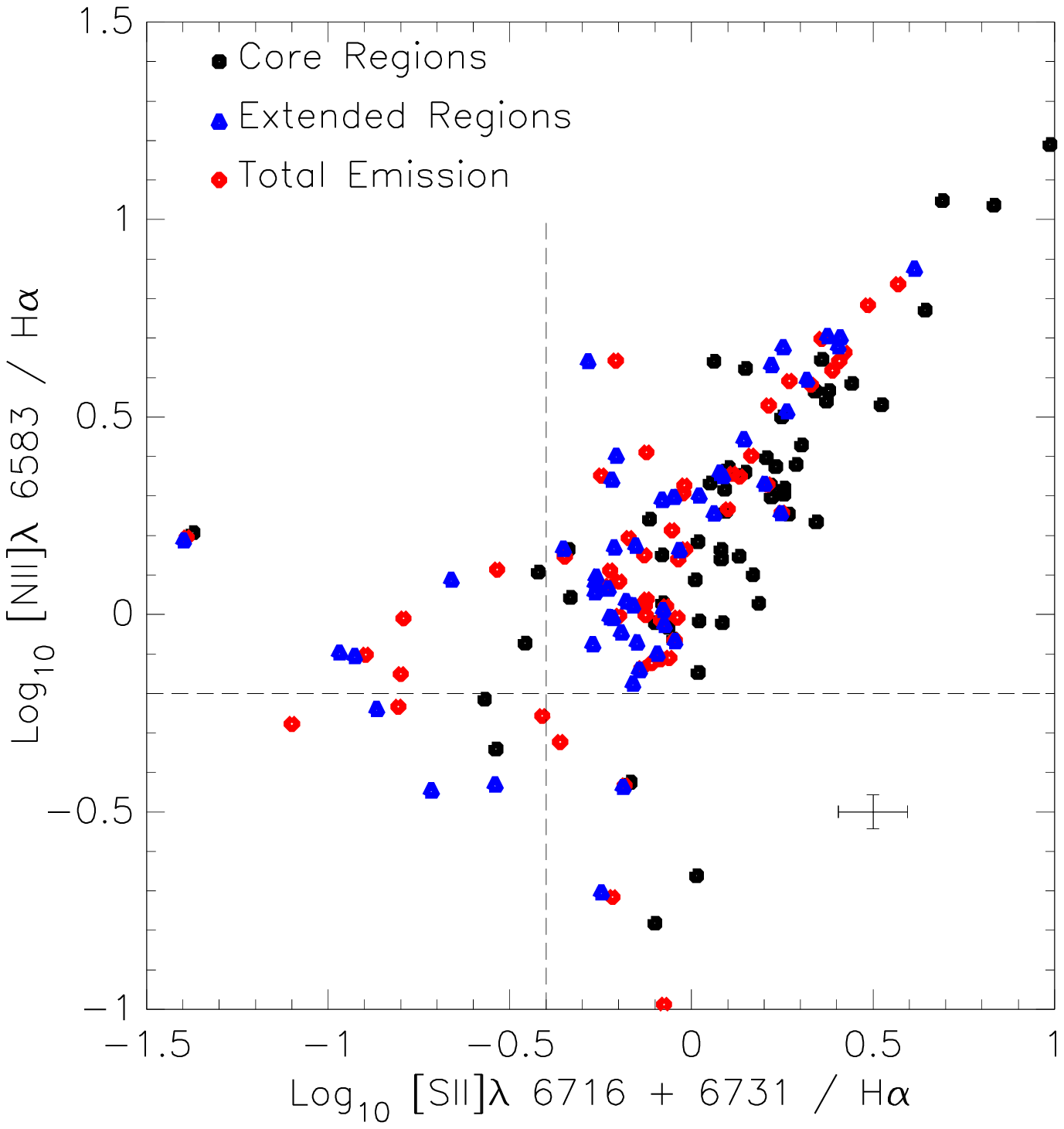, width=9cm}
\caption{A comparison of the [SII]$_{\lambda 6716 + 6731}$/H$\alpha$ and [NII]$_{\lambda 6583}$/H$\alpha$ ratios for all objects in the sample which had detections of all four lines.  The points show the core regions (black squares), extended regions (blue triangles) and total emission (red diamonds) for each object. The dashed lines indicate the level above which empirical evidence suggests ionisation is caused by non star formation processes.  The majority of sources fall within the upper right hand quadrant, suggesting that non stellar processes dominate the excitation.}
\label{fig:NIISII}
\end{figure}

In Figure \ref{fig:NIISII} we compare the [NII]$_{\lambda 6583}$/H$\alpha$ and [SII]$_{\lambda 6716 + 6731}$/H$\alpha$ ratios.
Once again the dashed lines represent the separation between high--energy ionisation 
and low energy ionisation regions.  The majority of the points fall above both cut offs 
again (with only 8/73 cores, 9/73 extended regions and 12/73 totals falling below the cut offs) 
suggesting that the gas is excited by more energetic processes.  Interestingly the 
points above both separating values show a much tighter correlation than is seen when 
either ratio is compared with [OI]$_{\lambda 6300}$/H$\alpha$. The reason for this 
is not clear however, we do note that [OI]$_{\lambda 6300}$ is the weakest line used
for this analysis and as such has larger measurement errors than the other lines. This 
may be responsible for introducing additional scatter in the other comparison plots 
but we note that there is considerably more scatter below the cut off values in 
Figure \ref{fig:NIISII}, which is comparable to that seen in the other comparison plots.

Using the results from all three of these plots we can categorise the 73 objects in 
our sample (separated into total emission, core emission and extended emission) based 
on all three line ratios. We classify star forming objects as those whose ratios 
fall below the cut off values for all three ratios, AGN like objects (those excited by 
AGN or other high ratio producing mechanisms) have all three 
ratios above the cut off values, ambiguous objects have at least one ratio above and 
at least one below the cut offs and unclassified objects have one line missing (due to 
absorption, falling outside the spectral coverage or just being too weak to detect) 
preventing analysis using all 
three ratios.  According to this scheme our sample contains 5 objects whose total emission 
is consistent with star formation, separating the cores and extended emission shows 
that only 3 cores are dominated by star formation while the number of extended regions 
rises to just 6. By contrast, 46 objects have a total emission consistent with AGN like 
excitation, with 47 cores and 41 extended 
regions also showing AGN like excitation.  We find that 18 objects have ambiguous 
ratios in their total emission, with 10 cores and 16 extended regions also being 
ambiguous when analysed separately.  Finally we find the emission of 4 total regions, 
13 cores and 10 extended regions cannot be classified due to having at least one line 
missing from their spectrum.

Some care must be taken when interpreting these ratios, in particular the presence of 
any extreme outliers.  Due to the range in redshift some objects in the sample showed 
atmospheric absorption features in the region of their spectra containing these principle 
lines.  For some objects this was very apparent as a dip of the spectra to negative 
values. However, this was not always the case especially in systems with strong lines.  
Fortunately, the relatively narrow absorption feature and large redshift range mean that 
this problem only affects a handful of objects in the sample. We do not include objects 
affected by this absorption in our plots thus allowing us to draw 
general conclusions from the overall statistics.   
Importantly, the majority of the points fall above the transition values between 
low and high--energy ionisation regions for all lines suggesting that the gas  
ionisation is not dominated by star formation within the BCG.  This is consistent with our BPT 
analysis presented earlier and with previous studies of line ratios in cluster cores
\citep{jf88,cf92}.  

AGN are expected to be present in the cores of BCGs \citep{rus13,hla13} and are thus a 
potential source of the ionisation seen, especially in the core regions.
However, it is important 
to note that the extended regions of the ionised nebula in clusters can reach out 
over 10s of kiloparsec from the core of the BCG.   This would require the AGN to remain 
the dominant ionising radiation source over many orders of magnitude in scale. Shocks 
could be present through out the gas and can produce the line ratios seen.  However, 
comparing to the shock models of \citet{all08} our objects would require shocks over a 
larger range of velocities (100--1000\,km\,s$^{-1}$) to 
fully explain all the line ratios seen.  We would expect high velocity 
shocks to be visible as regions of enhanced line width.  While the central regions of 
our objects do show line widths (FWHM) sufficiently broad to be consistent with such 
shocks most objects have FWHM on the order of $\sim$100--200\,km\,s$^{-1}$ in their 
extended regions suggesting that the fast shocks needed to produce the line ratios 
seen are not present here. The line ratios also do not agree well with the shock 
models of \citet{all08}.  Of the seven objects with [OIII] and H$\beta$ detections 
only one has an [OIII]/H$\beta$ ratio high enough to be consistent with a shock in 
excess of 300\,km\,s$^{-1}$.  From Figure \ref{fig:linerats} we see that the majority 
of the log$_{10}$([NII]/H$\alpha$ ratios fall within 0.0--0.5 which are consistent with 
those expected for shocks ($\sim$ 0.0--0.4) however, the log$_{10}$([OI]/H$\alpha$) 
ratios are mostly below -0.3 making them inconsistent with the values expected for 
shocks ($\sim$ -0.3--0.2).
The log$_{10}$([SII]/H$\alpha$) ratios for the majority of extended regions (-0.5--0.0) 
are consistent with those expected for shocks ($\sim$ -0.5--0.0) but the core regions 
are not (0.0--0.4). As such there is no consistent evidence that shocks play a major 
role in the excitation
of the gas.

Other sources of excitation 
within cluster cores \citep[particle heating for example,][]{fer09,fab11} are therefore
required to contribute significantly to the excitation of the gas. This can 
explain how the line ratios can 
remain relatively consistent throughout the gas, over a range of different scales, 
linewidths and average line ratios. This also explains how the ionised gas in Abell 2566 
shows evidence of AGN like ionisation despite being offset from the BCG 
(and thus, far from the AGN) and having narrow line widths.  While AGN, shocks and star
formation may not be the main source of the ionisation they are likely present in at 
least some of our objects leading to local variations of line ratios within a given 
object and can thus explain why some objects do show local variations while others 
do not.  In particular the trend towards higher average line ratios in more compact 
objects seen in Figure \ref{fig:N2Hkpc} suggest that while other mechanisms are 
likely dominating the excitation of the gas globally, the AGN still plays a role.

%% file: conclusions_r2.tex
\section{Discussion}
\label{chap:sum}






\subsection{What fraction of line emitting BCGs are highly disturbed}

Our analysis showed that the extent of the line emitting gas in cluster cores 
varies widely from system to system.  A few objects are barely resolved 
by VIMOS, their extent dominated by the seeing of the observations with their line 
emission concentrated at the centre of the 
BCG.   From the full sample of 73 objects only 
Abell 194 appeared to be truly compact (extent $\le$ seeing) in all directions.  A 
further two were compact along a single axis but showed extent along the other.
However, the majority of the objects had a true extent on the order of 5 to 20 kpc with 
a few extending out to 50~kpc (S780, Z348 etc.) to where the line emission far outshines 
the stellar continuum.

From the full sample we identify five distinct morphological states.  The compact 
objects described above account for the smallest fraction.  While it is possible that 
these 
objects do have structure, it would require higher resolution observations taken 
in better seeing conditions to detect it.  Offset objects in which the peak and majority 
of the line emission is centred off the BCG account for 4 of the 73 objects.
Two of these objects were studied in detail in \citet{ham12} and it was found that the 
offset line emission was co-spatial with the peak of X-ray emission from the ICM. Another
$\sim$15\% (10/73) of the sample showed plumes of H$\alpha$ emission which had a bright 
peak 
centred on the BCG but also showed a clear extent in one preferential direction.  For 
one object (Abell~3880) this plume appears to lead to another object within the VIMOS 
filed of view. However, we note that this object shows spectral absorption lines at 
z=0 clearly identifying it as a foreground star.  One possible explanation for plumes is 
that they are caused by a cooling wake similar to that proposed by \citet{fab01} as the 
origin of the long filaments in Abell 1795.  While in Abell 1795 the resulting H$\alpha$ 
emission appears as long filaments that we do not see in our plumes, we note that the 
thickness of the filaments in Abell 1795 is significantly smaller to our mean seeing and 
all our plumed objects are of comparable or greater redshift than Abell 1795.  As such 
the seeing may be blending the filaments together in our observations producing the 
plumes that we see.  A dedicated study of these objects combining high resolution X-ray 
and H$\alpha$ observations and comparing to Abell 1795 would be needed to confirm this.

Disturbed objects have line emission that peaks close 
to the centre of the BCGs continuum peak but also shows extended non-uniform structures 
at lower surface brightness.  There are 13 such objects ($\sim$18\%) present in the sample 
which suggests that they are not common amongst the BCG population.  It is possible 
that their prevalence in previous studies is a result of choosing to observe the most 
extreme systems.  Indeed the most common type of morphology seen in our sample are 
quiescent, simple elliptical systems which are concentrated towards the centre of the 
BCG.  More than 60\% of the sample (45/73) show a Quiescent morphology suggesting that 
major events (such as interactions) which can disturb the gas are relatively rare and that 
many systems remain undisturbed for long periods of time.  

One caveat that must be considered however, is that the observations of the sample are 
not able to resolve the filamentary structures seen in narrow band imaging of objects 
such as NGC~5044 \citep{sun12} and NGC~1275 \citep{con01}.  Such filaments are thought 
to be created when gas is uplifted through AGN feedback as the expanding radio lobe 
entrains the gas. The filaments in NGC~5044 appear as an extended, non-uniform 
low surface brightness emission region in our IFU observations and make the 
system appear disturbed.  It is possible that several of the other disturbed systems 
in our sample are similar to NGC~5044 with filamentary structures throughout their extended 
emission. 
We do identify the presence of filaments of H$\alpha$ emission in six objects,
Abell 3378, Abell 3998, Abell 85, RXCJ 0120.9-1351, RXCJ 1304.2-3030 and S 805. While it 
is possible these are related to the filaments seen in narrow-band observations we note 
that the apparent thickness of these structures is substantially higher than that of 
those observed with narrow-band filters.  While this may be an effect of the poorer 
seeing of our observations these systems would have solitary long filaments, more like 
those of Abell~1795 \citep{mcv09} than NGC~1275 or NGC~5044.


The kinematics of the ionised gas is also an important indicator 
of how disturbed the system is.  The mean velocity maps produced by the 
fitting routine allow us to directly examine the line of sight velocity structure of 
each object.  The velocity 
structure of the ionised gas in two thirds of the sample 
appears highly organised within the central regions with most 
objects showing a continuous velocity gradient across the brightest region of the 
H$\alpha$ 
emission which is typically coincident with the BCG.

\begin{figure}
\psfig{figure=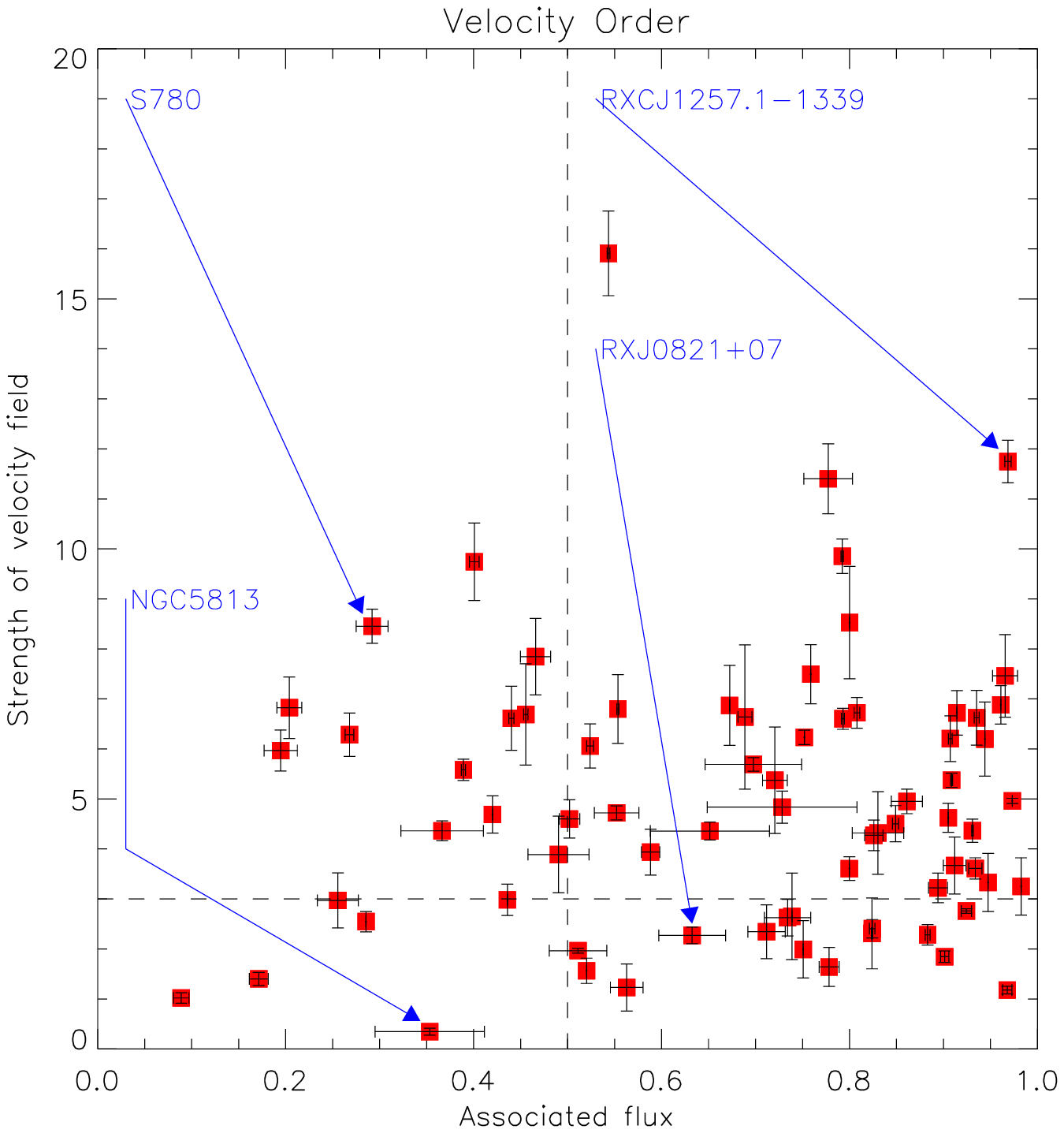,width=9cm}
\caption[Comparison of the Strength of the ordered structure in the velocity field with the flux associated to it]{A comparison of the strength of the ordered structure in the velocity field (measured in standard deviations of the residuals) against the flux associated to the velocity field as a fraction of the total flux.  The horizontal dashed line separates the strength of the velocity fields at 3 $\sigma_{vel-resid}$ while the vertical dashed line separates the associated flux at 50\%. One object from each quadrant is labelled to allow comparison with the velocity maps. More than half our objects fall into the upper right hand quadrant suggesting that the majority of their H$\alpha$ emission is associated with a strong velocity field.}
\label{fig:velorder}
\end{figure}

To test the significance of this ordered velocity structure
we produced a simple model velocity field for each object following a single 
velocity gradient through the peak of the H$\alpha$ emission and adjusted the PA and 
gradient to minimise the residuals with the H$\alpha$ velocity map.  We define the 
disorder in the velocity map to be the standard deviation of these residuals 
($\sigma_{vel-resid}$). We calculate the 
strength of the velocity field as the peak--to--peak velocity divided by 
$\sigma_{vel-resid}$ and the region of the line emission associated with 
the velocity gradient as having residuals less than $\sigma_{vel-resid}$/3.  We compare 
the strength of the velocity field with the fraction of the H$\alpha$ flux 
associated to it in Figure \ref{fig:velorder}.  More than half our 
objects (40 out of 73) fall in the upper right hand region of this plot (velocity field
strength greater than 3 and more than half the H$\alpha$ flux associated with the 
velocity gradient) suggesting that the majority of their H$\alpha$ emission is 
associated with a strong, ordered velocity field. We note that taking a velocity field 
strength of 2 as the dividing limit places roughly 2/3rds of our objects in this 
category consistent with our visual inspection.  This may not be an unrealistic 
assumption given that we use a very simple model for the velocity field and it is 
likely a more sophisticated model would have lower residuals and thus the value of 
the peak--to--peak velocity divided by the standard deviation would be higher. 

The more extended low surface 
brightness emission present in some systems does not seem to share this ordered line of 
sight velocity structure, showing a much more random and disturbed distribution of 
velocities.  We would expect to see increased line widths in regions of the gas which 
is turbulent and 
disturbed. However, the outer regions in most systems appear to have low line widths 
which are relatively 
uniform suggesting that this is not the case.  It is important to note that
in systems with a large line of sight velocity shift, the coarse spatial sampling 
can result in artificially broadened lines but we determine this effect will be 
negligible given the 
line widths and velocity gradients measured (see Figure \ref{fig:sigvel} and the related 
discussion).

A number of systems (9/73) showed multiple velocity components in their line emission.  
Typically these additional components are very broad line emission (8/73) in the 
very central regions of the BCG on scales comparable to the seeing and are most likely 
emission from the nuclear regions where gas is feeding onto the AGN.  Indeed this is 
expected as the presence of strong radio sources within the BCGs of many clusters suggest 
that cold gas is actively being accreted in many systems.  More interestingly however, 
two of our 73 objects showed multiple narrow line 
components similar to those found in Abell~1664 \citep{wil06,wil09}.  These objects were 
Abell~3574 and NGC~5044 and the two are very different. NGC~5044 shows a double 
component in its core regions (in addition to a broad line) while Abell~3574 has a 
double component ring of gas which surrounds
its core.  The full interpretation of these double component systems 
requires more data but suggests the ionised gas may be much more disturbed in a 
small fraction ($\sim$3\%) of systems.

\subsection{What role does the BCG play in the cooling of gas from the ICM?}

In \citet{ham12} we studied several objects in detail and determined that 
(for that subsample at least) the offset H$\alpha$ emission was coincident with the 
peak of the X-ray emission from the ICM. Our analysis of the X-ray -- optical line 
offsets in Section \ref{sec:offset} would appear to confirm this result as at most 
3 of the 24 objects studied showed a potential significant offset of the H$\alpha$ 
emission from the peak of the X-rays.
The analysis of the structure of the full sample 
presented in this paper does identify several more objects which show an offset between the 
H$\alpha$ and continuum peaks.  \citet{ham12} were able to associate the offset peak 
with evidence of gas sloshing  \citep{jon10,jon11,zuh10,zuh11,bla11} in the ICM as 
well as continued cooling at the position 
offset from the BCG.  However, in these additional systems the 
multi-wavelength data necessary to make this determination (deep X-ray imaging and/or 
resolved CO detections) were not available so we cannot say if this is true for all 
offset systems.  Despite this it is apparent that offset line emission, while still 
rare (4/73 systems), is more common than initially thought and may thus play a more 
significant role in the evolution of the cluster core. 

In a small number of objects the stellar continuum was sufficiently bright to allow 
us to extract the stellar kinematics 
from the BCG.  Comparing the maps of the stellar kinematics to the gas kinematics 
it is apparent that the line of sight velocity structure 
of the stellar component is negligible, in stark contrast to the high velocity, 
ordered structures seen in the gas component.  This suggests that the 
kinematics of the line emitting gas are decoupled from those of the stellar population 
of the BCG implying that cooling gas is not necessarily related to the 
BCG even when cooling is occurring at the BCGs location. 

This raises the question, if the gas is relaxed and stable why is it not forming stars 
and producing a stellar velocity field that matches that seen in the gas?
Hydra-A has one of the most extreme H$\alpha$ velocity fields in our sample, the gas   
is contained in a rotating disc that is known to be forming stars at a rate of 
2--3~M$\odot$~yr$^{-1}$ \citep{don11,hof12}.  If the disc is responsible for feeding 
the AGN \citep{ham14} then it must have been stable for the lifetime of the outburst 
\citep[$\sim$10$^8$~yr for an accretion rate of 0.1--0.25~M$\odot$~yr$^{-1}$ and a total accreted mass of 10$^7$~M$\odot$][]{wis07}.  
Assuming the star formation rate is constant during this time then the total mass of 
stars the disc will form is 3$\times$10$^8$~M$\odot$, which is a negligible fraction of 
the typical stellar mass of a BCG (10$^{11}$--10$^{12}$~M$\odot$).  
Thus the stellar velocity field we observe is 
dominated by the bulk of the stellar component formed in previous generations of star 
formation, which need not share the current velocity field of the gas.

\subsection{What role does the cold gas play in fuelling feedback?}

Mechanical feedback from AGN outbursts within BCGs is commonly believed to be the 
main contributing factor preventing the catastrophic cooling of gas in cluster 
cores.  A requirement for this process to work effectively is a mechanism of 
self-regulation to maintain the balance between cooling gas and feedback. Thus a 
link must be established between the gas cooling (on kpc scales) and its supply 
to the super massive black hole which results in the activation of the AGN once
cooling reaches a critical threshold.  In \citet{ham14} we show that one system (Hydra-A) 
in our sample has kinematics in the ionised gas that clearly trace the presence of an 
extended ($\sim$ 5 kpc) disc of cold, molecular and atomic gas in the core of the 
system. Such a disc can potentially channel gas to the centre of the system relatively quickly allowing it to 
fuel the AGN.  The confirmation of the disc in Hydra-A was made possible thanks to 
the system being relatively close by, having a large velocity shift and being viewed at 
close to edge on.  
Our analysis of the velocity gradients suggest that this sample is consistent 
with a sample of rotating discs similar to the one seen in Hydra-A and Figure 
\ref{fig:velorder} suggests that $>$\,50\% of systems have the bulk of their 
H$\alpha$ emission associated with this velocity gradient.  However, 
confirming the presence of a disc within each individual system requires a more 
sophisticated, case--by--case analysis (which will be presented in paper II, 
 Hamer {\em et al.} in prep.). 

From the velocity dispersion maps it is apparent that the linewidths are typically 
broadest towards the peak of the H$\alpha$ emission which is consistent with this 
position having the longest line of sight through the gas and thus sampling the widest 
distribution of velocity components.  We note that the FWHM in the central regions is 
typically on the order of the velocity shift across the whole system suggesting it 
originates from the extended gas in the ionised nebula.  However, 8 of the 73 systems show 
some evidence of very broad (FWHM on the order of 2355 km s$^{-1}$ or more) components 
localised (on scales of the seeing) to their central regions.  This can be 
interpreted as emission from an AGN suggesting that the AGN in these systems are actively
accreting. However, we note that the broad component is typically very 
weak and the scales of the seeing are much greater than those expected for an AGN 
making this difficult to confirm without deeper, higher resolution observations.
 
\subsection{The source of the ionisation}

The maps of the [NII]$_{\lambda 6583}$/H$\alpha$ ratio produced in Section 
\ref{chap:results} show a wide variation in value 
between objects. More interestingly perhaps the ratio is seen to vary within a given 
object suggesting that the source of the ionising radiation may not be uniformly 
distributed.  Likewise it shows no trend radiating from a single point suggesting that 
the dominant source of the ionisation is not a single localised source but is more 
homogeneously spread throughout the cluster core.  
Given the extended nature of these 
sources, and the variation seen within different regions of individual objects, it seems 
unlikely that an AGN positioned at the centre of the BCG could be responsible for the 
ionisation throughout the nebula. However, the trend that more compact objects typically 
show higher line ratios than extended objects does suggest that AGN still play a role in 
the excitation of the gas even if it is not dominant.   We therefore conclude that 
other, more diffuse, sources must also be significantly contributing to the excitation 
of the gas. 

Star formation is (along with AGN) the most commonly invoked excitation mechanism and 
has the potential to be distributed throughout the BCG.  For the few objects (just 7, less 
than 10\% of our sample) in which we had access to the H$\beta$ and [OIII] lines and we 
perform a BPT analysis \citep{bpt81} to look for evidence of excitation by star formation.
This analysis suggests that the gas in all of these objects is predominantly 
ionised by non stellar processes and is, as such, inconsistent with excitation by star 
formation.  Analysis with other spectral indicators (the [NII/H$\alpha$], [SII]/H$\alpha$
and [OI]/H$\alpha$ emission line ratios) also indicates that non stellar processes 
dominate the ionisation of the gas in most systems.  These line ratio diagnostics are all
consistent and rule out star formation as the dominant excitation mechanism in the vast 
majority of objects in our sample. 

Shocks also seem to be an appropriate candidate for a distributed excitation mechanism 
however, we note that the velocity dispersion maps do not show evidence of shocks of 
sufficient velocity in the extended regions (seen as enhanced line widths) to produce the 
line ratios seen. Comparison of the measured line ratios with those predicted by shock 
models also shows no consistent evidence that shocks are a major contributor to the
gas excitation for the objects in our sample.  The most likely candidate then is 
particle heating caused by a reconnection of the gas phases \citep{fer09,fab11}. This 
can also explain the presence of ionised emission in offset objects which have bright 
line emission coincident with their X-ray peaks \citep{ham12} but are displaced from the
bulk of the BCGs stellar component.  We note however, that AGN still play a role in 
the excitation of the gas and indeed other sources of excitation, such as cosmic rays, 
shocks and star formation, are likely also contributing to the overall
ionisation state of the gas. All of these processes, and their interaction, need to be 
considered to fully understand the excitation of the gas.


\section{Conclusions}
\label{sec:con}
We have analysed the morphology, kinematics and excitation of the optical line emitting
nebula in a sample of 73 BCGs with the VIMOS IFU.  Our goal has been to develop a more 
complete understanding of the nature of these ionised nebula, what role they play in the 
physical processes affecting the cluster core and what they can tell us about the 
interaction of the BCG with the ICM.  Throughout this paper we present a number of 
results that can be used to address these issues which we will now summarise.

\begin{enumerate}[1)]
\item We classify our objects into five categories based on the morphology of their optical line emitting nebula.  While we do find that some objects appear disturbed (13/73), the majority of our sample have elliptical, centrally concentrated morphologies (45/73) suggesting that the gas is in a relatively quiescent, undisturbed state (Section \ref{sec:morph}).
\item We detect a number of systems in which the peak of the line emission is offset from the BCG.  Some of these objects where already known and they are typically found in systems in which the ICM is ``sloshing'' \citep{ham12} suggesting that the gas in the ICM has been disturbed.  The number of these objects in our sample is small (4/73) suggesting that such major disturbances are rare (Section \ref{sec:offset}).
\item The extent of the nebulae varies quite considerably between objects but can extend out over several tens of kpc in some objects.  While previous narrow band studies suggest we might expect to see complex filamentary structures in these extended regions we are unable to resolve the filamentary structures in most objects. However, in some systems we do detect isolated filaments which may be related and only detectable because they are isolated and thus not blurred with other filaments as a result of the seeing (Section \ref{sec:morph}).
\item The majority of our objects show gas with a high line of sight velocity in the form of a velocity shift that runs across the ionised nebula.  This velocity structure is visible in the velocity maps (Section \ref{sec:velstruct}) but it is also visible in the channel maps (Section \ref{sec:chan}) where the position of the gas can be seen to vary smoothly with channel.
\item The velocity structure of the gas in most systems appears quite ordered and typically varies little away from the main velocity shift that runs across the nebula (Section \ref{sec:velstruct}).  This is especially apparent when we consider the strength of this velocity shift against the fraction of the H$\alpha$ flux that it represents 
(Figure \ref{fig:velorder}) and we find that in 40/73 of our objects more than half the 
H$\alpha$ flux is associated with a strong velocity shift.
\item When measuring the gradient of the velocity shifts observed we find that the gradient shows a clear relationship with the separation of the maximum and minimum velocities but none with the magnitude of the velocity difference.  The relationship we measure is inconsistent with random motions suggesting that the measured velocity along the velocity shift is a function of distance from the BCG (Section \ref{sec:velstruct}).
\item The velocity dispersion maps of the gas show that the linewidth typically peaks towards the centre of the nebula but is otherwise low ($\sim$ 100--200\,km\,s$^{-1}$) in the more extended regions (Section \ref{sec:veldisp}).
\item We show that the total H$\alpha$ luminosity and extent of the line emitting nebula are well correlated.  This coupled with the fact the velocity of the gas appears related to the distance from the core of the BCG suggests that the magnitude of the velocity shift should be related to the H$\alpha$ luminosity.  By assuming that differences in the velocity shift at a given luminosity are caused by inclination we are able to reproduce an inclination distribution consistent with what is expected for a random sample such as this (Section \ref{sec:stats}).
\item By measuring the stellar kinematics from the NaD absorption feature we show that the stars do not have a strong velocity shift, or any obvious velocity structure, like those seen in the gas.  This suggests that the velocity field of the stars is dominated by random motions and is thus decoupled from the gas component (Section \ref{sec:starkin}).
\item The maps of measured line ratio ([NII]/H$\alpha$) show a considerable amount of variation within each nebula.  However, importantly they show no obvious trend or ordered structure suggesting that the source of the excitation is not localised to a specific region of the nebula (Section \ref{sec:excit}).
\item By comparing the global [NII]/H$\alpha$ ratio for each object to the extent of the line emitting nebula we find a loose trend that objects with a smaller extent have higher line ratios. This implies that more compact objects are preferentially excited by more energetic sources suggesting that AGN may be playing some role in the excitation of the gas, at least in the central regions of the BCG (Section \ref{sec:excit}).
\item We perform line ratio diagnostics on the gas to determine the source of the excitation and find that for most objects in our sample it is inconsistent with star formation. We are also unable to find any consistent evidence that shocks are playing a significant role in the excitation (Section \ref{sec:bpt}).
\end{enumerate}

These results allow us to draw several conclusions about the nature of the ionised gas 
within cluster cores. The fact that the majority of objects have a H$\alpha$ morphology 
which suggests the gas is relatively quiescent (Point 1) suggests that the gas is 
typically not highly disturbed within most systems.  Similarly while many objects do 
show gas at high velocities relative to the BCG the structures of the high velocity 
components appear well ordered varying smoothly from blue shifted to red shifted 
across the nebula (Points 4 and 5). If the gas were disturbed then we might expect to 
see evidence of shocks, caused by moving gas colliding with gas moving at different 
velocities, in the velocity dispersion maps.  However, the velocity dispersion maps 
do not show evidence of enhanced line widths away from the central regions (Point 7) 
suggesting that random motions are not significant within the extended nebula.  We 
do see a number of systems with disturbed morphologies but these objects account for 
less than 20\% (13/73) of our sample (Point 1).  These objects also typically show a 
more disordered velocity structure in the extended, low surface brightness gas (e.g. 
Abell~3581, NGC~5044, RXJ~0338+09). We also identify several objects 
with offset emission similar to those studied in \citet{ham12} (Point 2). While the 
nature of these objects clearly suggests they are disturbed they account for a very 
small fraction of the overall population (just 4/73 of our objects).  
While a few of these disturbed systems did show evidence of a possible interaction with 
another cluster member (e.g. RXJ 0338+09) it was not a common factor in all disturbed 
systems. However, we cannot rule out the possibility that the gas was disturbed by 
a previous fly--by interaction with a galaxy that has since left the VIMOS field of view.
Despite these systems the majority of the BCGs in our sample have line emitting nebula 
which are quiescent and ordered 
in both their morphology and kinematics suggesting that the gas is relaxed rather than 
disturbed within the bulk of the population.

The identification of several more objects with offset emission similar to those studied 
in \citet{ham12} (Point 2) suggests that such offsets may be more common than previously 
thought.  It is possible that the multiple velocity components seen in the cores of 
objects like Abell~1664 and NGC~5044 may be a related phenomena but a more detailed 
analysis of these objects is needed to confirm such a link.  If these additional offset 
systems are similar to those presented in \citet{ham12} then it suggests that sloshing 
may be present (albeit at different levels) in all systems and strengthens the 
conclusion that the BCG is not required for cooling gas to condense from the ICM despite 
the strong spatial correlation with the cluster core \citep{per98,san09a,hud10}. The fact 
that the gas does not share the kinematics of the stars (Point 9) also suggests that 
the stellar component of the BCG and the gas that make up the nebula are not linked and 
that the gas need not have cooled at its current location within the BCG. The 
large extent of some of the objects in our sample and the filaments seen may also 
support this (Point 3).  While these structures may be the result of gas which has been 
removed from the BCG (either entrained by radio jets or blown out by winds) the fact that 
we see gas cooling away from the BCG in the offset systems suggests a different 
possibility.  These filaments and extended regions could be the result of gas which 
has cooled from the ICM away from the BCG and is now falling towards it. Our current 
data do not allow us to test this hypothesis but with new instrumentation (MUSE, ALMA 
etc.) it should be possible.

The shift from positive to negative velocities across the brightest regions of the 
H$\alpha$ seen in the majority of the systems in our sample (Point 5) is similar to 
the velocity structure found in Hydra-A \citep{ham14}.  We also note that the velocity 
gradients show evidence of being dependant only on the distance over which they are 
measured 
(Point 6).  Such a velocity structure 
could be the result of a disc like rotation similar to that seen in Hydra-A and our 
estimations of the inclinations from the magnitude of the velocity shift at a given 
luminosity (Point 8) are consistent with the expected inclinations of a random sample 
of discs. This suggests that kpc scale discs may be common in cluster cores however, we 
note that other phenomena can also produce velocity structures 
similar to those seen (bi--polar outflows for example) so a more detailed study is 
needed to confirm the presence of discs.  If discs are common this would provide a 
potential link between the gas cooling on kpc scales and the fuelling of the central 
black hole.  The fact that the stellar and gas kinematics are decoupled (Point 9) seems 
to support this interpretation. If the gas is stable as it appears to be, then it should 
form stars which over the lifetime of the cluster would produce a stellar velocity 
field matching that of the gas.  Thus gas discs such as the one seen in Hydra-A must be 
a transient phenomena and the 
gas must be disturbed or consumed before it can form a sufficient number of 
stars to affect the overall stellar velocity field. The main questions remaining then 
are whether the velocity structure seen in the gas within most BCGs relates to a disc 
and if it does, is the disc jet alignment seen in Hydra-A \citep{ham14} unique to that 
system or common to all BCGs with gas discs. We will address 
these questions in the next paper in this series by presenting a detailed analysis of the 
kinematics of each individual object and comparing to tracers of the AGN outbursts at 
radio and X-ray wavelengths.

From our study of the gas excitation we conclude that a distributed process is the main 
component of the gas ionisation as the line ratio maps show significant variation within 
a given object but no obvious trend from a given point to suggest a localised 
source
(Point 10). Despite this the global line ratios for each object are higher in more 
compact objects (Point 11) which suggests that an energetic source at the centre of the 
emission (most likely an AGN) is contributing to the excitation of the gas but is not the 
dominant component in the most extended regions. Through the use of line ratio 
diagnostics we can rule out star formation as the dominant distributed excitation source 
(Point 12).  The line ratio diagnostics also show no consistent evidence for shocks 
(Point 12) and the narrow linewidths seen in the extended emission suggest that the 
fast shocks needed to excite the gas to the levels seen are not present (Point 7). As such 
we must appeal to particle heating \citep{fer09,fab11} to explain the excitation in the 
extended nebula, though AGN (and likely other processes) still play a role.

This work represents a significant step forward in the study of the interplay between 
the Brightest Cluster Galaxy, the cold gas in the cluster core and the cooling of the 
hot gas in the ICM.  We have shown that despite the high velocity of the ionised gas 
it typically forms simple and ordered spatial and kinematics 
structures.  Such simple structures clearly favour a situation in which the gas has 
cooled from the surrounding medium and has remained undisturbed by interactions with 
other cluster members for some time.